\documentclass[fleqn,usenatbib]{mnras}

\usepackage[T1]{fontenc}

\DeclareRobustCommand{\VAN}[3]{#2}
\let\VANthebibliography\thebibliography
\def\thebibliography{\DeclareRobustCommand{\VAN}[3]{##3}\VANthebibliography}

\usepackage{graphicx}	
\usepackage{amsmath}	
\usepackage{amssymb}	

\usepackage{color}

\usepackage{url}

\hypersetup{draft}

\title[\textit{Kepler} triples with EDV]{Von Zeipel -- Lidov -- Kozai cycles in action: \textit{Kepler} triples with eclipse depth variations: KICs 6964043, 5653126, 5731312 and 8023317}

\author[Borkovits et al.]{
T.~Borkovits$^{1,2,3,4,5}$\thanks{E-mail: borko@electra.bajaobs.hu},
S.~A.~Rappaport$^6$,
S.~Toonen$^7$,
M.~Moe$^8$,
T.~Mitnyan$^{1,2}$,
I.~Cs\'anyi$^1$\\
$^{1}$Baja Astronomical Observatory of University of Szeged, H-6500 Baja, Szegedi \'ut, Kt. 766, Hungary\\
$^2$ ELKH-SZTE Stellar Astrophysics Research Group, H-6500 Baja, Szegedi \'ut, Kt. 766, Hungary\\
$^{3}$Konkoly Observatory, Research Centre for Astronomy and Earth Sciences,  H-1121 Budapest, Konkoly Thege Mikl\'os \'ut 15-17, Hungary\\
$^4$ ELTE Gothard Astrophysical Observatory, H-9700 Szombathely, Szent Imre h. u. 112, Hungary \\
$^5$ MTA-ELTE Exoplanet Research Group, H-9700 Szombathely, Szent Imre h. u. 112, Hungary \\
$^6$ Department of Physics, Kavli Institute for Astrophysics and Space Research, M.I.T., Cambridge, MA 02139, USA\\
$^7$ Anton Pannekoek Institute for Astronomy, University of Amsterdam, 1090 GE Amsterdam, The Netherlands \\
$^8$ University of Arizona, Steward Observatory, 933 N. Cherry Ave., Tucson, AZ 85721, USA 
}
\date{Accepted XXX. Received YYY; in original form ZZZ}

\pubyear{2022}

\begin{document}
\label{firstpage}
\pagerange{\pageref{firstpage}--\pageref{lastpage}}
\maketitle

\begin{abstract}
We report the results of the photodynamical analyses of four compact, tight triple stellar systems, KICs 6964043, 5653126, 5731312, 8023317, based largely on \textit{Kepler} and \textit{TESS} data. All systems display remarkable eclipse timing and eclipse depth variations, the latter implying a non-aligned outer orbit.  Moreover, KIC 6964043 is also a triply eclipsing system. We combined photometry, ETV curves, and archival spectral energy distribution data to obtain the astrophysical parameters of the constituent stars and the orbital elements with substantial precision. KICs 6964043 and 5653126 were found to be nearly flat with mutual inclinations $i_\mathrm{mut}=4\fdg1$ and $12\fdg3$, respectively, while KICs 5731312, 8023317 ($i_\mathrm{mut}=39\fdg4$ and $55\fdg7$, respectively) are found to lie in the high $i_\mathrm{mut}$ regime of the von Zeipel-Kozai-Lidov (ZKL) theorem.  We show that currently both high inclination triples exhibit observable unusual retrograde apsidal motion.  Moreover, the eclipses will disappear in all but one of the four systems within a few decades.  Short-term numerical integrations of the dynamical evolution reveal that both high inclination triples are currently subject to ongoing, large amplitude ($\Delta e\sim0.3$) inner eccentricity variations on centuries-long timescales, in accord with the ZKL theorem. Longer-term integrations predict that two of the four systems may become dynamically unstable on $\sim$ Gyr timescales, while in the other two triples common envelope phases and stellar mergers may occur.  Finally we investigate the dynamical properties of a sample of 71 KIC/TIC triples statistically, and find that the mutual inclinations and outer mass ratios are anti-correlated at the 4$\sigma$ level. We discuss the implications for the formation mechanisms of compact triples.

\end{abstract}

\begin{keywords}
binaries:eclipsing -- binaries:close -- stars:individual: KIC\,6964043 -- stars:individual: KIC\,5653126 -- stars:individual: KIC\,5731312 -- stars:individual: KIC\,8023317
\end{keywords}



\section{Introduction}
\label{sect_intro}

The advent of planet-hunter space telescopes such as \textit{CoRoT} \citep{auvergneetal09}, \textit{Kepler} \citep{boruckietal10}, \textit{TESS} \citep{ricker15} has additionally opened up a new window for the detection of exotic multiple stellar systems.  This especially includes triple systems and quadruples which are both `compact' and `tight'.  For simplicity, considering only hierarchical triple systems, by `compactness' we mean that the outer orbital period does not exceed, let's say $P_\mathrm{out}=1000$\,days, while a triple system is said to be `tight' for an outer to inner period ratio $P_\mathrm{out}/P_\mathrm{in}\leq100$.\footnote{Note, however, that these numerical values were chosen somewhat arbitrarily, and we do not mean to attempt to impose strict limits.}  Tightness is important because the magnitude of the gravitational three-body perturbations in a hierarchical triple stellar system, relative to the Keplerian motion, is primarily determined by the ratio of semi-major axes of the inner and outer orbits.  This ratio, with the use of Kepler's third law, can easily be converted to the ratio of the much easier-to-observe orbital periods.  On the other hand, compactness, which is closely connected to the physical dimensions of the system, strongly influences the timescale of three- (multiple-) body perturbations. Moreover, tightness is also important in determining the relations of the magnitudes of other kinds of perturbations, e.g., tidal and relativistic, to the dynamical ones, since these other perturbations primarily scale with the semi-major axis, i.e., the compactness of the system.  Finally in this regard, compactness is also an important parameter because, according to our present knowledge, compact multiple systems form and evolve in a different manner than their wider counterparts (\citealp[see, e.g.][in the sense of their formation]{tokovinin21}, and \citealp[][in regard of their later evolution]{toonenetal20,toonenetal22}).

Hierarchical triple systems exhibit periodic gravitational perturbations in three different, well separable timescales. (i) The shortest period ones act on the timescale of the inner orbit ($P_\mathrm{in}$), and have a relative amplitude on the order of the ratio $(P_\mathrm{in}/P_\mathrm{out})^2$ and thus, this class of perturbations is hardly observable. (ii) The medium timescale class of periodic perturbations have characteristic periods on the order of the period of the outer orbit ($P_\mathrm{out}$) and relative magnitude of $P_\mathrm{in}/P_\mathrm{out}$ (i.e. longer in period and larger in amplitude than the short-period ones by a factor of $P_\mathrm{out}/P_\mathrm{in}$).  (iii) Finally, the long-period (or, according to another nomenclature, the `apse-node' type) perturbations have a timescale related to $P_\mathrm{out}^2/P_\mathrm{in}$ and relative amplitude of unity\footnote{Unity is to be meant in that sense, that during the timescale of the long-period perturbations, the numerical value of a given orbital element may take any numerical value within its physical range.} (i.e., yet longer/larger than the previous perturbations, with a second factor of $P_\mathrm{out}/P_\mathrm{in}$.)

In terms of the formation, evolution, and future fate and stability of hierarchical stellar systems, the largest amplitude, long-period perturbations are by far the most important.  It is not surprising, therefore, that theoretical studies of the stellar third-body problem have focused almost exclusively on long-period perturbations \citep[see, e.g, the recent review of][and further references therein]{naoz16}.

Perhaps the most spectacular consequence of these long-period perturbations occurs in the case of highly inclined inner and outer orbital planes.  The eccentricity of the inner orbit may then alternate periodically between small values (including zero) and nearly unity and, in parallel with this, the mutual inclination of the orbits also varies substantially in an anticorrelated manner. In recent years these effects are frequently referred as `von Zeipel-Lidov-Kozai effects' (hereafter ZLK) or `ZLK oscillations or, cycles' in the honour of its first investigators \citep{vonzeipel10,lidov62,kozai62}.\footnote{These early authors did not deal with the hierarchical three-body problem in the context of stellar triple systems, but, instead studied the motions of small bodies in triple systems (e.g. perturbations of Jupiter from comets and asteroids orbiting around the Sun -- von Zeipel; Kozai; or perturbations of the Moon for Earth-orbiting artificial satellites -- Lidov; \citealp[see the recent monograph of][]{itoohtsuka19}). The first analytic works on this theorem, dedicated directly to the stellar three body problem was published by \citet{harrington68} and \citet{soderhjelm82}.} 

The original ZLK theorem takes into account only the lowest order quadrupole, long-period perturbations, and is strictly valid only if the orbital angular momentum of the triple system is stored exclusively in the outer orbit, which is also circular.  This asymptotic quadrupole ZKL theorem has only one degree of freedom and, hence, is integrable; therefore, the variations of the inner eccentricity ($e_\mathrm{in}$), dynamical argument of periastron ($\omega_\mathrm{in}^\mathrm{dyn}$), and mutual inclination $(i_\mathrm{mut})$ can be given in closed form.  Based on the analytic form of the perturbation equations, it can readily be shown that the behaviour of the ZKL cycles may be essentially divided into two mutual inclination regimes which are separated by the value of $|\cos i_\mathrm{mut}|=\sqrt{3/5}$, i.e. $i_\mathrm{mut}\approx39\fdg23$ and its retrograde counterpart\footnote{Within the framework of the original ZKL theorem the prograde and retrograde configurations are equivalent.}.  Below this mutual inclination (i.e. in the low mutual inclination regime) there may occur only small-amplitude variations in the inner eccentricity and the mutual inclination as well. Furthermore, in the case of an initially circular inner orbit, it remains circular at all times (of course, only insofar as the approximation used remains valid).  

In the high mutual inclination case, depending on the initial conditions, (i.e., the value of the inner binary eccentricity, mutual inclination and dynamical argument of pericenter at a given instant), the inner eccentricity and the mutual inclination may oscillate with large amplitudes, while the apsidal line may exhibit either circulation or libration.  As an extreme example, an originally circular inner orbit, perturbed by a tertiary orbiting perpendicular to its orbit (i.e. $i_\mathrm{mut}=90\degr$), may have its eccentricity periodically grow to near unity and then shrink back to zero while, at the same time, the mutual inclination oscillates between $\cos^2i_\mathrm{mut}=0$ and $\cos^2i_\mathrm{mut}=3/5$. Taking into consideration the further octupole terms of the perturbation function, and allowing the third body to have any arbitrary eccentricity, e.g., \citet{fordetal00} and \citet{naozetal13} have shown that the problem will no longer be analytically integrable, and the behaviour of the orbital elements may be more complex. For example, large eccentricity variations may already occur in the case of lower mutual inclination triples, the relative orientations of the orbital planes may show flip-flop phenomena between prograde and retrograde orientations, the semi-major axis of the inner orbit may alternate between libration and circulation, and so forth.

Although the large amplitude ZLK eccentricity cycles are evidently the most important long-period, or secular, effects in the formation and evolution of triple star systems, these are often barely observable in action. The main reason is that, due to their typically long timescale, one cannot expect measurable eccentricity variations from spectroscopic, photometric or, astrometric observations\footnote{At least within the accuracy limits of instruments that are commonly available for observing hierarchical triple stellar systems.} on timescales of years or decades.  Fortunately, however, in the case of some other orbital elements, the variations driven by the very same mechanism (i.e., third-body perturbations) may lead to readily observable effects even on `astronomer career'  timescales. These 
long-period perturbations are (i) the third-body forced apsidal motion (in the case of an eccentric inner binary), and (ii) the orbital plane precession (or, nodal regression) in those triple systems where the inner and outer orbital planes are not aligned. (Hence the other name of `apse-node timescale perturbations'.) 

Eclipsing binaries (EBs) that belong to tight and compact hierarchical triple and multiple stellar systems may be ideal for the detection of both of these effects.  In particular, they can provide important information on the non-alignment of the orbital planes, or, oppositely, the flatness of the given systems. This is the case because the orbital plane precession, in the case of an EB, leads to a variation in the eclipse depths and, even to the disappearance of former eclipses or, in the opposite sense, the occurrence of eclipses in formerly non-eclipsing binaries. The characteristic time-scale of this effect for triples can be estimated by the following simple expression \citep{soderhjelm75}:
\begin{equation}
P_\mathrm{node}=\frac{4}{3}\frac{1+q_\mathrm{out}}{q_\mathrm{out}}\frac{P_\mathrm{out}^2}{P_\mathrm{in}}\left(1-e_\mathrm{out}^2\right)^{3/2}\left|\frac{C}{G_\mathrm{out}}\cos i_\mathrm{mut}\right|^{-1},
\label{Eq:Pnode}
\end{equation}
where $q_\mathrm{out}$ represents the outer mass ratio, $P_\mathrm{in,out}$ denote the periods of the inner and outer orbits, $e_\mathrm{out}$ stands for the outer eccentricity, $C$ is the (constant) total orbital angular momentum of the triple, while $G_\mathrm{out}$ stands for the orbital angular momentum stored in the outer orbit and, finally, $i_\mathrm{mut}$ the mutual (relative) inclination of the two orbital planes. 

From an observational point of view, however, the problem is that, although the stability of a hierarchical triple stellar system allows for very tight configurations with $P_\mathrm{out}/P_\mathrm{in} \approx 5$ \citep[see, e.g.][]{mardlingaarseth01}, in the vast majority of known triple or multiple systems that contain an EB, this ratio is larger than 100. Moreover, because the periods of the outer stellar components are typically longer than 1 year, the timescale of the precession, and therefore the variation of the eclipse depths, may reach several centuries or, millennia, in most of the systems. Therefore, one might expect eclipse depth variations (EDVs) that occur rapidly enough to be observable during the $\approx1.5$ century-long history of EB observations, only in the most compact, tightest, and relatively rare \citep{tokovinin14} triple systems. 

Consequently, it is hardly surprising that before the era of the long-duration, ground-based and space-borne photometric surveys, EDVs were (serendipitously) detected in only fewer than a dozen EBs \citep[see the recent review of][]{borkovits22}. In contrast to this, \citet{kirketal16} listed 43 EBs in the original \textit{Kepler} sample which exhibit EDVs. Though some of their systems were probably false positives (for a number of different reasons), on the other hand, they did not list some EBs with evident EDVs (e.g., neither KIC 6964043, nor KIC 5731312, two of our four systems that are analysed in this paper). Therefore, we may confidently state that the number of the EBs with known EDVs had more than tripled by the end of the four-year-long primary mission of the \textit{Kepler} space telescope. The majority of these EDV-exhibiting EBs in the \textit{Kepler}-field also display rapid apsidal motion.  Moreover, they have readily detectable medium-period eclipse timing variations (ETV) which are, again, forced by the dynamical effects (DE) of the third body (besides, and instead of, the usual, well-known light-travel time effect -- LTTE, \citealp{rappaportetal13,borkovitsetal15,borkovitsetal16}).  Therefore, these systems offer excellent opportunities to determine their complete geometric and dynamical configurations with unexpectedly high precision through complex, photodynamical analyses. In this manner, these triple stellar systems may serve as actual case studies for later, in-depth investigations of the formation, as well as past, present and future evolutions of hierarchical triple star systems.

In Sect.~\ref{sec:targets} we introduce the four triples selected for our present investigations, and we discuss the prior results on these systems that are relevant in the context of the present study. Then the description of the observational data and their preparation can be found in Sect.~\ref{sec:Obsdata}, while the applied photodynamical method is summarized in Sect.~\ref{sec:dyn_mod}.  In Sect.~\ref{sec:discussion} we discuss the results of the photodynamical analyses, and study the future dynamical and astrophysical evolutions of each system based on these results. Moreover, the implications of our present and former findings about compact triple star systems for the formation processes of multiple star systems are also discussed on a statistical basis. Finally, general concluding remarks are made in Sect.~\ref{sec:conclusions}.

\begin{table*}
\centering
\caption{Main properties of the three systems from different catalogs}
\begin{tabular}{lcccc}
\hline
\hline
Parameter & 6964043 & 5653126 & 5731312 & 8023317 \\
\hline
RA (J2000) & $19:44:03.474$ & $19:58:48.479$ & $19:53:00.434$ & $19:19:52.879$ \\  
Dec (J2000)& $+42:25:20.09$ & $+40:53:46.51$ & $+40:54:12.77$ & $+43:49:13.84$ \\  
$G^b$ & $15.5868\pm0.0003$  & $13.1434\pm0.0001$ & $13.8246\pm0.0001$ & $12.8453\pm0.0002$ \\
$G_{\rm BP}^b$ & $16.0472\pm0.0029$ & $13.5364\pm0.0007$ & $14.3641\pm0.0008$ & $13.2064\pm0.0005$ \\
$G_{\rm RP}^b$ & $14.9591\pm0.0014$ & $12.5802\pm0.0004$ & $13.1272\pm0.0005$ & $12.3194\pm0.0005$ \\
B$^a$ & $16.307 \pm 0.068$ & $14.159\pm0.026$ & $15.030\pm0.066$ & $13.819\pm0.036$ \\
V$^a$ & $16.007 \pm 0.137$ & $13.225\pm0.069$ & $14.035\pm0.103$ & $12.990\pm0.069$ \\
g$'$  & $16.007\pm0.071^c$ & $13.709\pm0.014^d$ & $14.584\pm0.001^e$ & $13.329\pm0.031^d$ \\
r$'$  & $15.533\pm0.048^c$ & $13.139\pm0.035^d$ & $13.659\pm0.001^e$ & $12.810\pm0.045^d$ \\
i$'$  & $15.276\pm0.083^c$ & $13.028\pm0.093^d$ & $13.069\pm0.001^e$ & $12.702\pm0.028^d$ \\
J$^f$ & $14.231 \pm 0.027$ & $11.957\pm0.021$ & $12.242\pm0.022$ & $11.682\pm0.023$ \\
H$^f$ & $13.758 \pm 0.026$ & $11.710\pm0.022$ & $11.729\pm0.023$ & $11.395\pm0.020$ \\
K$^f$ & $13.705 \pm 0.043$ & $11.632\pm0.020$ & $11.598\pm0.018$ & $11.340\pm0.011$ \\
W1$^g$ & $13.658\pm0.025$  & $11.543\pm0.023$ & $11.559\pm0.024$ & $11.310\pm0.023$ \\
W2$^g$ & $13.680\pm0.031$  & $11.572\pm0.021$ & $11.611\pm0.023$ & $11.359\pm0.021$ \\
W3$^g$ & $13.061$          & $11.595\pm0.122$ & $12.926        $ & $11.163\pm0.094$ \\
W4$^g$ & $ 9.614$          & $8.897$        & $9.458$          & $9.442$ \\
$T_{\rm eff}$ (K)$^a$ & $5445\pm122$ & $6980\pm118$ & $4885\pm60$ & $5638\pm103$ \\
Distance (pc)$^h$ & $ 2162\pm142 $ & $1296\pm32$ & $351\pm3$ & $753\pm7$ \\ 
$[M/H]^a$ & $-$ & $0.435\pm0.111$ & $-0.372\pm0.057$ & $0.224\pm0.016$ \\ 
$E(B-V)^a$ & $0.130 \pm 0.008$ & $0.213\pm0.014$ & $0.073\pm0.024$ & $0.035\pm0.002$ \\
$\mu_\alpha$ (mas ~${\rm yr}^{-1}$)$^b$ & $-2.95\pm0.03$ & $0.38\pm0.02$ & $1.69\pm0.03$ & $-4.25\pm0.01$ \\ 
$\mu_\delta$ (mas ~${\rm yr}^{-1}$)$^b$ & $-1.69\pm0.04$ & $-1.12\pm0.02$ & $5.58\pm0.03$ & $-1.09\pm0.01$ \\ 
\hline
\label{tbl:mags}  
\end{tabular}

\textit{Notes.}  (a) TESS Input Catalog (TIC v8.2) \citep{TIC8}. (b) Gaia EDR3 \citep{GaiaEDR3}. (c) AAVSO Photometric All Sky Survey (APASS) DR10, \citep{APASS10}, \url{https://www.aavso.org/download-apass-data}. (d) AAVSO Photometric All Sky Survey (APASS) DR9, \citep{APASS}, \url{http://vizier.u-strasbg.fr/viz-bin/VizieR?-source=II/336/apass9}. (e) The Kepler-INT survey \citep{kisdr2}. (f) 2MASS catalog \citep{2MASS}.  (g) WISE point source catalog \citep{WISE}. (h) \citet{bailer-jonesetal21}. \\
Note also, that for the SED analysis in Sect.~\ref{sec:dyn_mod} the uncertainties of the passband magnitudes were set to $\sigma_\mathrm{mag}=\mathrm{max}(\sigma_\mathrm{catalog},0.030)$ to avoid the strong overdominance of the extremely accurate Gaia magnitudes over the other measurements.
\end{table*}

\section{Selected systems}
\label{sec:targets}

\begin{figure*}
\includegraphics[width=0.5\textwidth]{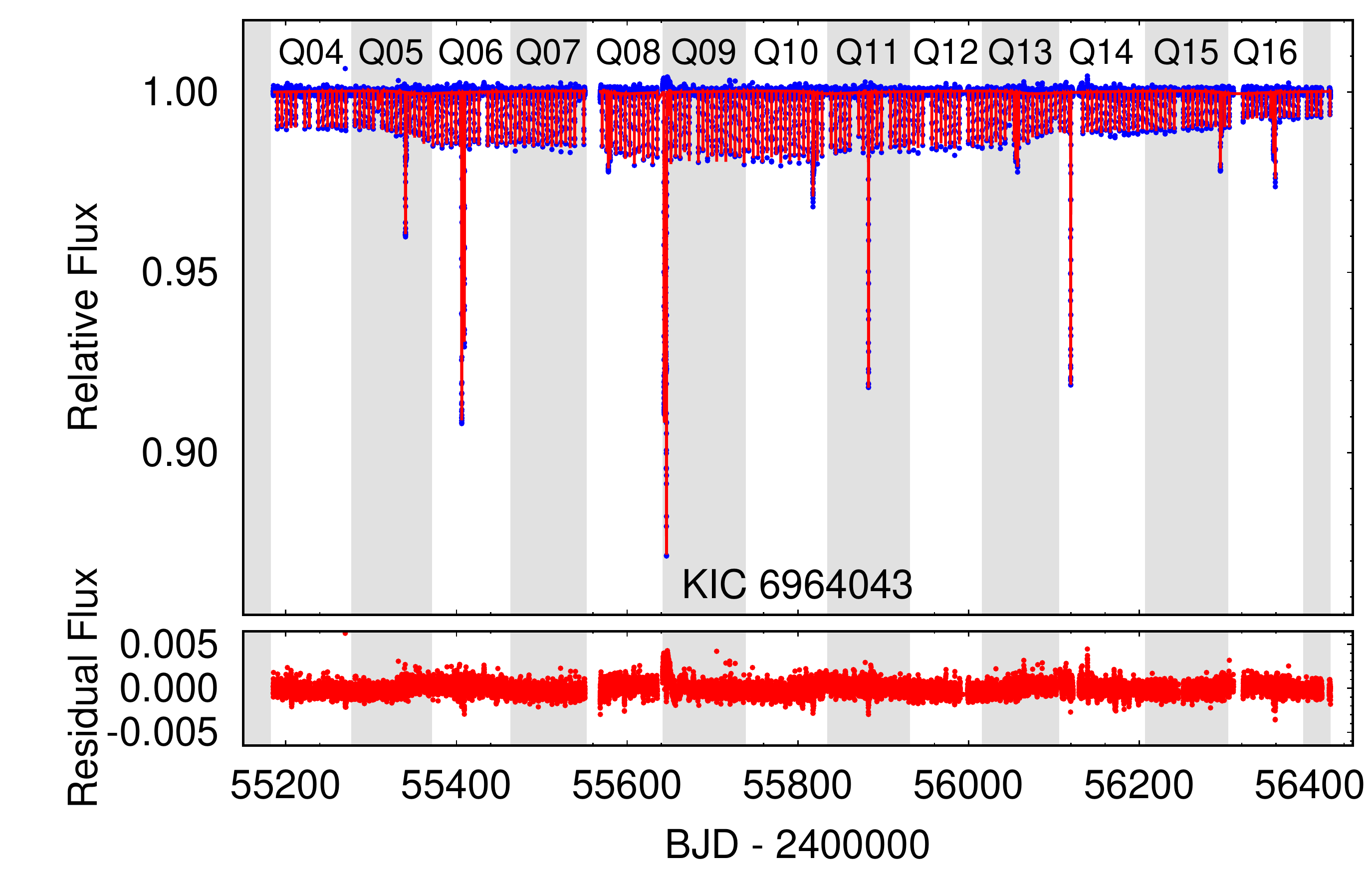}\includegraphics[width=0.5\textwidth]{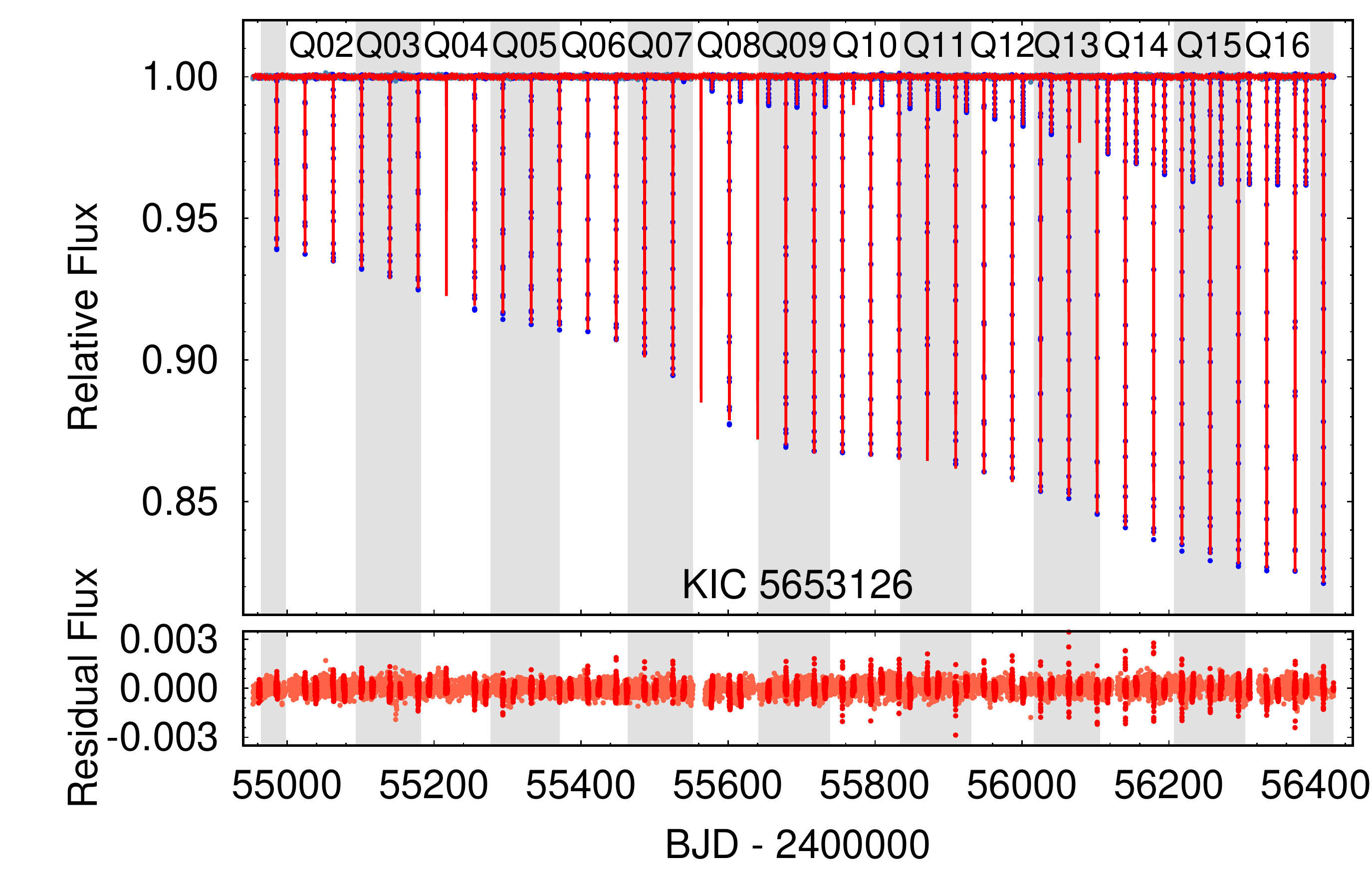}
\includegraphics[width=0.5\textwidth]{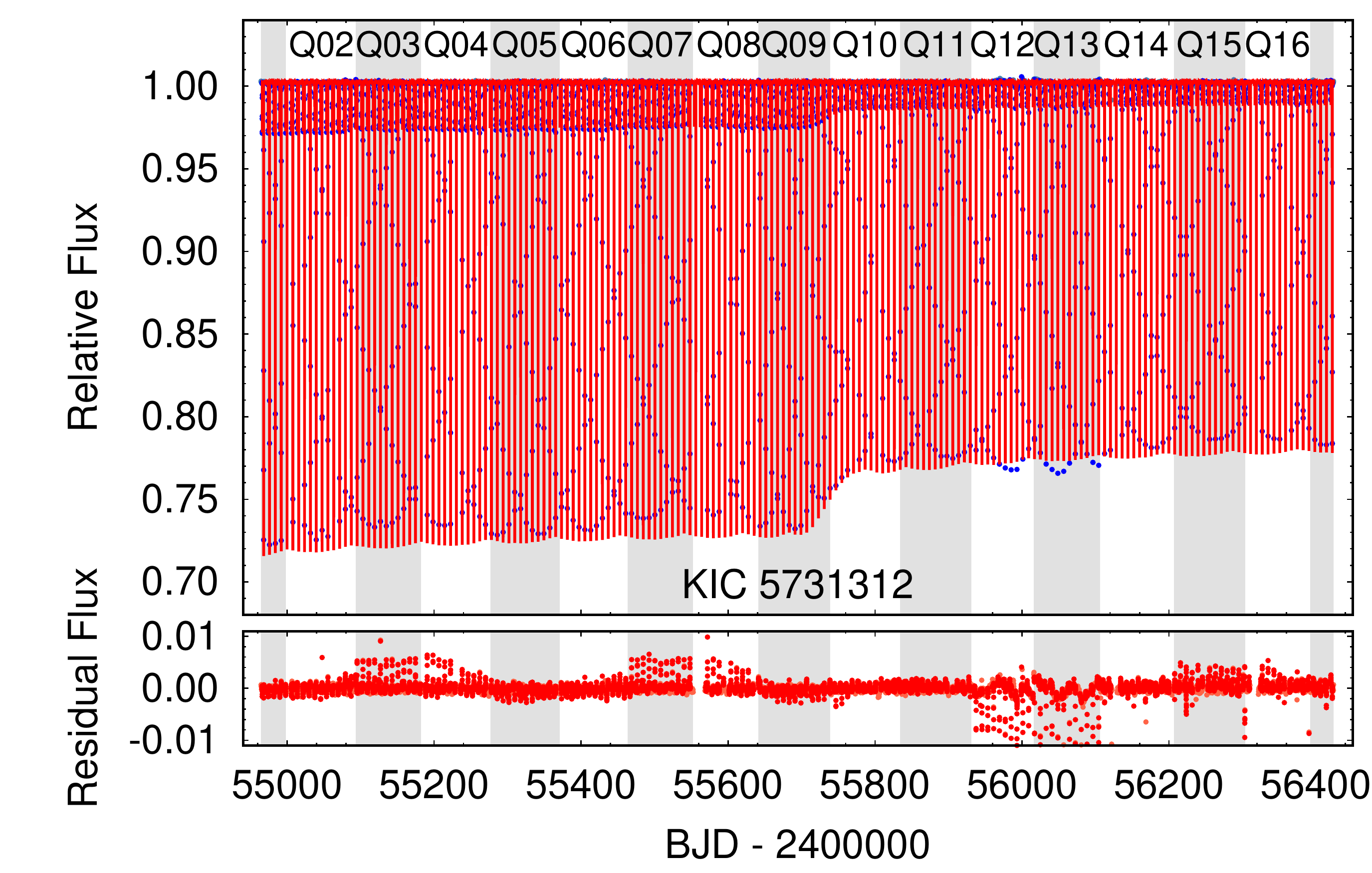}\includegraphics[width=0.5\textwidth]{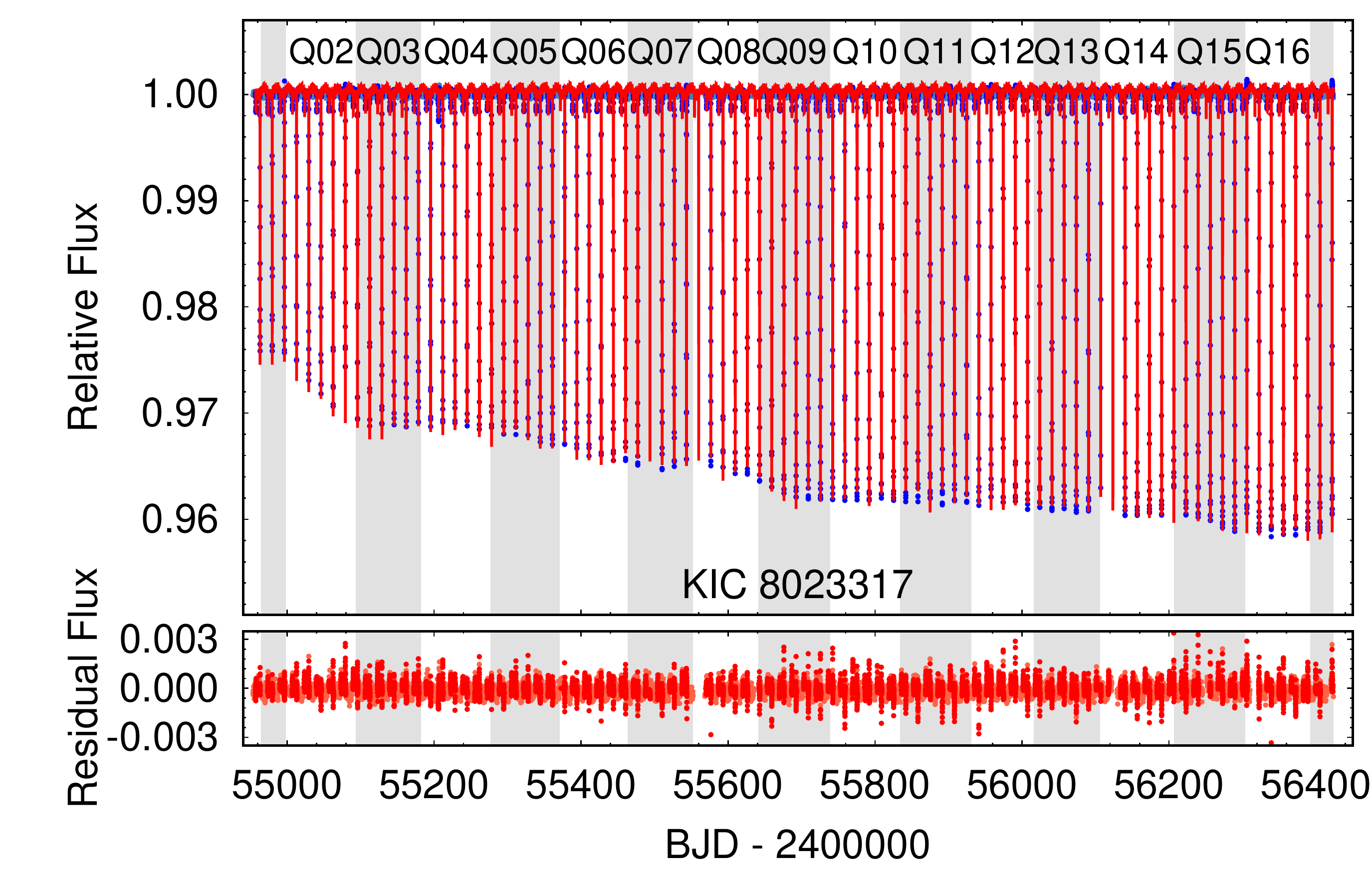}  
 \caption{Four-year long \textit{Kepler} long cadence lightcurves of the four investigated systems with the photodynamical model solution. Small blue circles are the data points while the continuous red curves are the photodynamical models. Note, the alternating gray-white stripes help to distinguish the observing quarters.}
\label{fig:keplerlcs}
\end{figure*}

In a previous work \citet{borkovitsetal15} analysed 26 such eccentric EBs in the original \textit{Kepler}-field for which perturbations by a close tertiary star dominated the eclipse timing variations (ETV). Amongst these triples, 17 were found to exhibit EDVs to differing degrees, indicating that the tertiary star has an inclined orbit relative to the plane of the inner binary.  We have selected four targets of these 17 systems for a detailed analysis. According to the previous analytic ETV analysis of \citet{borkovitsetal15}, two of the four systems belong to the low mutual inclination regime and, hence, one can expect rapid, but small amplitude medium and long period variations in the orbital elements.  By contrast the other two systems belong to the rare class of highly inclined close triples, with $i_\mathrm{mut}$ close to the inclination limit of the asymptotic, quadrupole ZKL theory.  Therefore, they may be ideal for direct observational detections of such exotic perturbation effects, as, e.g. rapid, strong eccentricity variations and/or retrograde (or, librating) apsidal motion. The main cataloged parameters of the systems are tabulated in Table~\ref{tbl:mags}, while their four-year-long \textit{Kepler} lightcurves are plotted in Fig.~\ref{fig:keplerlcs}. Below we summarize the main properties found from their prior results for these four systems individually. Because we are mainly interested in the consequences of the inclined inner and outer orbital planes, we list the systems throughout this paper in the order of increasing value of the mutual inclination ($i_\mathrm{mut}$) parameter as it is obtained from the present study.

\begin{figure*}
\includegraphics[width=0.33\textwidth]{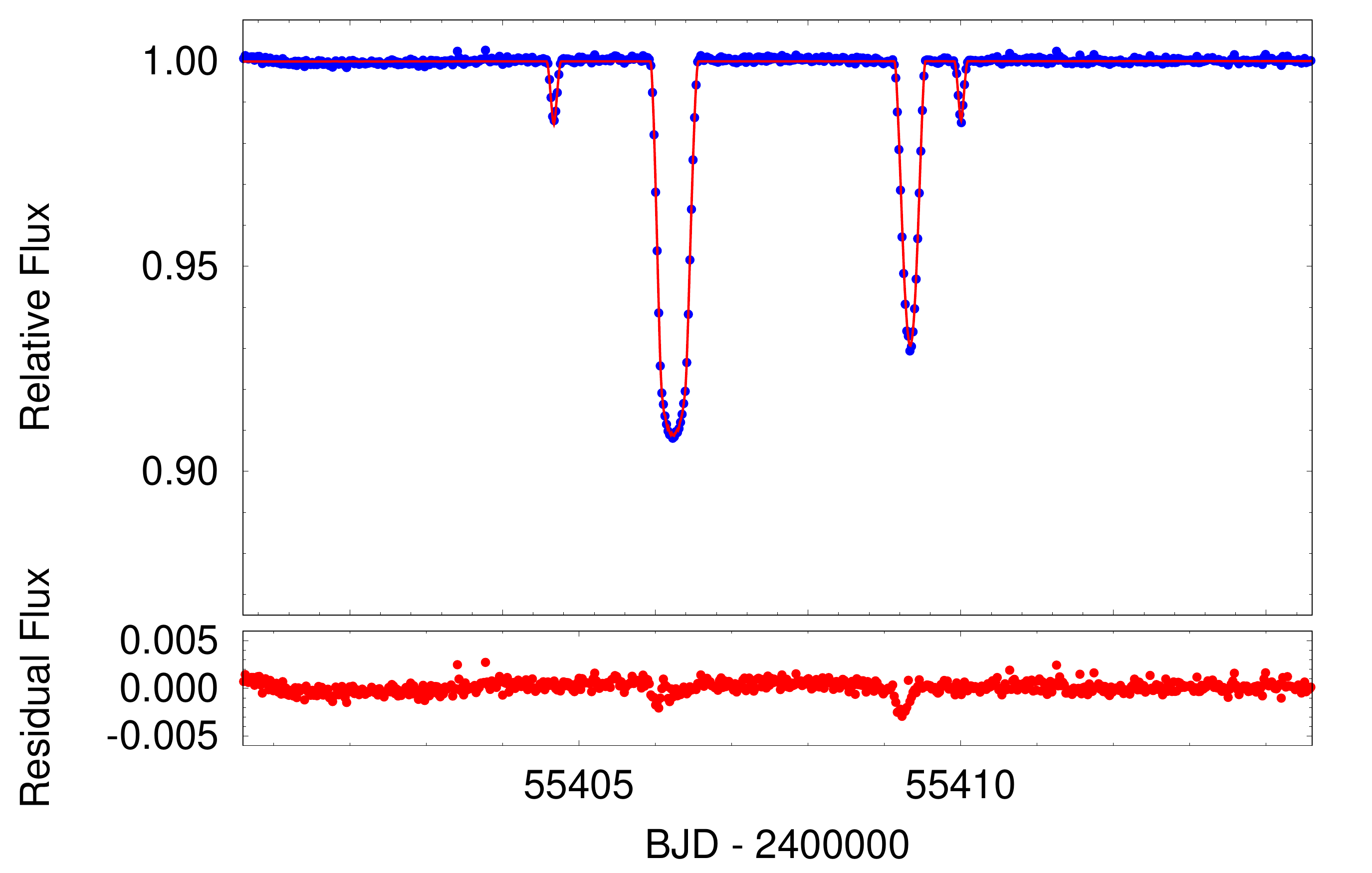}\includegraphics[width=0.33\textwidth]{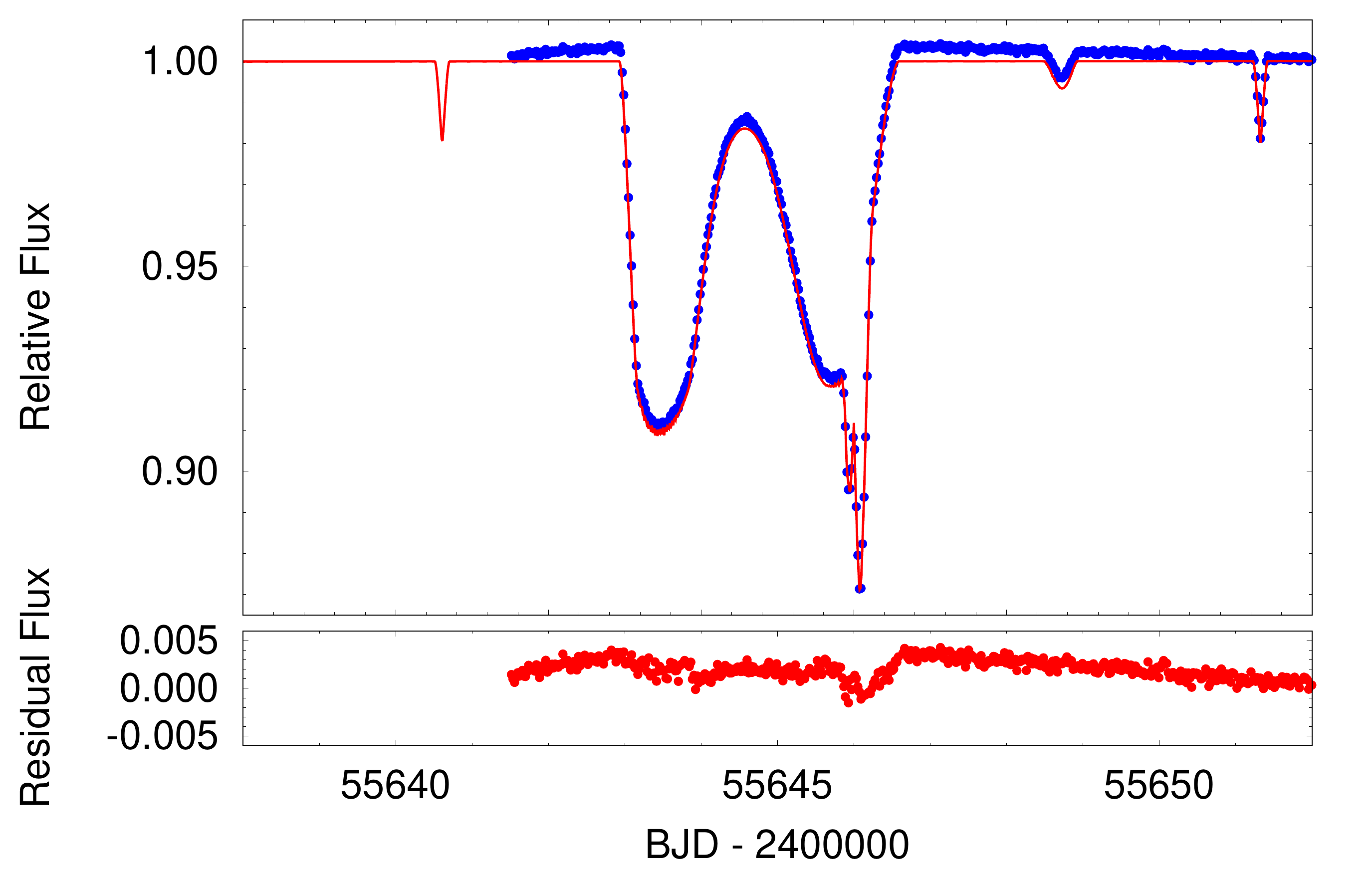}\includegraphics[width=0.33\textwidth]{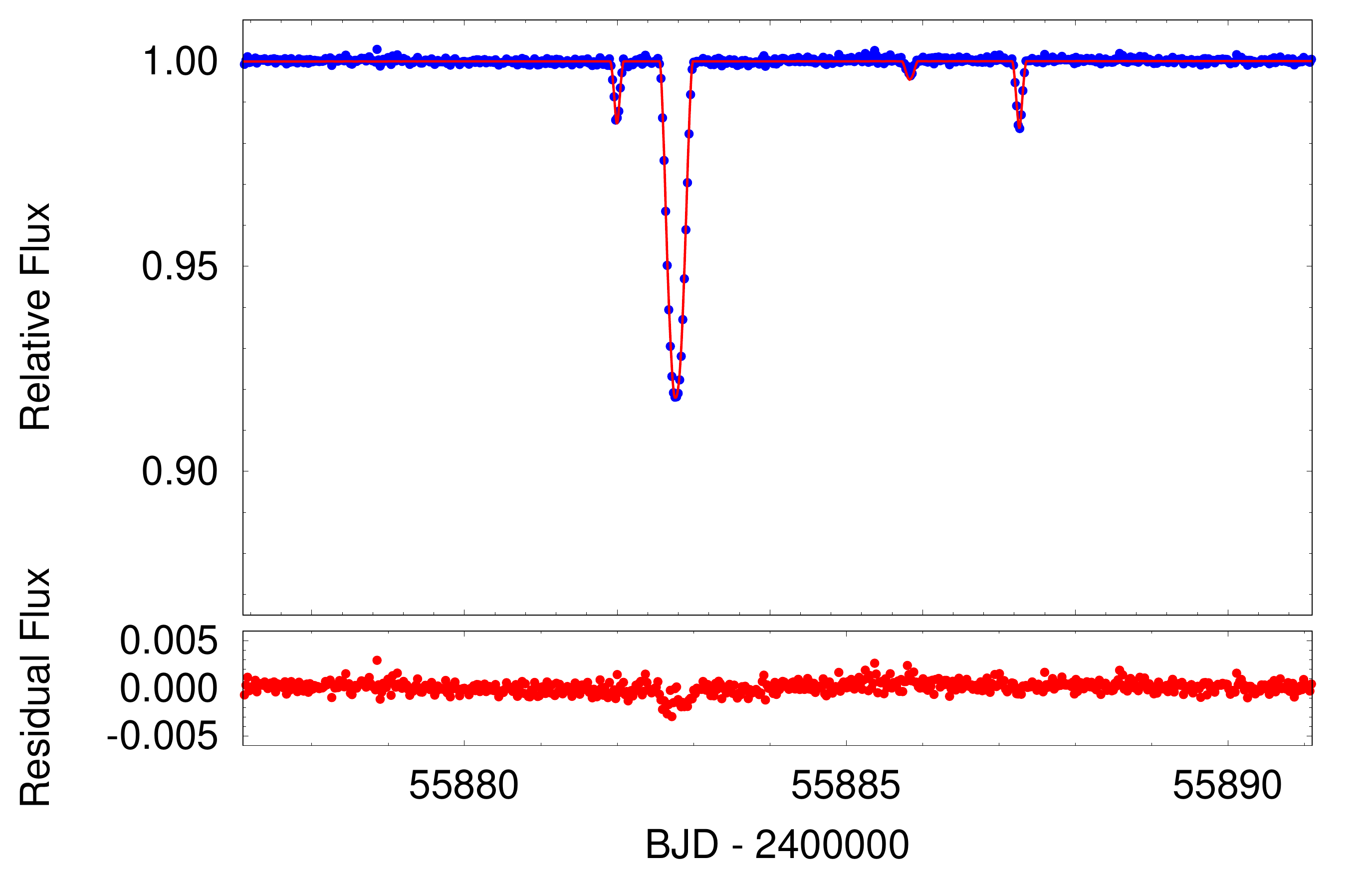} 
\includegraphics[width=0.33\textwidth]{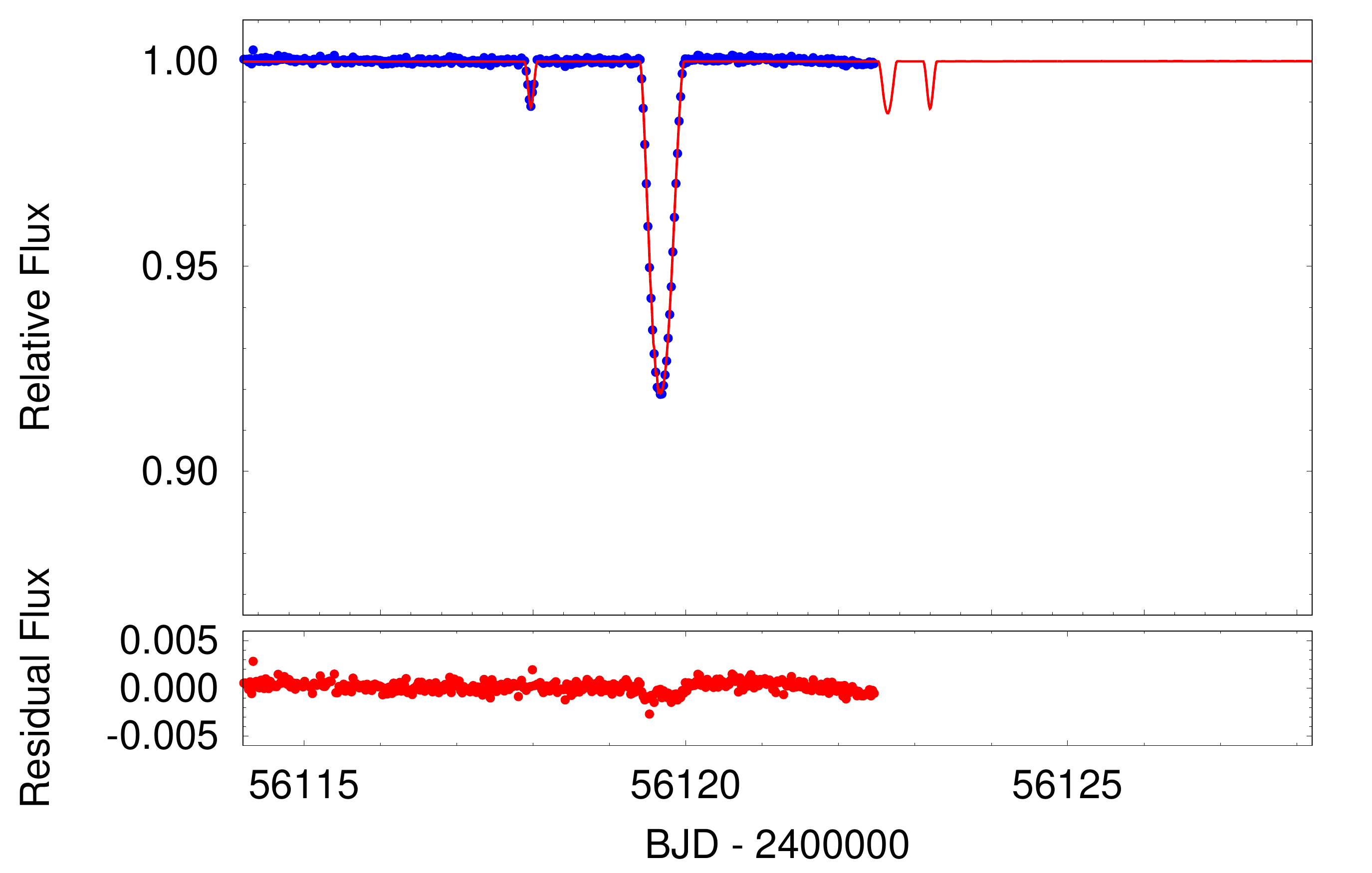}\includegraphics[width=0.33\textwidth]{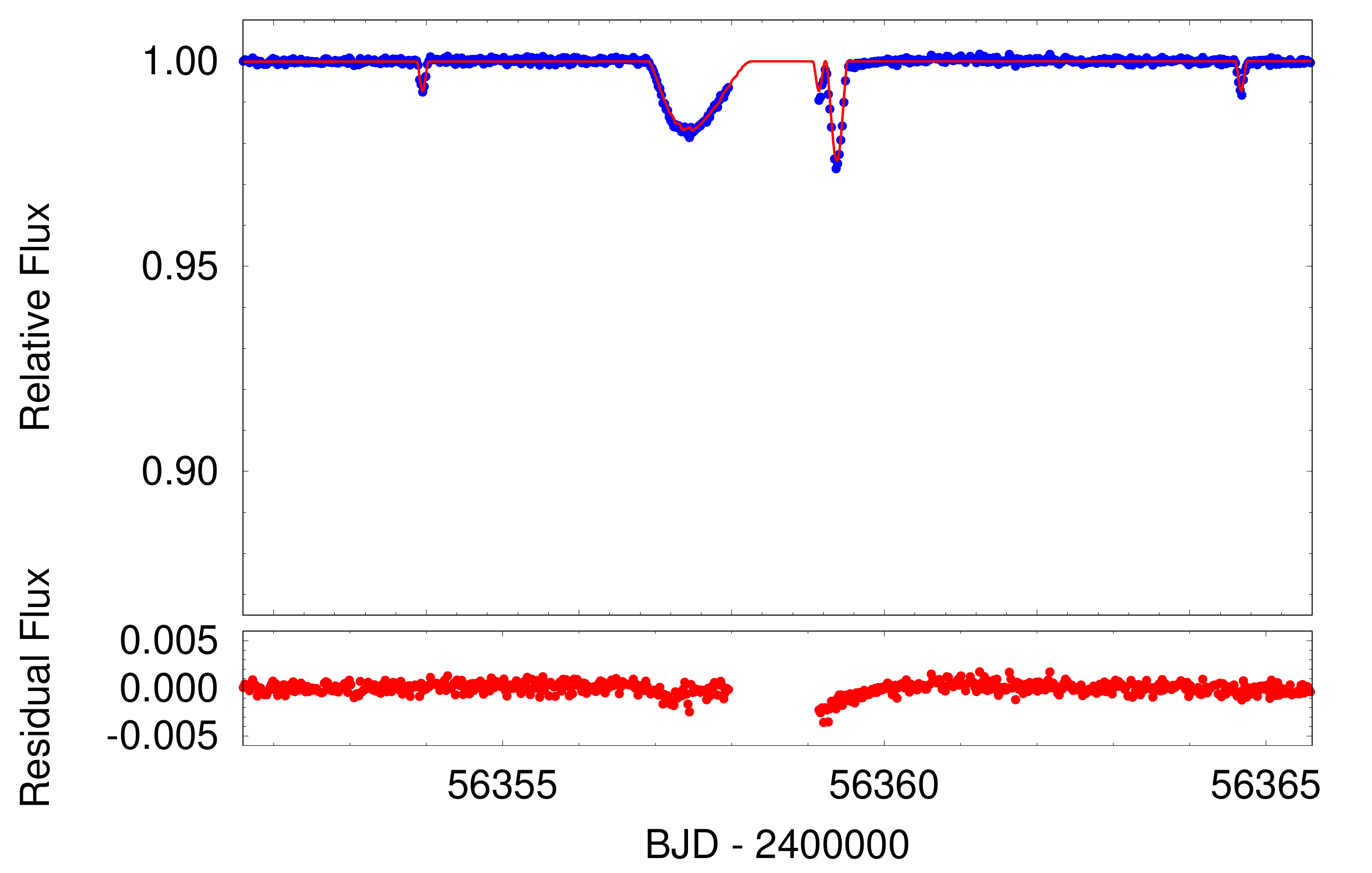}  
 \caption{The `primary' third-body, or outer eclipses of KIC~6964043 observed with \textit{Kepler}. During these events the two red dwarfs of the inner binary transit (either partially or, totally) in front of the disk of the outer, most massive, G-star component. Blue dots represent the observed LC fluxes, while red lines represent the best-fitting photodynamical model.}
\label{fig:K6964043E3primaries}
\end{figure*}

\begin{figure*}
\includegraphics[width=0.33\textwidth]{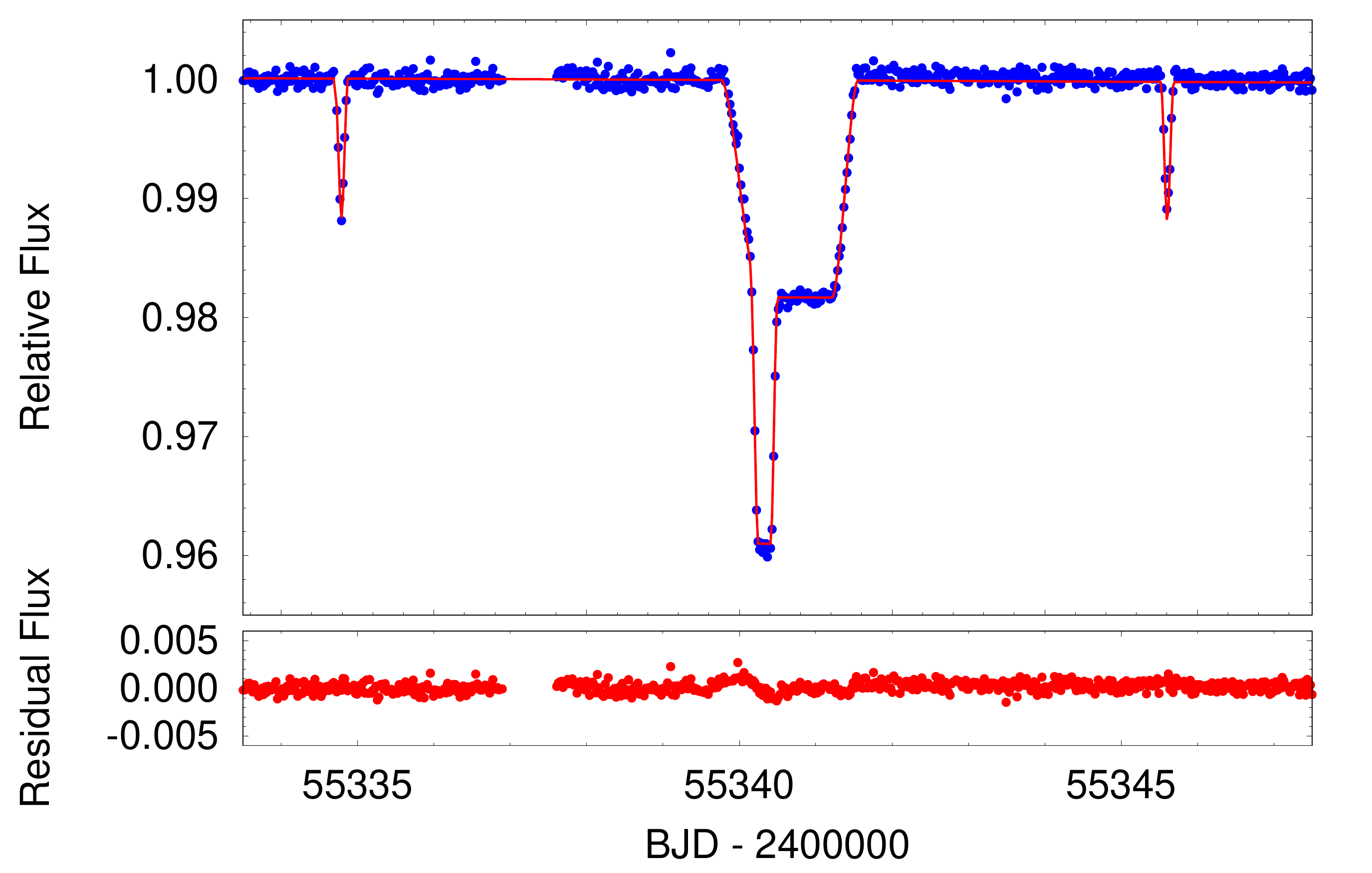}\includegraphics[width=0.33\textwidth]{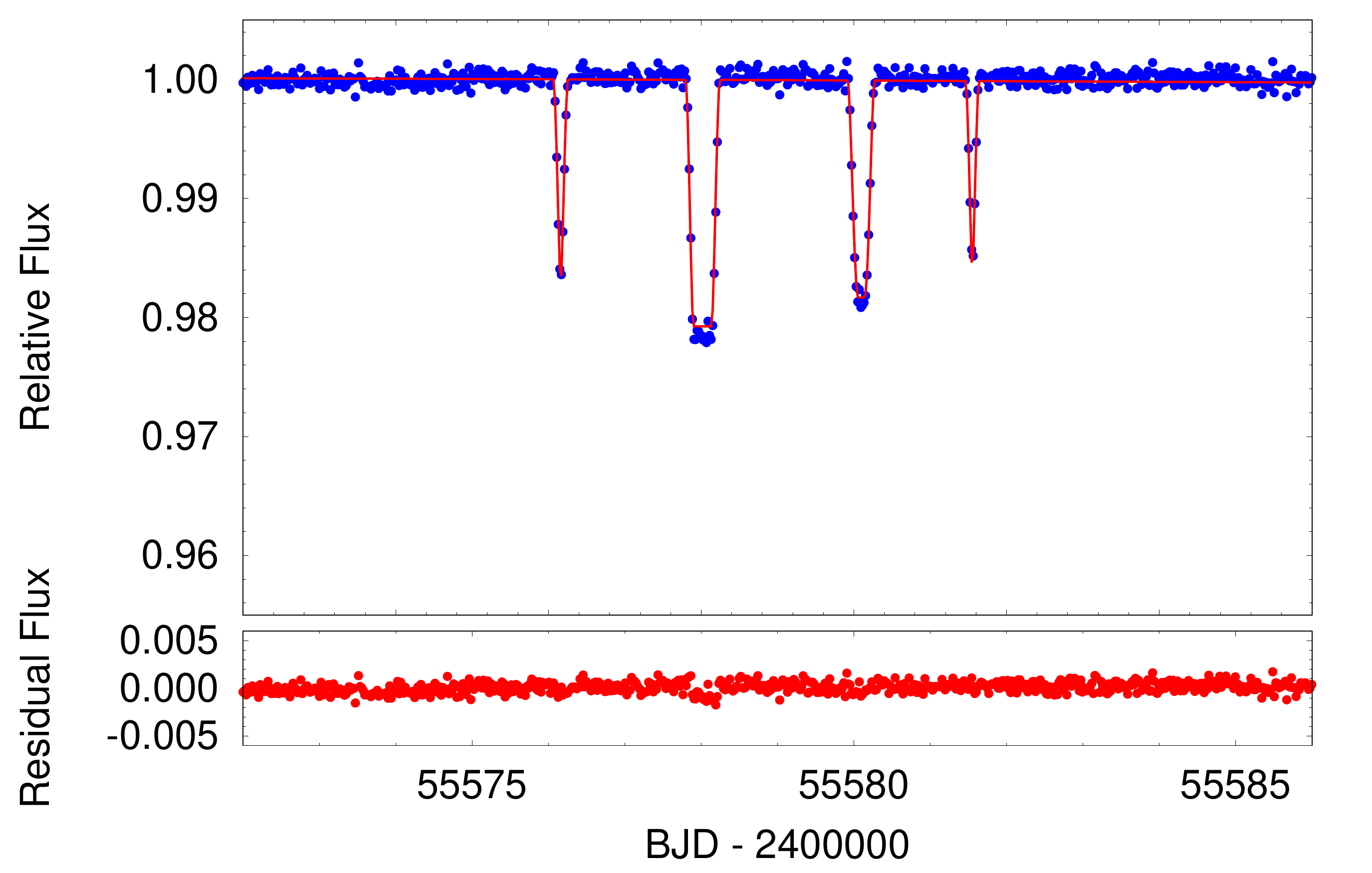}\includegraphics[width=0.33\textwidth]{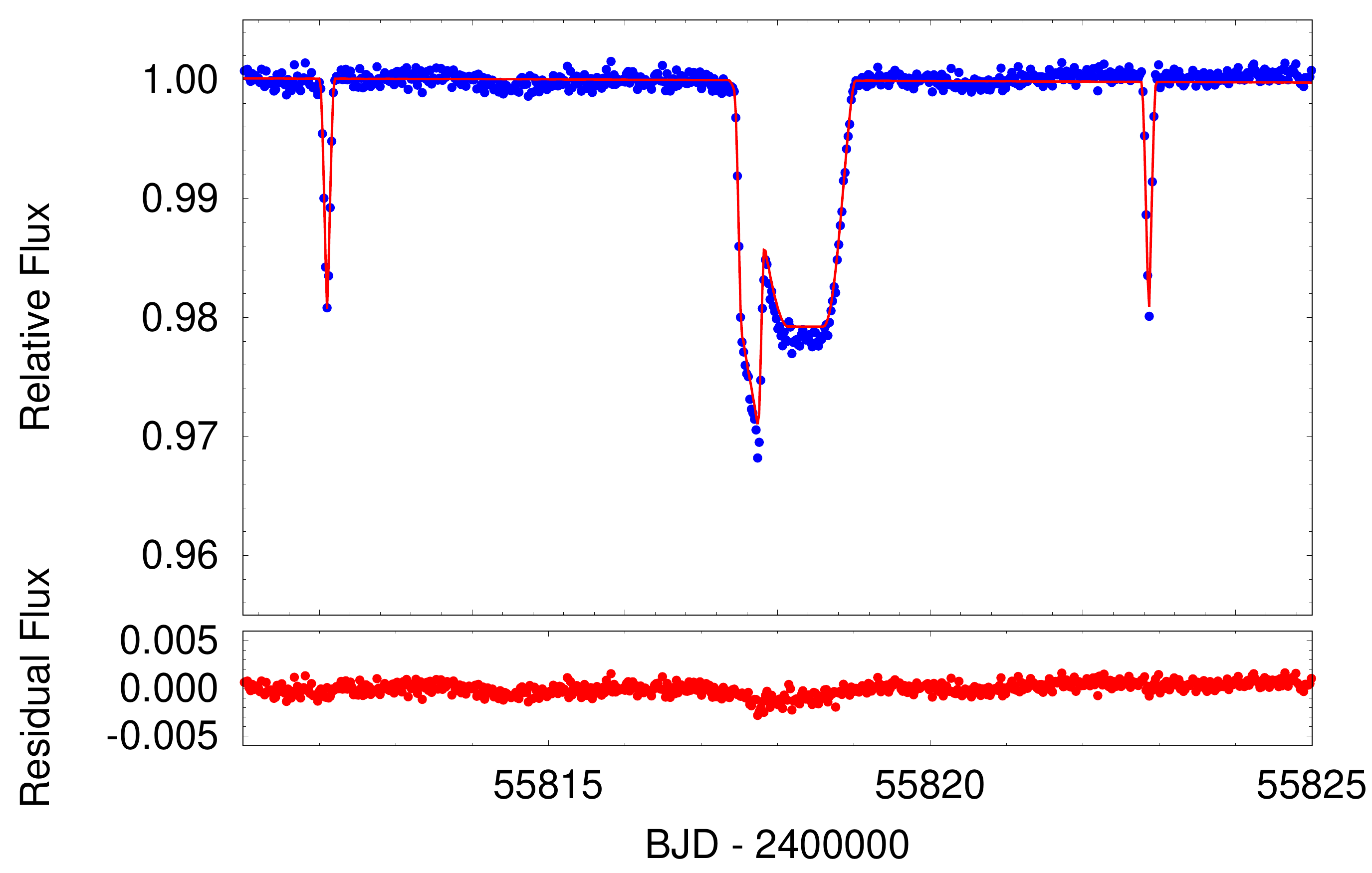} 
\includegraphics[width=0.33\textwidth]{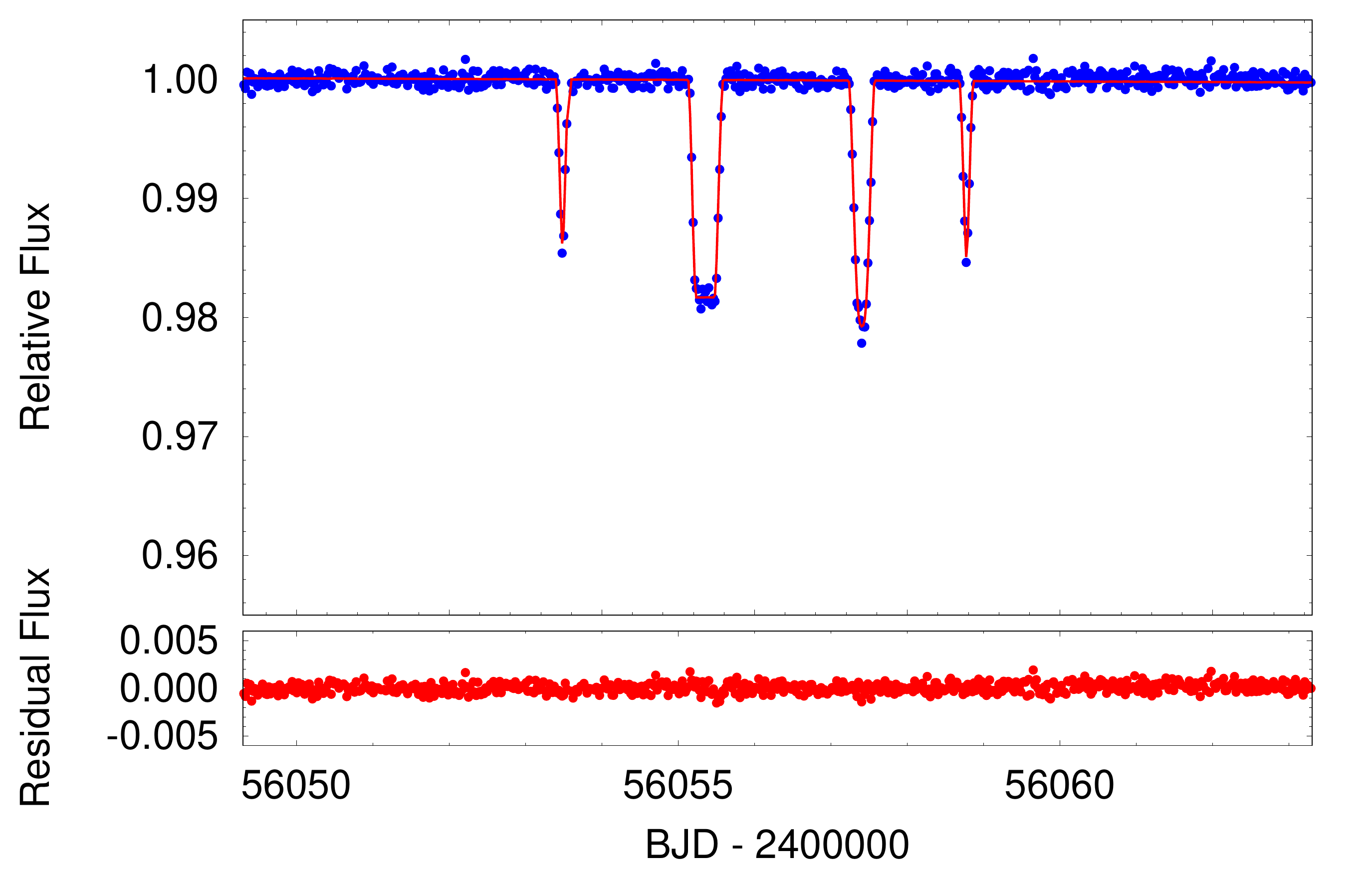}\includegraphics[width=0.33\textwidth]{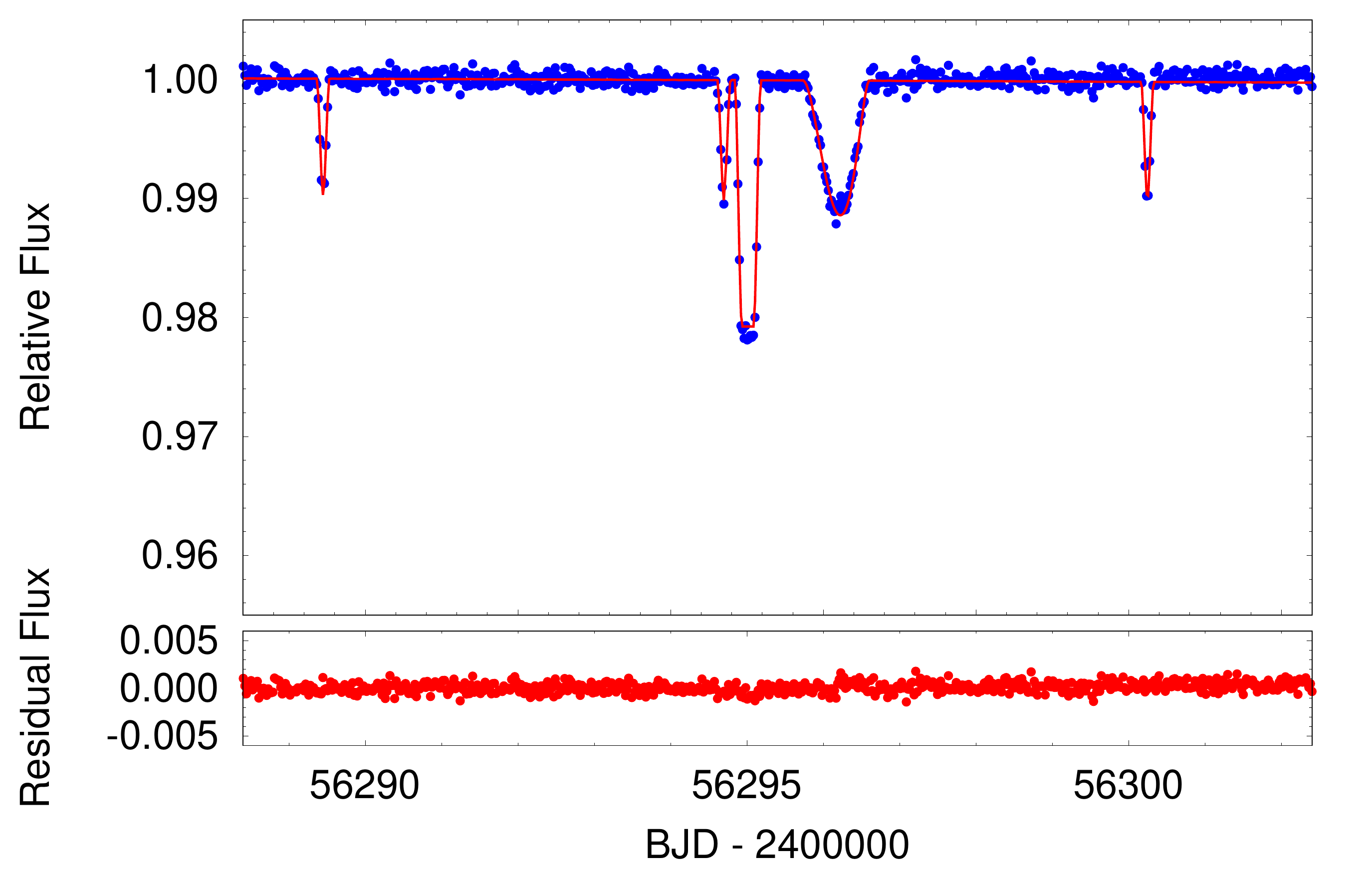}  
 \caption{The `secondary' third-body, or outer eclipses of KIC~6964043 observed with \textit{Kepler}. During these events the two red dwarfs of the inner binary are eclipsed (either partially or, totally) by the outer, most massive, G-star component. Blue dots represent the observed LC fluxes, while red lines represent the best-fitting photodynamical model.}
\label{fig:K6964043E3secondaries}
\end{figure*}

\textit{KIC 6964043 = KOI-1351} ($P_1=10.73$\,d; $P_2=239$\,d; $P_2/P_1=22.3$) was observed by \textit{Kepler} in long cadence (LC) mode only from the fourth quarter (Q4) of the primary mission to its end (Q17). Moreover, it was also observed in 1-min short cadence (SC) mode in Q12. It is the only triply eclipsing triple system in our sample. These extra, third-body eclipses are the largest amplitude, longest duration features of the lightcurve (see upper left panel of Fig.~\ref{fig:keplerlcs} and Figs.~\ref{fig:K6964043E3primaries},~\ref{fig:K6964043E3secondaries}). The regular inner eclipses are quite shallow (with depths of $\sim1-2$\%). The inner EB exhibits growing eclipse depths during the first few quarters and then, after a short period of constancy, the eclipse depths continuously decrease. This is the most compact and tightest triple in our sample. \citet{borkovitsetal15} have investigated the ETV of the EB and pointed out that it is clearly dominated by DE, i.e., third-body perturbations over the LTTE. Therefore, besides the orbital elements and mass ratio of the outer orbit, they were able to determine the full spatial configuration of the triple and, hence, the mutual inclination and got $i_\mathrm{mut}=19\degr\pm2\degr$\footnote{Note, that in contrast to the other three systems where our new results confirm the mutual inclinations obtained from the prior pure ETV analysis of \citet{borkovitsetal15}, in the case of this triple the present, more complex study (see later in Sect.~\ref{sec:dyn_mod}) resulted in a much lower value of $i_\mathrm{mut}=4\fdg1\pm0\fdg1$.} 
Moreover, they estimated the periods of both the apsidal motion and the nodal regression to be about 26-27\,yr. KIC 6964043 was also observed with \textit{TESS} in full-frame image (FFI) mode in Sectors 14 and 15, but these observations exhibit neither regular inner eclipses, nor third-body eclipses, which is in accord with our photodynamical model, as will be discussed below, in Sect.~\ref{sec:discussion}.

\begin{figure*}
\includegraphics[width=0.50\textwidth]{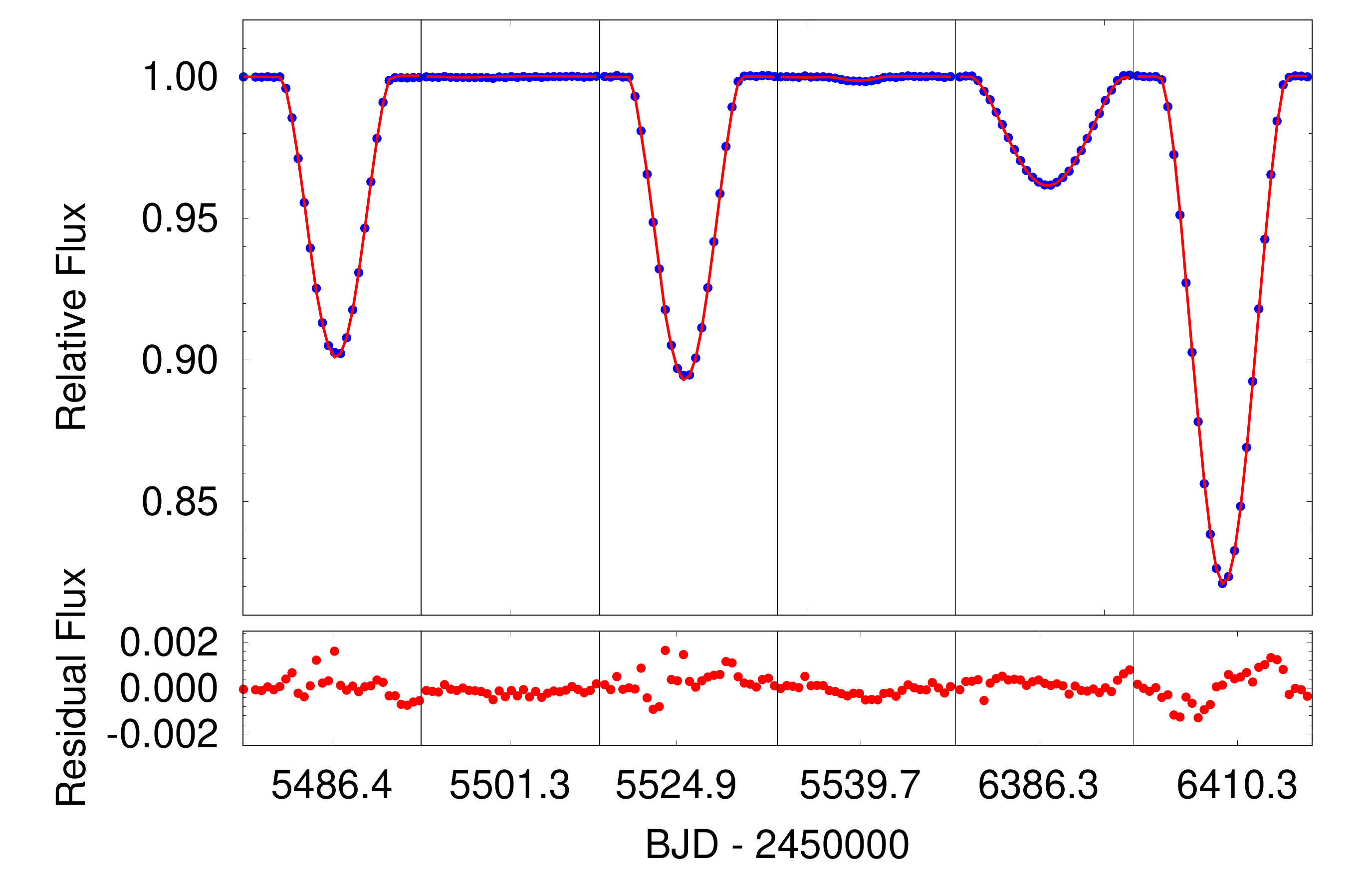}\includegraphics[width=0.50\textwidth]{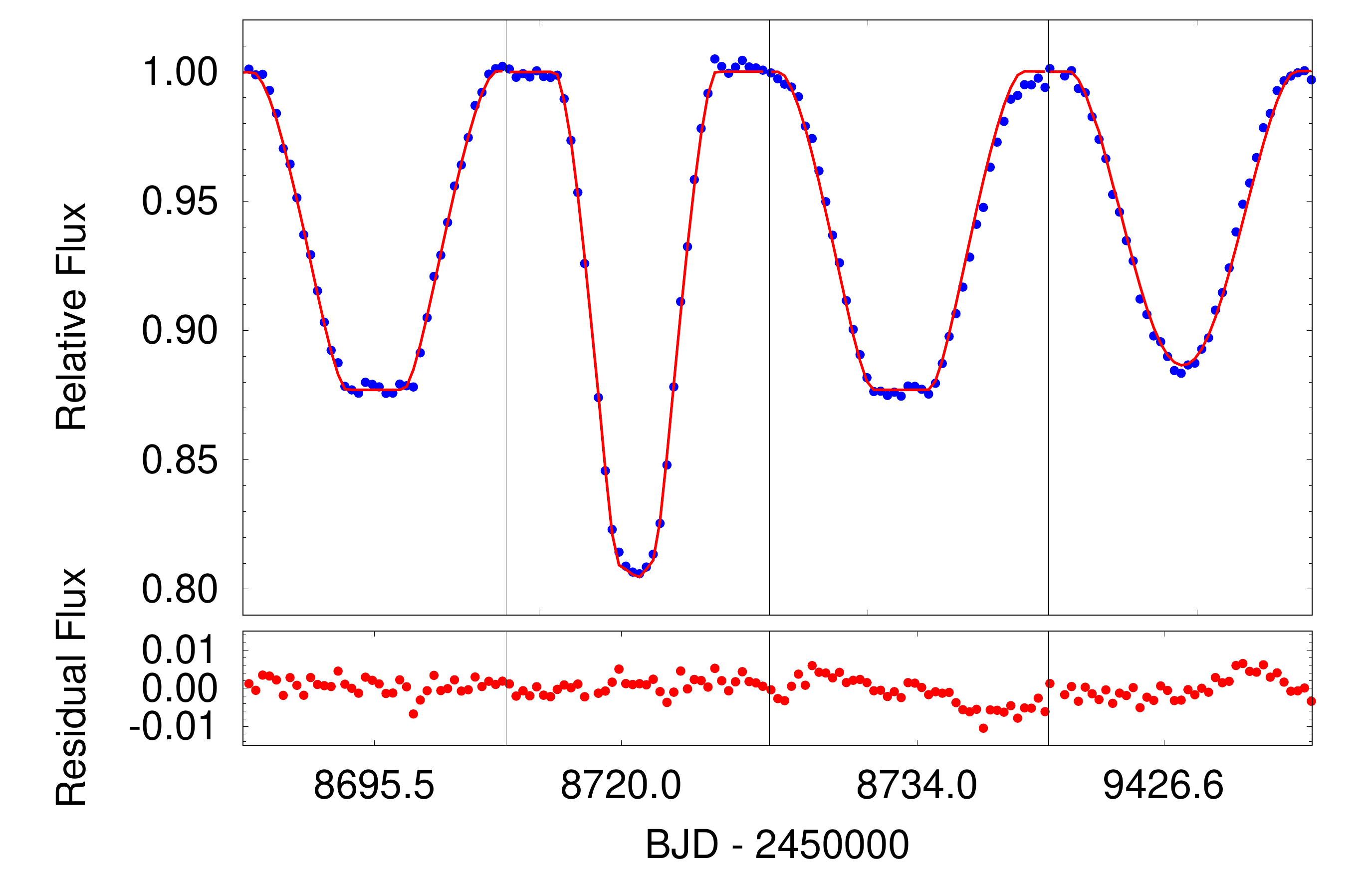} 
 \caption{Some characteristic primary and secondary eclipses of KIC\,5653126 in the \textit{Kepler} (left) and \textit{TESS} lightcurves (right) overplotted with the photodynamical model solution. All the strips have durations of 0.6 and 0.8\,days in the left and right panels, respectively. The left panel shows that between the two consecutive primary eclipses around BJDs\,2\,455\,486.3 and 2\,455\,524.9 no secondary eclipse can be detected, but just one orbital period later, the very first extremely shallow secondary eclipse occurs at $\sim$2\,455\,539.7. The last secondary and primary eclipses observed by \textit{Kepler} are also plotted in the last two segments of the left panel. As one can see, by that time, the durations of the secondary eclipses exceeded that of the primary ones, which is a consequence of the slower orbital motion closer to apastron. In the right panel all the eclipses (three secondary and one primary minima) observed with \textit{TESS} are plotted. Note, the secondary eclipses in sectors 14 and 15 are clearly flat-bottomed, implying total eclipses, while the S41 secondary eclipse was, again, V-shaped, i.e., a partial eclipse.}
\label{fig:K5653126lcfits}
\end{figure*}

\textit{KIC 5653126 = KOI-6612} ($P_1=38.51$\,d; $P_2=968$\,d; $P_2/P_1=25.1$) has by far the longest inner period of all the triples in our sample. It was observed throughout the original \textit{Kepler}-mission ($Q0-Q17$) in LC mode, while SC data are also available for quarters $Q9-Q17$. The EB exhibits continuously increasing eclipse depths. In the first seven quarters only primary eclipses were observable. The first, very shallow, grazing secondary eclipse with a depth of $\sim0.002$\% can be identified at BJD~2\,455\,539.7 (near the end of $Q7$). During the second half of the data train both kinds of eclipses can be clearly detected (see upper right panel of  Fig.~\ref{fig:keplerlcs} and left panel of Fig.~\ref{fig:K5653126lcfits}). \citet{borkovitsetal15} found a mutual inclination of $i_\mathrm{mut}=11\degr\pm1\degr$ and apsidal motion and nodal periods on the order of 2-3 centuries. The system was observed also by the \textit{TESS} spacecraft in FFI-mode during Sectors 14, 15 and 41, where the observations led to the detection of one additional primary, and three secondary eclipses. The 2019 summer (S14,15) observations show flat-bottomed eclipses, while the somewhat shallower 2021 (S41) secondary eclipse is again V-shaped, indicating that the eclipse depths are now decreasing after an interval of total eclipses (right panel of Fig.~\ref{fig:K5653126lcfits}).

\begin{figure*}
\includegraphics[width=0.50\textwidth]{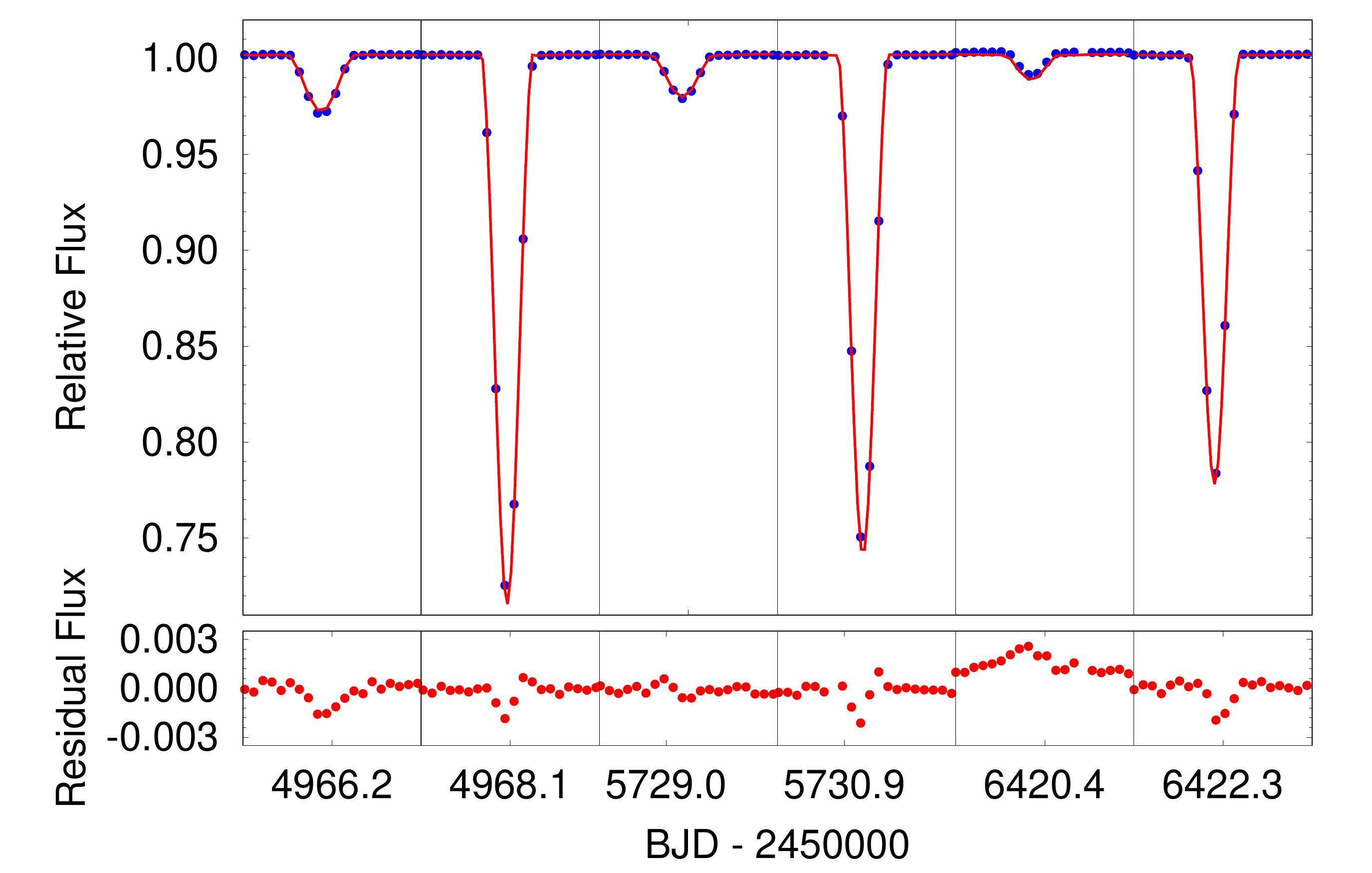}\includegraphics[width=0.50\textwidth]{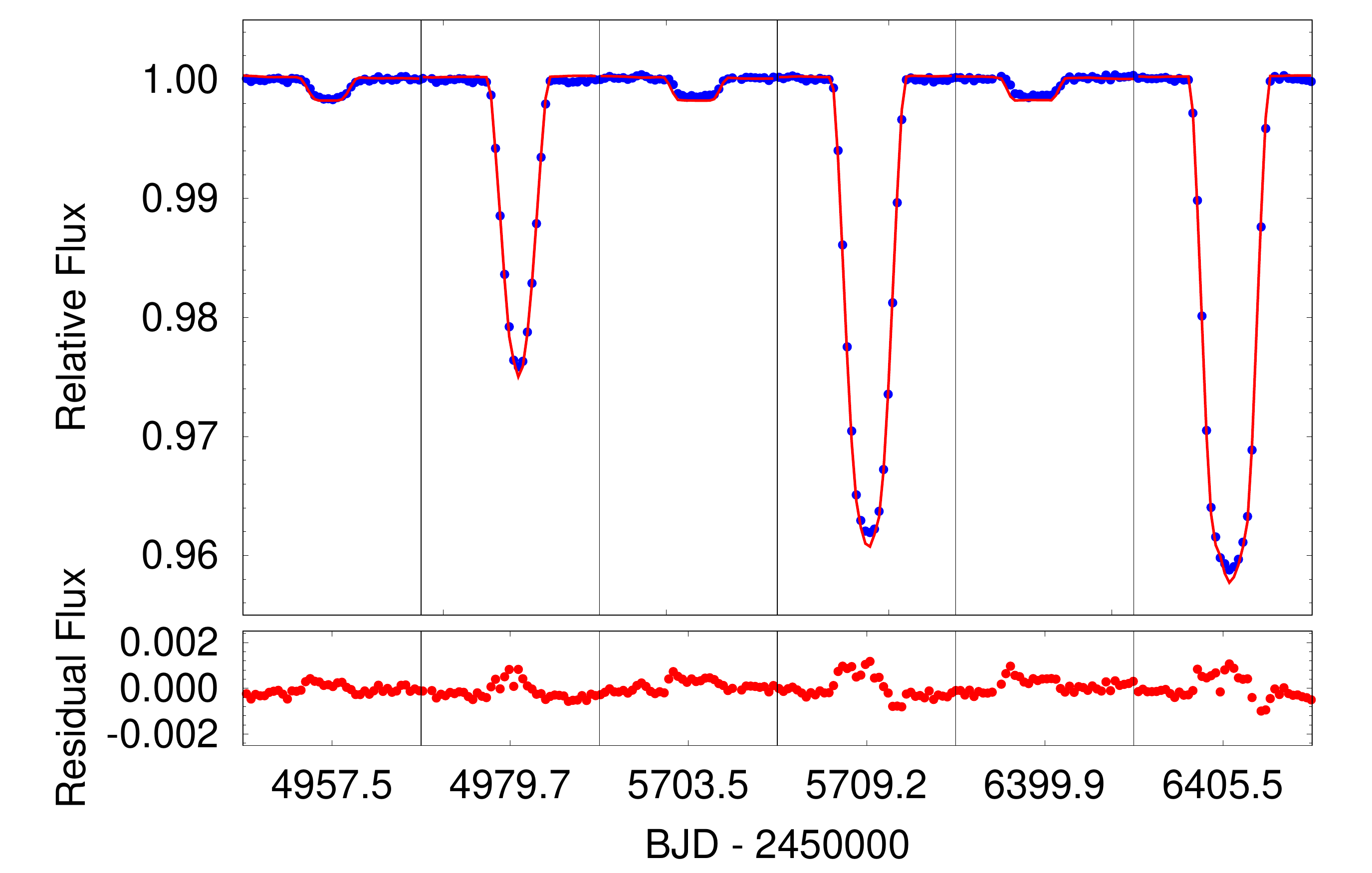}
 \caption{Secondary and primary eclipses of KICs\,5731312 (left) and 8023317 (right) at the beginning, middle and end of the four-year-long \textit{Kepler} observations. The lengths of the lightcurve sections are 0.4 and 0.8\,days in the left and right panels, respectively.}
\label{fig:K57313128023317lcfits}
\end{figure*}

\textit{KIC 5731312 = KOI-6621} ($P_1=7.95$\,d; $P_2=911$\,d; $P_2/P_1=114.6$) is the shortest inner period triple in our sample. Similar to KIC~5653126, it was observed throughout the original \textit{Kepler}-mission ($Q0-Q17$) in LC mode. One quarter ($Q10$) of SC data are available as well. During the \textit{Kepler} observations this system exhibited the deepest primary eclipses amongst the four systems. During the first half of the observations, only a small smooth decrease in the amplitude of the eclipses can be observed (which is nearly obscured by a wave-like pattern which, however, is only a beating effect between the data sampling and the eclipsing periods). Then, between BJDs~2\,455\,700 and 800 a sudden decrease in eclipse depths occurred, which was then again followed by a slow smooth decrease. The much shallower secondary eclipses followed the same trend (see lower left panel of Fig.~\ref{fig:keplerlcs} and left panel of Fig.~\ref{fig:K57313128023317lcfits}). At the same time, the primary and secondary ETV curves also exhibited dramatic peaked variations, clearly indicating periastron passage of the eccentric third body. Note, despite the fact that the outer to inner period ratio exceeds one hundred, the ETV curves of this EB are clearly dominated by the DE over the LTTE. \citet{borkovitsetal15} and, in their slightly revised ETV-analysis \citet{borkovitsetal16}, also found that this triple system is probably quite inclined with a mutual inclination of $i_\mathrm{mut}=37\fdg8\pm0\fdg4$. In addition, their analytical model fit resulted in a currently retrograde apsidal motion. The EB was also observed with \textit{TESS} in Sectors 14, 15 and 41. By this time the secondary eclipses had disappeared, which is again in accord with our photodynamical results to be presented below. Moreover, in between the 2019 and 2021 space telescope observations, two additional primary eclipses were also observed at Baja Observatory, Hungary, within the framework of our photometric \textit{Kepler} and \textit{TESS} EB follow-up programme.

\textit{KIC 8023317 = KOI-6049} ($P_1=16.58$\,d; $P_2=611$\,d; $P_2/P_1=36.9$) displays slowly and continuously increasing primary eclipse depths throughout the $Q0-Q17$ \textit{Kepler} LC data. (SC data were also gathered during quarters $Q5.2$ and $Q17.1$.)  The very shallow secondary eclipses, however, have constant depths during the 4-year-long data set (see lower right panel of Fig.~\ref{fig:keplerlcs} and right panel of Fig.~\ref{fig:K57313128023317lcfits}). This is due to the fact that the flat bottom of the secondary eclipses clearly indicates that during these events, the smaller fainter and colder secondary star was totally eclipsed. The EDV for the first time was reported by \cite{rappaportetal13} who also made the first ETV analysis of the system. The system was also included in the more sophistical ETV analyses of \citet{borkovitsetal15,borkovitsetal16}. All three investigations led to similar results indicating that KIC 8023317 is one of the most inclined triple systems with $i_\mathrm{mut}=49\fdg5\pm0\fdg6$ amongst the EBs observed with \textit{Kepler}. Moreover, similar to KIC~5731312, the analytic solutions of \citet{borkovitsetal15,borkovitsetal16} led to the conclusion that the apsidal motion of this EB is, again, retrograde. This triple was reobserved with \textit{TESS} during Sectors 14, 40 and 41. These data exhbit five additional, U-shaped primary eclipses, however, the much smaller SNR of the \textit{TESS} lightcurves does not allow for a secure detection of the very low-amplitude secondary eclipses.


\section{Observational data and its preparation}
\label{sec:Obsdata}

Our analysis is primarily based on the photometric data gathered by the \textit{Kepler} spacecraft during its four-year-long primary mission. We use exclusively the same LC (time resolution of 29.4\,min) data sets, that were used in our previous work on the comprehensive ETV analyses of hundreds of \textit{Kepler} EBs \citep{borkovitsetal16}. The majority of these lightcurves were downloaded from the Villanova website\footnote{\url{http://keplerebs.villanova.edu/}} of the third revision of the Kepler Eclipsing Binary Catalog \citep{kirketal16,abdulmasihetal16} while, in the case of KIC\,5731312, the missing quarters 12 and 13 were downloaded directly from the MAST database\footnote{\url{https://archive.stsci.edu/missions-and-data/kepler}}.

Regarding the more recent \textit{TESS} observations, as mentioned above, all four systems were measured in 30-min cadence mode during Year 2 observations.  Additionally, three of the systems (with the exception being KIC\,6964043) were also observed in the 10-min cadence mode during the first sectors of Year 4 observations. We processed the original \textit{TESS} FFIs using a convolution-based differential photometric pipeline implemented in the {\sc FITSH} package \citep{pal12}. These raw lightcurves, however, exhibited large amounts of scattered light and, moreover, due to the large pixel size of \textit{TESS}, the signal of KIC\,5653126 was blended with that of the 4.81-day-period EB KIC\,5738698 and, therefore, needed disentanglement. This disentanglement was carried out with a principal component analysis (PCA) as follows.

We used the built-in routines from the {\sc lightkurve} Python package \citep{lightkurve}. First, differential images centered on our targets were constructed by the previously mentioned {\sc FITSH} pipeline subtracting a median image from the raw images. These differential images were then read in as a $60 \times 60$ pixel-sized \textit{TargetPixelFile} object and the raw differential lightcurve was calculated by summing up the residual fluxes in a $3 \times 3$ pixel-sized aperture around the target. After that, assuming that all of the signal originated from our target is located inside the chosen aperture, all the other pixels outside this aperture were used to constrain a \textit{DesignMatrix} object with the vectors that contain information about scattered light, spacecraft motion and other leftover extrinsic light variabilities (e.g. the signal of a close variable star partly measured in the aperture). From this \textit{DesignMatrix}, we supplied the first 5-7 principal components to {\sc lighkurve}'s \textit{RegressionCorrector} in order to construct the noise model and remove the underlying systematic noise from the raw lightcurves, and, in the case of KIC\,5653126, also to deblend the signal of KIC\,5738698 from the lightcurve. As a final step, the relative fluxes of the detrended lightcurves were referenced to the zero-point values calculated from the $G_{\rm RP}$ magnitudes\footnote{This is a rather good approximation due to the significant overlap between the passbands of \textit{TESS} and Gaia. \citep{ricker15,jordi10}} and were normalized to unity.

Moreover, in the case of KIC\,5731312 we included in the analysis two additional primary eclipses that were observed by us at Baja Astronomical Observatory (Hungary) between the Year 2 and Year 4 \textit{TESS} observations. 

In preparing these datasets for the photodynamical analyses, first we determined accurate eclipse times from the \textit{TESS} and ground-based follow up observations in the same manner as the \textit{Kepler}-observed eclipse times were determined earlier for our previous analytic ETV analyses in \citet{borkovitsetal16}. Thus, we obtained 4 (1 primary and 3 secondary) new mid-eclipse times for KIC\,5653126 and, moreover, 11 and 5 primary eclipse times for KICs\,5731312 and 8023317, respectively. In the case of KIC\,6964043 no further eclipses were detected in the \textit{TESS} data. We list all of these eclipse times together with the former \textit{Kepler} data in the tables of Appendix~\ref{app:ToMs}. Moreover, the ETV curves, together with the best photodynamical fits (see below, in Sect.~\ref{sec:dyn_mod}) are plotted in Fig.~\ref{fig:ETVswithfit}.

For the photodynamical analysis we took some additional preparatory steps with the lightcurves. First, to obtain a more uniform sampling, we converted the Year 4 10-min cadence \textit{TESS} data into 30-min bins. Second, in order to save computational time and also to give the eclipses themselves  (which carry the vast bulk of both the relevant astrophysical and dynamical information) a higher statistical weight we kept only small sections of the lightcurves around the primary and secondary eclipses of the inner binaries. (Note, more strictly speaking, we kept the small sections of the lightcurves around the minima of the sky-projected distances of the stellar disks of the inner binaries, irrespective of whether eclipses have actually occurred or not.) Moreover, in the case of KIC\,6964043 we kept some additional few-day-long sections around the third-body eclipses, as well.

Finally we note that the lightcurves also exhibit some low-level systematics, which might have either an instrumental origin or be caused by stellar variability.  Due to their very low amplitude we did not attempt to remove these effects. However, in the future, analyses using a simultaneous fit of Gaussian Processes or wavelet-based \citep{csizmadiaetal21} noise models can remove some of these systematic behaviours and thereby slightly improve the fit and error estimations.

\begin{figure*}
\center
\includegraphics[width=0.45\textwidth]{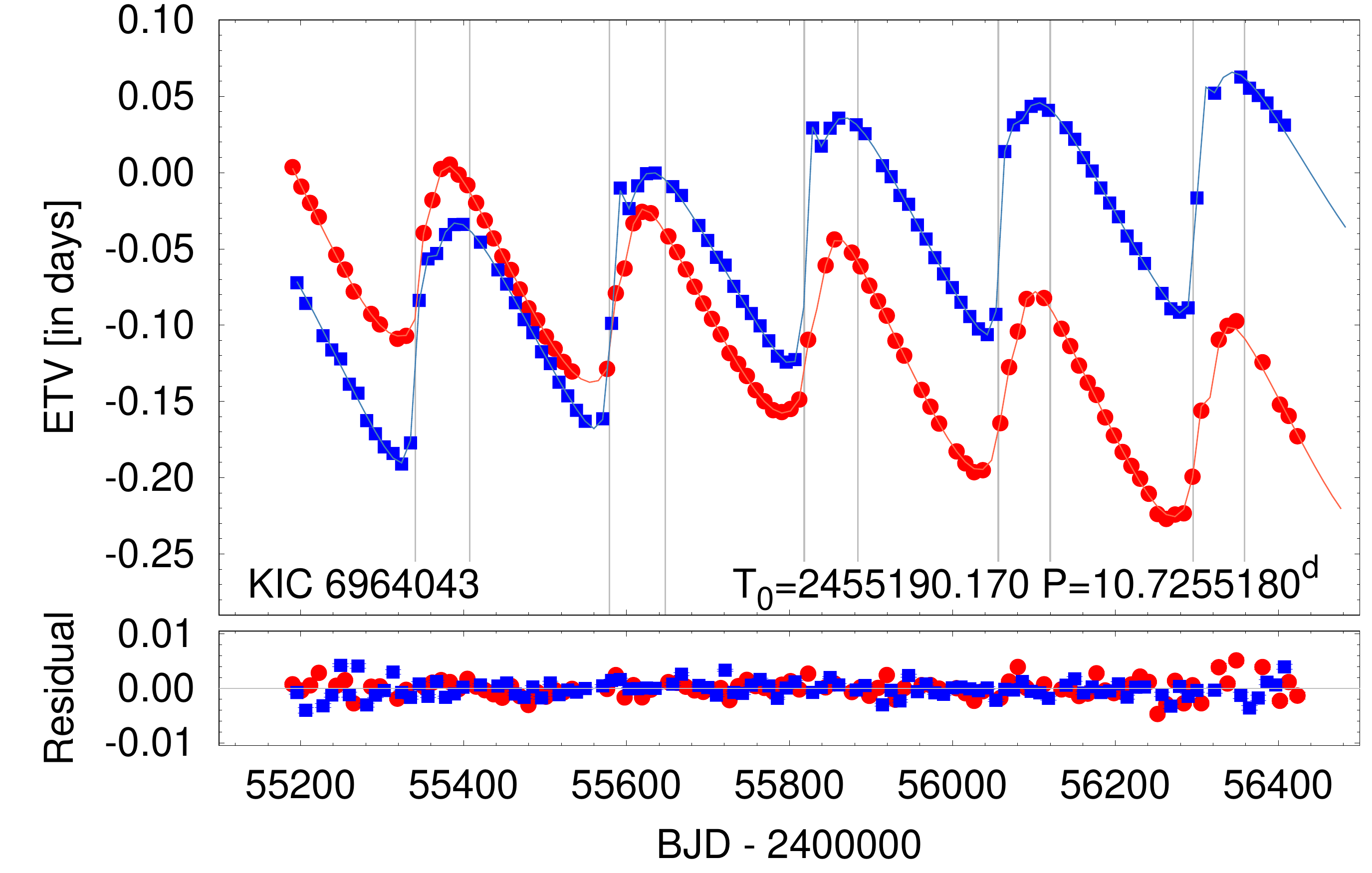}\includegraphics[width=0.45\textwidth]{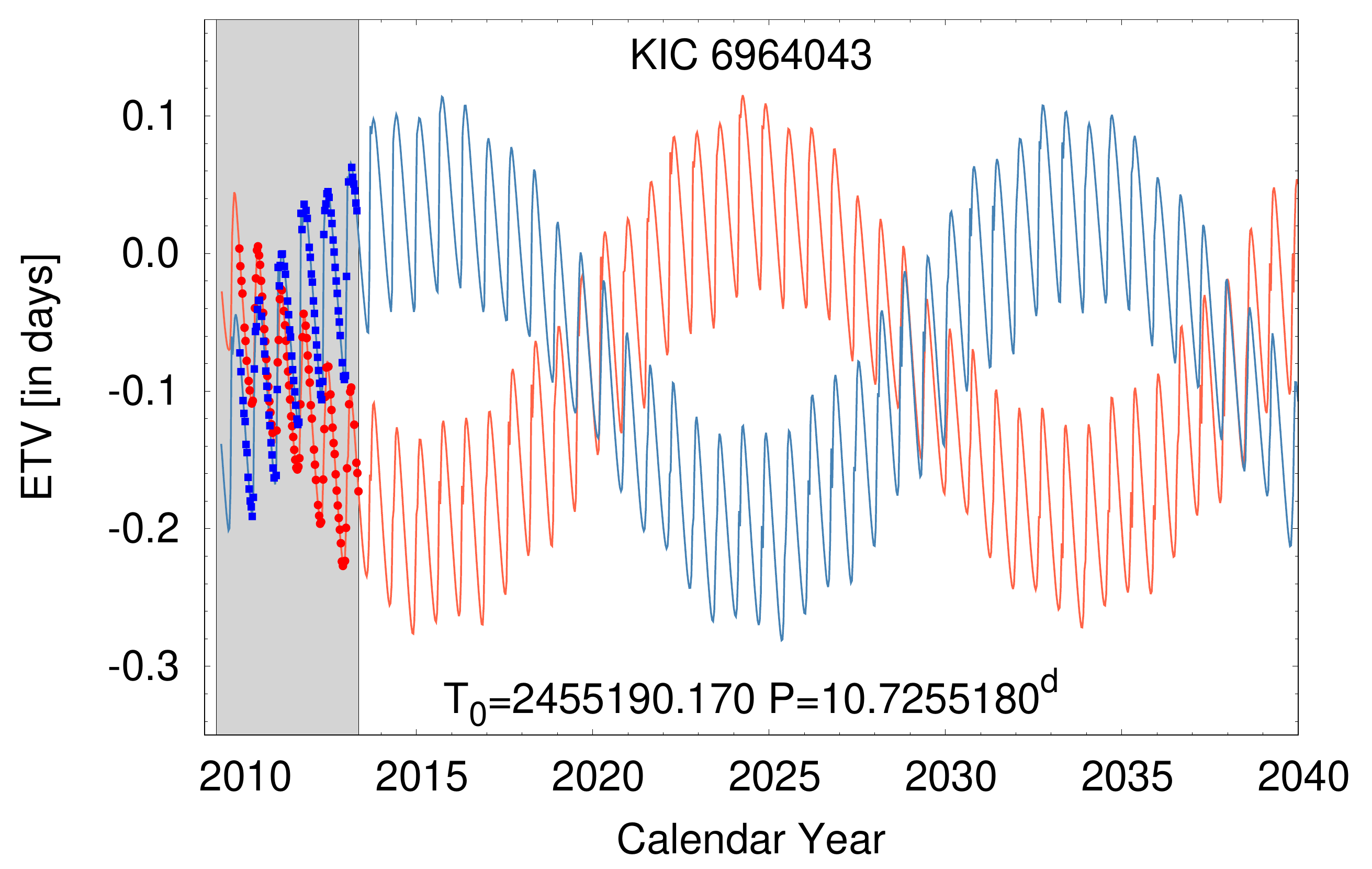} 
\includegraphics[width=0.45\textwidth]{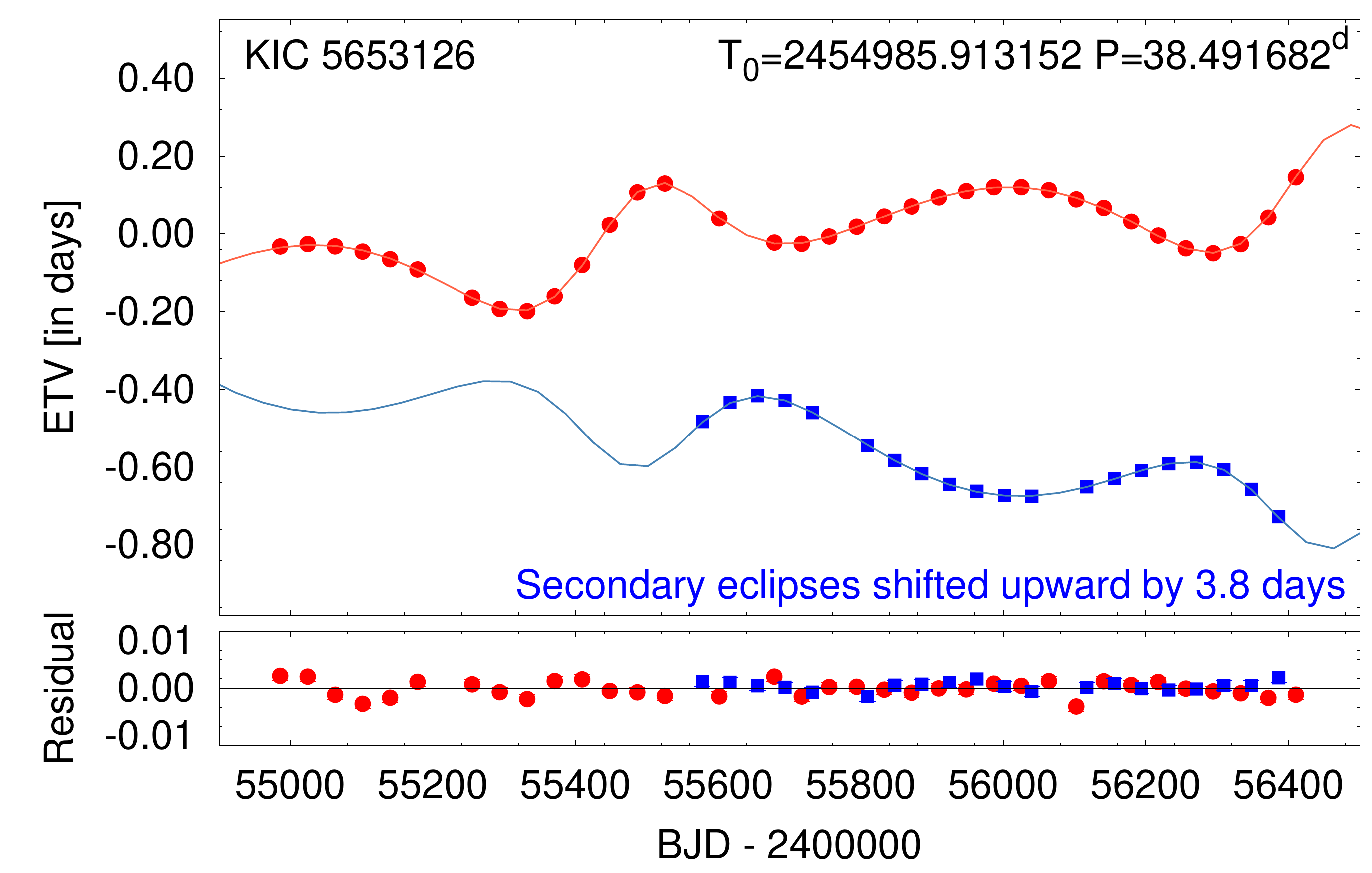}\includegraphics[width=0.45\textwidth]{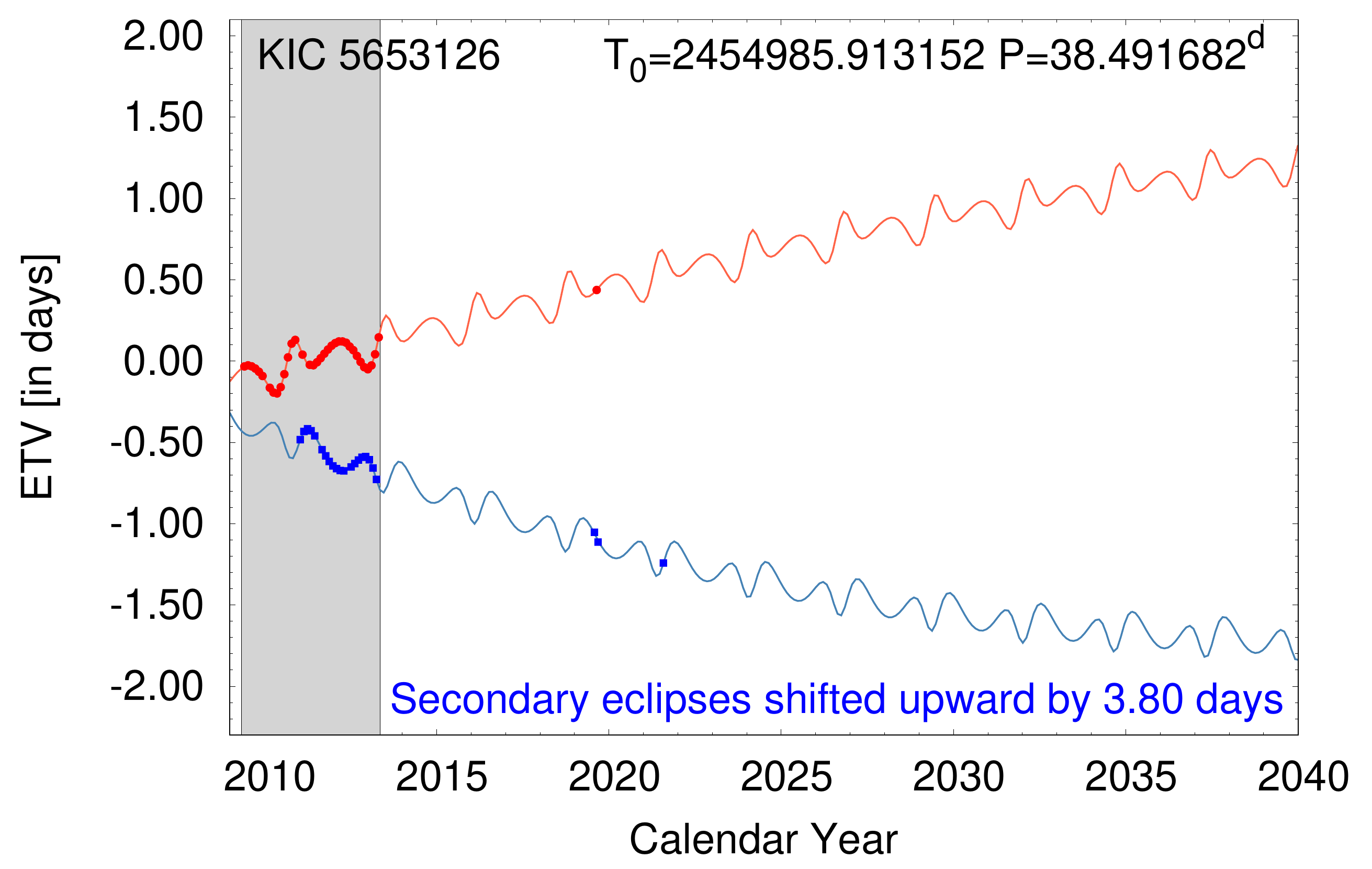} 
\includegraphics[width=0.45\textwidth]{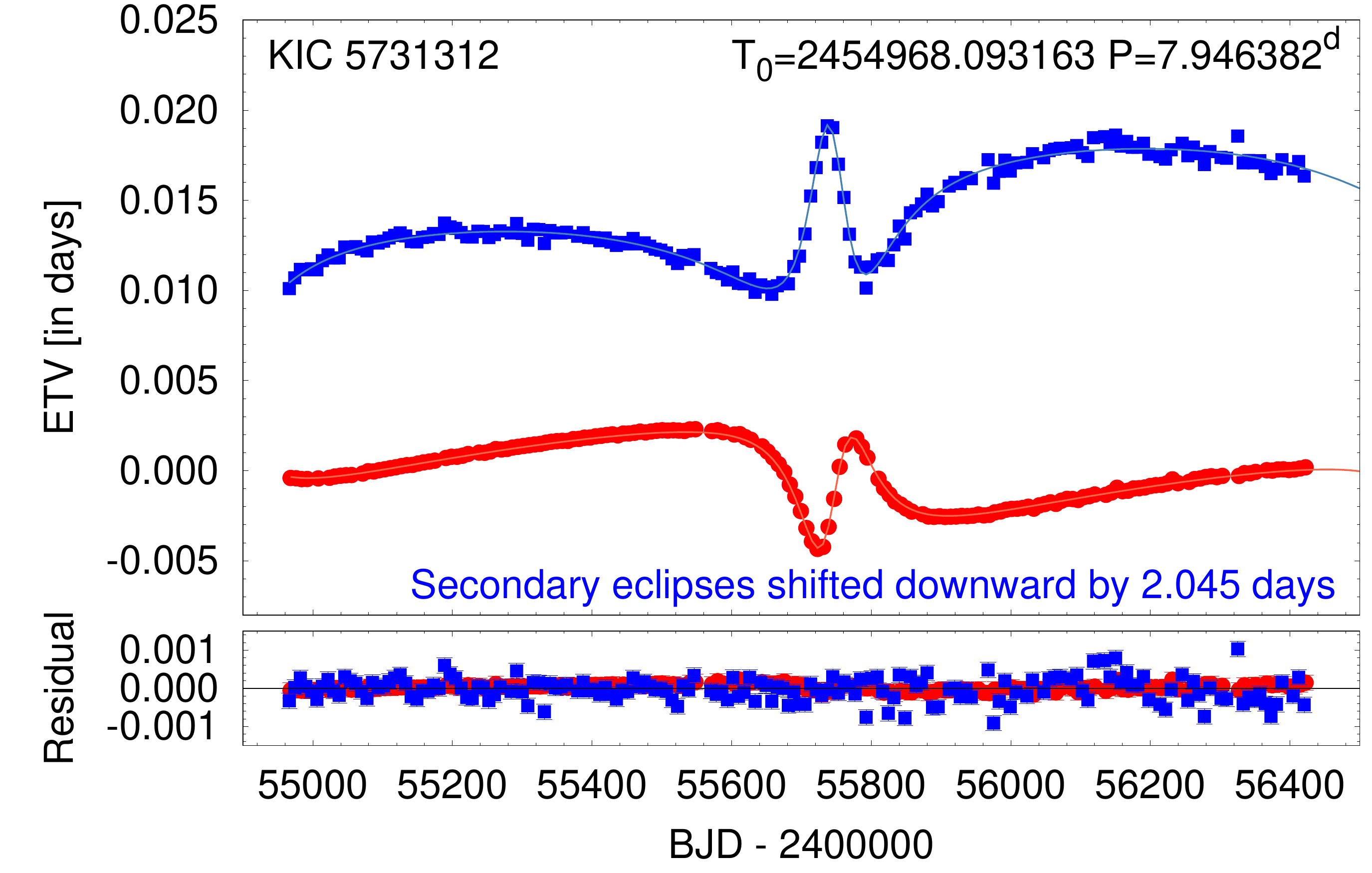}\includegraphics[width=0.45\textwidth]{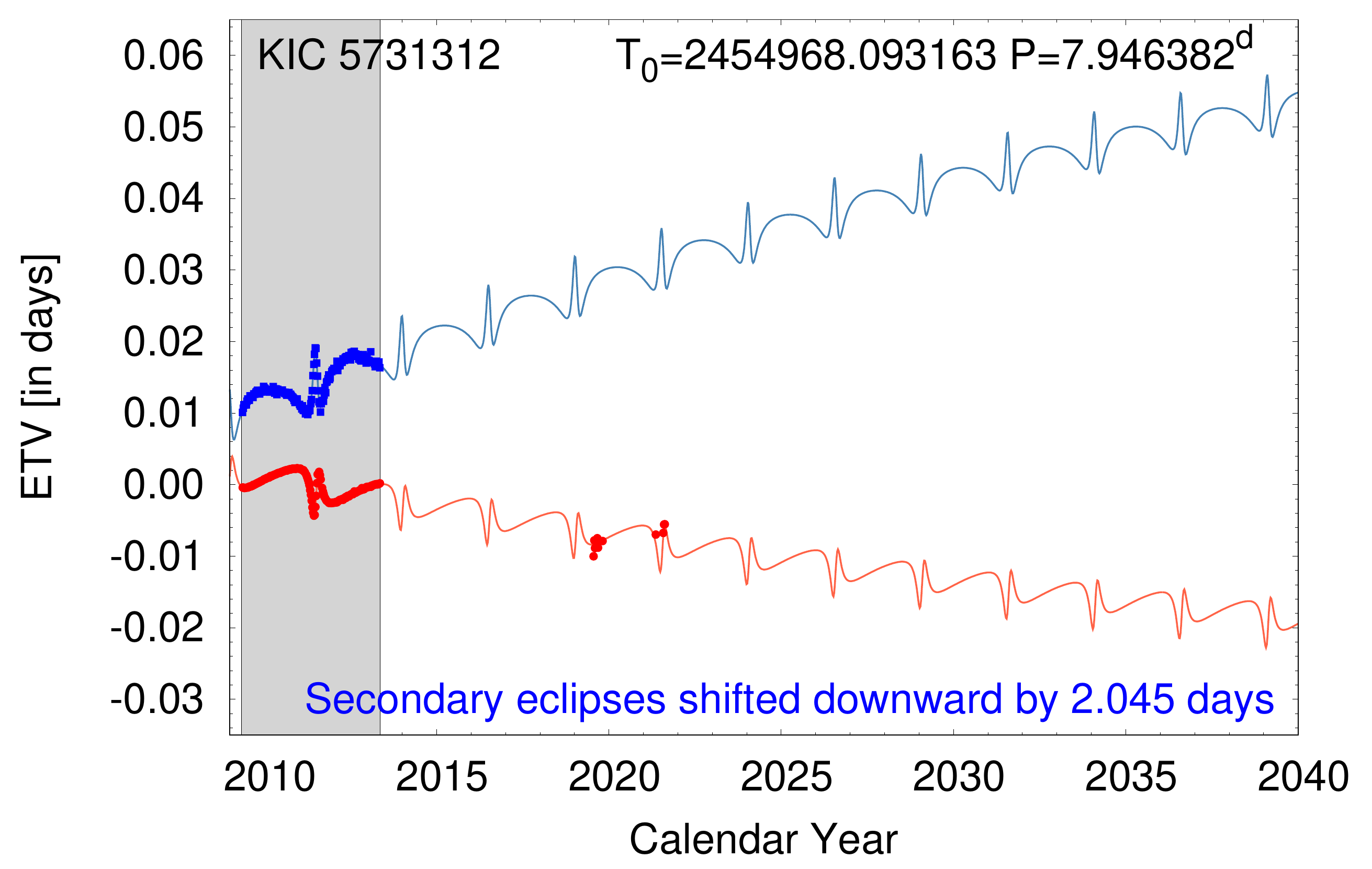} 
\includegraphics[width=0.45\textwidth]{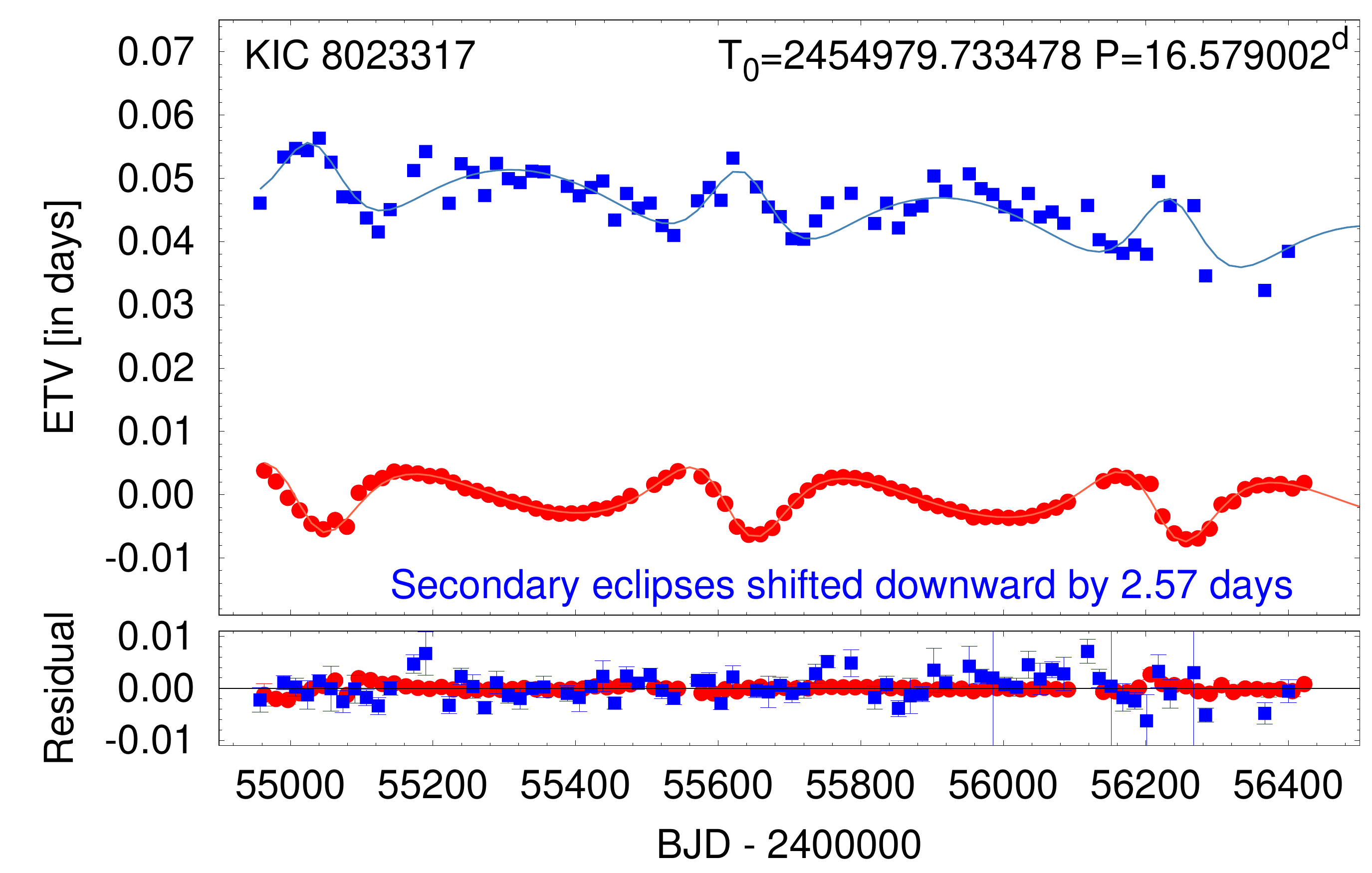}\includegraphics[width=0.45\textwidth]{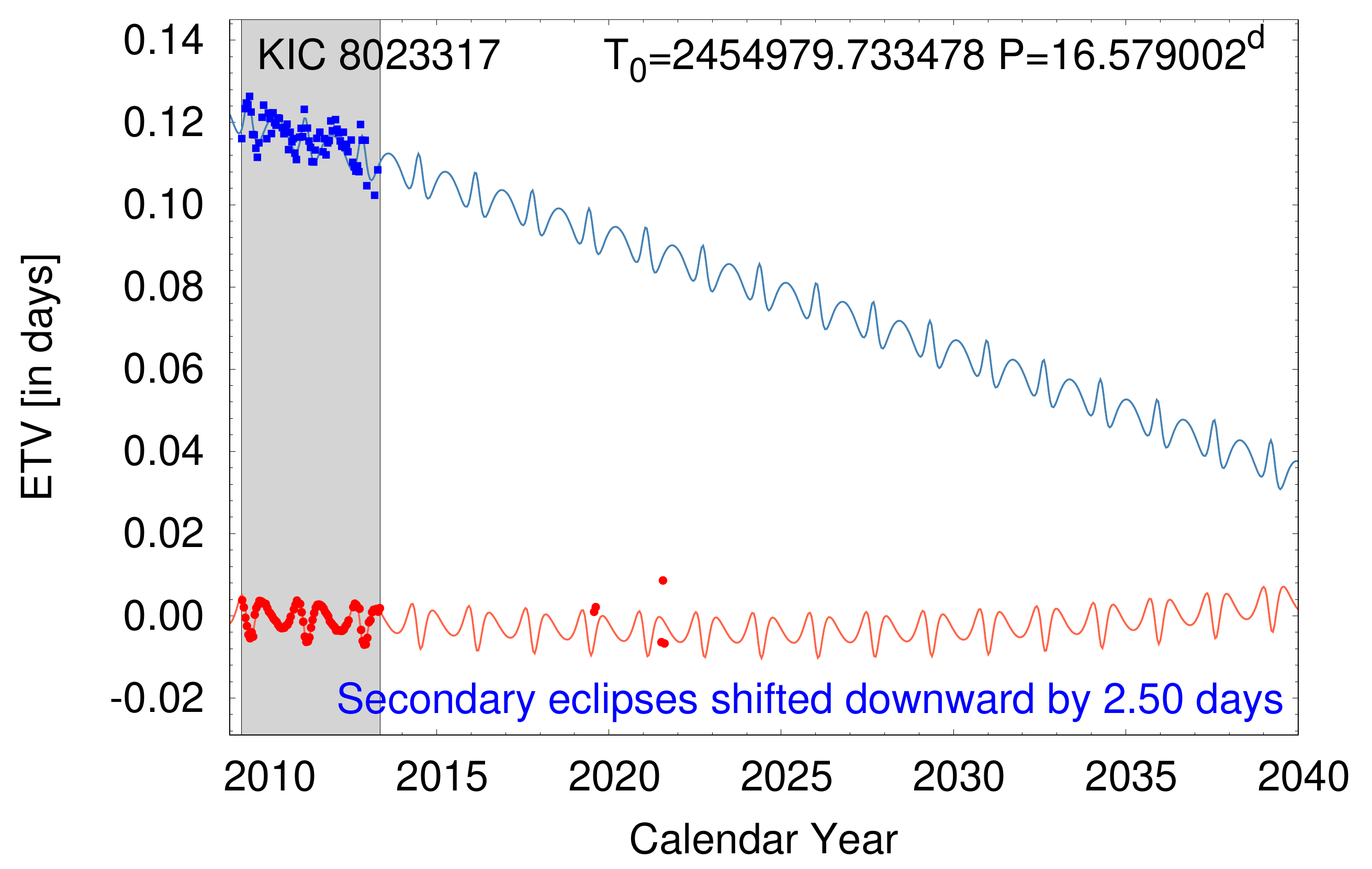} 
\caption{Eclipse timing variations of KICs 6964043 (upper), 5653126 (2nd row), 5731312 (3rd row) and 8023317 (bottom). Left panels represent the timing variations during the \textit{Kepler} observations, while right panels exhibit longer time intervals up to 2040. In these panels gray shaded area shows the nominal interval of the original \textit{Kepler} mission. (Note, in those parts of the curves, where no eclipses would be observed due to the low orbital inclinations, the times of the minimal sky-projected distances of the stellar disks were calculated.) Red and blue dots represent the primary and secondary binary eclipses, respectively, while the correspondingly colored curves connect the best-fitting photodynamical model ETV points. In the upper left panel the solid vertical gray lines indicate the times of the third-body eclipses detected by \textit{Kepler} in the triple system KIC 6964043. The lower parts of the left panels show the residuals from the model fit.}
\label{fig:ETVswithfit}
\end{figure*}  

\section{Photodynamical modeling}
\label{sec:dyn_mod}

For the four triple systems, we have carried out a photodynamical analysis with the software package {\sc Lightcurvefactory} \citep[see, e.g.][and references therein]{borkovitsetal19a,borkovitsetal20a}. As described in these earlier papers, the code contains (i) a built-in numerical integrator to calculate the gravitationally (three- or four-body effects), tidally, and (optionally) relativistically perturbed coordinates and velocities of the three (four) stars in the system; (ii) emulators for multiband light curves (allowing multiple eclipses), ETV curves, and radial velocity (RV) curves (if available), (iii) built-in, tabulated \texttt{PARSEC} grids {\citep{PARSEC} for interpolating fundamental stellar parameters (e.g. radii and effective temperatures) as well as compositie spectral energy distribution (SED) model for the three (four) stars, and (iv) an MCMC-based search routine for fitting the parameters. The latter utilizes an implementation of the generic Metropolis-Hastings algorithm \citep[see e.g.][]{ford05}. The use of this software package and the consecutive steps of the entire analysis process have been previously explained in detail as the code was applied to a variety of multiple stellar systems \citep{borkovitsetal18,borkovitsetal19a,borkovitsetal19b,borkovitsetal20a,borkovitsetal20b,borkovitsetal21,borkovitsetal22,mitnyanetal20,rappaportetal22}. 

For the four current systems, in the absence of RV data, we combined a composite SED analysis and the use of precalculated \texttt{PARSEC} grids as proxies to determine the masses of the constituent stars. As a consequence, such a photodynamical model solution is no longer astrophysically model-independent.  Nonetheless, in our previous work \citep{borkovitsetal22} we have shown that for compact, strongly dynamically interacting triples (as is the case for the currently investigated systems), such a solution (which we call `MDN' - astrophysical model-dependent solution without RV data) results in fundamental stellar parameters (masses, radii) within 5-10\% to the results of an `MIR' (astrophysical model-independent solution with RV data) solution.  The geometric and dynamical parameters (i.e. the orbital elements and, also the mass ratios), as well as dimensionless quantities like the fractional radii and temperature ratios of the stars remain practically unchanged between the MDN and MIR solutions.

In the case of these `MDN' runs the adjusted parameters were as follows:
\begin{itemize}
\item[(i)] Orbits: Three of six orbital-element related parameters of the inner, and six parameters of the outer orbits, i.e., the components of the eccentricity vectors of both orbits $(e\sin\omega)_\mathrm{in,out}$, $(e\cos\omega)_\mathrm{in,out}$, the inclinations relative to the plane of the sky ($i_\mathrm{in,out}$), and moreover, three other parameters for the outer orbit, including the period ($P_\mathrm{out}$), the longitude of the node relative to the inner binary's node ($\Omega_\mathrm{out}$), and, in the case of the triply eclipsing system KIC~6964043, the sixth parameter was the time of the first (inferior) conjunction of the tertiary star observed in the \textit{Kepler} data ($\mathcal{T}_\mathrm{out}^\mathrm{inf})$ (which corresponds to the first observed third-body eclipse), while for the other three systems the periastron passage times ($\tau_\mathrm{out}$) were adjusted. 
\item[(ii)] Stars: Three stellar mass related parameters: the mass of the most massive component (being either $m_\mathrm{Aa}$, the primary of the inner pair or, $m_\mathrm{B}$, the tertiary star), and the inner and outer mass ratios ($q_\mathrm{in,out}$). Additionally, the metallicity of the system ([$M/H$]), the (logarithmic) age of the three coeval stars ($\log\tau$), and the interstellar reddening $E(B-V)$ toward the given triple were also varied. Moreover, due to the strong contamination of the \textit{TESS} lightcurve of KIC\,8023317, for this lightcurve only, the `extra light' contamination, $\ell_4$ parameter was also freely adjusted.
\end{itemize} 

A couple of other parameters were {\it constrained} instead of being adjusted or held constant during our analyses, as follows:
\begin{itemize}
\item[(i)] Orbits: The orbital period of the inner binary ($P_\mathrm{in}$) and its orbital phase (through the time of an arbitrary primary eclipse or, more strictly, the time of the inferior conjunction of the secondary star -- $\mathcal{T}^\mathrm{inf}_\mathrm{in}$) were constrained internally through the ETV curves. 
\item[(ii)] Stars: The radii and temperatures of the three stars were calculated with the use of three-linear interpolations from the precomputed 3D (metallicity; logarithmic age; stellar mass) \texttt{PARSEC} grids. Additionally, the distance of the system (which is necessary for the SED fitting) were calculated a posteriori at the end of each trial step, by minimizing the value of $\chi^2_\mathrm{SED}$.
\item[(iii)] Atmospheric parameters of the stars: we handled them in a similar manner as in our previous photodynamical studies. We utilized a logarithmic limb-darkening law \citep{klinglesmithsobieski70} for which the passband-dependent linear and non-linear coefficients were interpolated in each trial step via the tables from the original version of the {\tt Phoebe} software \citep{Phoebe}. We set the gravity darkening exponents for all late type stars to $\beta=0.32$ in accordance with the classic model of \citet{lucy67} valid for convective stars and held them constant. Note, however, that the choice of this parameter has only minor consequences, since the stars in the present study are close to spheroids.
\end{itemize}

\begin{table*}
 \centering
\caption{Orbital and astrophysical parameters of the low mutual inclination triples KICs\,6964043 and 5653126 from the joint photodynamical lightcurve, ETV, SED and \texttt{PARSEC} isochrone solution. Note, besides the usual angular orbital elements (given in both the observational and dynamical frame of references), $i_\mathrm{inv}$ and $\Omega_\mathrm{inv}$ give the position of the invariable plane with respect to the tangential plane of the sky. 
The osculating orbital elements are given for epochs $t_0$, given in the first row.}
 \label{tab: syntheticfit_KIC69640435653126}
\begin{tabular}{@{}lllllll}
\hline
 & \multicolumn{3}{c}{KIC\,6964043} & \multicolumn{3}{c}{KIC\,5653126} \\
\hline
\multicolumn{7}{c}{orbital elements} \\
\hline
   & \multicolumn{2}{c}{Aa--Ab} & A--B  & \multicolumn{2}{c}{Aa--Ab} & A--B  \\
  \hline
  $t_0$ [BJD - 2400000]& \multicolumn{3}{c}{$55182.0$} & \multicolumn{3}{c}{$54953.0$} \\
  $P$ [days] & \multicolumn{2}{c}{$10.69787_{-0.00015}^{+0.00012}$} & $239.2519_{-0.0083}^{+0.0076}$  & \multicolumn{2}{c}{$38.44825_{-0.00028}^{+0.00028}$} & $971.39_{-0.25}^{+0.25}$ \\
  $a$ [R$_\odot$] & \multicolumn{2}{c}{$21.47_{-0.34}^{+0.47}$} & $209.8_{-3.6}^{+4.9}$ & \multicolumn{2}{c}{$69.77_{-0.35}^{+0.26}$} & $675.8_{-3.8}^{+3.0}$ \\
  $e$ & \multicolumn{2}{c}{$0.03687_{-0.00052}^{+0.00025}$} & $0.47806_{-0.00053}^{+0.00056}$ & \multicolumn{2}{c}{$0.2935_{-0.0021}^{+0.0015}$} & $0.1787_{-0.0011}^{+0.0011}$ \\
  $\omega$ [deg]& \multicolumn{2}{c}{$75.46_{-0.15}^{+0.15}$} & $311.41_{-0.10}^{+0.09}$ & \multicolumn{2}{c}{$304.36_{-0.22}^{+0.32}$} & $323.90_{-0.32}^{+0.36}$ \\ 
  $i$ [deg] & \multicolumn{2}{c}{$88.797_{-0.087}^{+0.111}$} & $89.811_{-0.015}^{+0.013}$ & \multicolumn{2}{c}{$87.128_{-0.018}^{+0.018}$} & $87.608_{-0.463}^{+0.526}$ \\
  $\mathcal{T}_0^\mathrm{inf}$ [BJD - 2400000]& \multicolumn{2}{c}{$55190.1678_{-0.0004}^{+0.0004}$} & $55340.658_{-0.015}^{+0.018}$ & \multicolumn{2}{c}{$54985.8390.1678_{-0.0007}^{+0.0007}$} & ... \\
  $\tau$ [BJD - 2400000]& \multicolumn{2}{c}{$55184.3510_{-0.0044}^{+0.0043}$} & $55110.444_{-0.050}^{+0.047}$ & \multicolumn{2}{c}{$54949.355_{-0.020}^{+0.030}$} & $55481.39_{-0.63}^{+0.65}$ \\
  $\Omega$ [deg] & \multicolumn{2}{c}{$0.0$} & $-4.023_{-0.086}^{+0.089}$ & \multicolumn{2}{c}{$0.0$} & $-12.344_{-0.175}^{+0.163}$ \\
  $i_\mathrm{mut}$ [deg] & \multicolumn{3}{c}{$4.138_{-0.061}^{+0.087}$} & \multicolumn{3}{c}{$12.348_{-0.152}^{+0.168}$} \\
  $\omega^\mathrm{dyn}$ [deg]& \multicolumn{2}{c}{$331.3_{-1.3}^{+1.7}$} & $27.2_{-1.4}^{+1.8}$ & \multicolumn{2}{c}{$212.4_{-2.1}^{+2.0}$} & $51.3_{-2.1}^{+2.1}$ \\
  $i^\mathrm{dyn}$ [deg] & \multicolumn{2}{c}{$3.620_{-0.055}^{+0.076}$} & $0.519_{-0.008}^{+0.009}$ & \multicolumn{2}{c}{$10.152_{-0.119}^{+0.129}$} & $2.197_{-0.036}^{+0.041}$ \\
  $\Omega^\mathrm{dyn}$ [deg] & \multicolumn{2}{c}{$284.2_{-1.8}^{+1.3}$} & $104.2_{-1.8}^{+1.3}$ & \multicolumn{2}{c}{$272.4_{-2.1}^{+2.4}$} & $92.4_{-2.1}^{+2.4}$ \\
  $i_\mathrm{inv}$ [deg] & \multicolumn{3}{c}{$89.685_{-0.017}^{+0.015}$} & \multicolumn{3}{c}{$87.514_{-0.383}^{+0.437}$} \\
  $\Omega_\mathrm{inv}$ [deg] & \multicolumn{3}{c}{$-3.506_{-0.062}^{+0.045}$} & \multicolumn{3}{c}{$-10.149_{-0.135}^{+0.127}$} \\
  \hline
  mass ratio $[q=m_\mathrm{sec}/m_\mathrm{pri}]$ & \multicolumn{2}{c}{$0.972_{-0.009}^{+0.008}$} & $0.866_{-0.007}^{+0.009}$ & \multicolumn{2}{c}{$0.736_{-0.006}^{+0.006}$} & $0.418_{-0.004}^{+0.004}$ \\
  $K_\mathrm{pri}$ [km\,s$^{-1}$] & \multicolumn{2}{c}{$50.06_{-0.77}^{+1.10}$} & $23.41_{-0.44}^{+0.74}$ & \multicolumn{2}{c}{$40.69_{-0.28}^{+0.22}$} & $10.52_{-0.11}^{+0.11}$ \\ 
  $K_\mathrm{sec}$ [km\,s$^{-1}$] & \multicolumn{2}{c}{$51.54_{-0.70}^{+1.04}$} & $27.09_{-0.37}^{+0.46}$ & \multicolumn{2}{c}{$55.28_{-0.37}^{+0.25}$} & $25.18_{-0.09}^{+0.09}$ \\ 
  \hline
  \multicolumn{7}{c}{Apsidal and nodal motion related parameters} \\
  \hline
$P_\mathrm{apse}$ [year] & \multicolumn{2}{c}{$28.56_{-0.15}^{+0.12}$} & $199.8_{-0.7}^{+0.8}$ & \multicolumn{2}{c}{$309_{-4}^{+3}$} & $1204_{-4}^{+3}$\\ 
$P_\mathrm{apse}^\mathrm{dyn}$ [year] & \multicolumn{2}{c}{$13.34_{-0.06}^{+0.05}$} & $22.21_{-0.07}^{+0.06}$ & \multicolumn{2}{c}{$131.4_{-0.7}^{+0.9}$} & $190_{-2}^{+2}$\\ 
$P_\mathrm{node}^\mathrm{dyn}$ [year] & \multicolumn{3}{c}{$24.99_{-0.08}^{+0.10}$}  & \multicolumn{3}{c}{$225_{-3}^{+3}$} \\
$\Delta\omega_\mathrm{3b}$ [arcsec/cycle] & \multicolumn{2}{c}{$1329_{-6}^{+7}$} & $4249_{-16}^{+16}$ & \multicolumn{2}{c}{$441_{-4}^{+6}$} & $2863_{-7}^{+10}$ \\ 
$\Delta\omega_\mathrm{GR}$ [arcsec/cycle] & \multicolumn{2}{c}{$0.447_{-0.014}^{+0.020}$} & $0.111_{-0.004}^{+0.005}$ & \multicolumn{2}{c}{$0.399_{-0.004}^{+0.003}$} & $0.0553_{-0.0006}^{+0.0005}$ \\ 
$\Delta\omega_\mathrm{tide}$ [arcsec/cycle] & \multicolumn{2}{c}{$0.026_{-0.004}^{+0.005}$} & ... & \multicolumn{2}{c}{$0.050_{-0.002}^{+0.001}$} & ... \\ 
  \hline  
\multicolumn{7}{c}{stellar parameters} \\
\hline
   & Aa & Ab &  B & Aa & Ab &  B \\
  \hline
 \multicolumn{7}{c}{Relative quantities} \\
  \hline
 fractional radius [$R/a$]  & $0.0274_{-0.0007}^{+0.0008}$ & $0.0266_{-0.0008}^{+0.0010}$  & $0.00944_{-0.00024}^{+0.00022}$ & $0.0315_{-0.0002}^{+0.0002}$ & $0.0188_{-0.0002}^{+0.0002}$  & $0.00190_{-0.00004}^{+0.00004}$ \\
 temperature relative to $(T_\mathrm{eff})_\mathrm{Aa}$ & $1$ & $0.9786_{-0.0075}^{+0.0069}$ & $1.3303_{-0.0159}^{+0.0175}$ & $1$ & $0.8877_{-0.0042}^{+0.0077}$ & $0.8824_{-0.0035}^{+0.0036}$ \\
 fractional flux [in \textit{Kepler}-band] & $0.0203_{-0.0012}^{+0.0012}$ & $0.0174_{-0.0009}^{+0.0010}$ & $0.9623_{-0.0020}^{+0.0017}$ & $0.6875_{-0.0093}^{+0.0098}$ & $0.1588_{-0.0050}^{+0.0048}$ & $0.1532_{-0.0061}^{+0.0073}$ \\
 fractional flux [in \textit{TESS}-band] & $-$ & $-$ & $-$ & $0.4995_{-0.0165}^{+0.01791}$ & $0.1227_{-0.0016}^{+0.0016}$ & $0.1084_{-0.0050}^{+0.0048}$ \\
 \hline
 \multicolumn{7}{c}{Physical Quantities} \\
  \hline 
 $m$ [M$_\odot$] & $0.588_{-0.026}^{+0.038}$ & $0.571_{-0.028}^{+0.038}$ & $1.003_{-0.055}^{+0.081}$ & $1.774_{-0.029}^{+0.024}$ & $1.303_{-0.016}^{+0.018}$ & $1.285_{-0.029}^{+0.025}$ \\
 $R$ [R$_\odot$] & $0.588_{-0.025}^{+0.031}$ & $0.571_{-0.027}^{+0.034}$ & $1.980_{-0.077}^{+0.087}$ & $2.198_{-0.017}^{+0.017}$ & $1.310_{-0.020}^{+0.022}$ & $1.281_{-0.037}^{+0.034}$ \\
 $T_\mathrm{eff}$ [K]& $4134_{-49}^{+45}$ & $4041_{-41}^{+44}$ & $5495_{-45}^{+42}$ & $7062_{-88}^{+53}$ & $6260_{-47}^{+59}$ & $6228_{-64}^{+47}$ \\
 $L_\mathrm{bol}$ [L$_\odot$] & $0.091_{-0.010}^{+0.013}$ & $0.079_{-0.010}^{+0.011}$ & $3.234_{-0.279}^{+0.240}$ & $10.822_{-0.626}^{+0.353}$ & $2.384_{-0.132}^{+0.103}$ & $2.225_{-0.201}^{+0.157}$ \\
 $M_\mathrm{bol}$ & $7.38_{-0.14}^{+0.12}$ & $7.53_{-0.14}^{+0.15}$ & $3.50_{-0.08}^{+0.10}$ & $2.18_{-0.03}^{+0.06}$ & $3.83_{-0.05}^{+0.06}$ & $3.90_{-0.07}^{+0.10}$ \\
 $M_V           $ & $8.27_{-0.14}^{+0.15}$ & $8.50_{-0.12}^{+0.16}$ & $3.62_{-0.08}^{+0.11}$ & $2.10_{-0.03}^{+0.07}$ & $3.80_{-0.05}^{+0.07}$ & $3.88_{-0.08}^{+0.11}$ \\
 $\log g$ [dex] & $4.668_{-0.017}^{+0.018}$ & $4.679_{-0.021}^{+0.022}$ & $3.846_{-0.014}^{+0.016}$ & $4.000_{-0.006}^{+0.007}$ & $4.317_{-0.009}^{+0.009}$ & $4.331_{-0.015}^{+0.016}$ \\
 \hline
\multicolumn{7}{c}{Global system parameters} \\
  \hline
$\log$(age) [dex] &\multicolumn{3}{c}{$9.934_{-0.107}^{+0.075}$} & \multicolumn{3}{c}{$9.016_{-0.038}^{+0.028}$} \\
$[M/H]$  [dex]    &\multicolumn{3}{c}{$-0.230_{-0.094}^{+0.148}$} & \multicolumn{3}{c}{$0.372_{-0.107}^{+0.059}$} \\
$E(B-V)$ [mag]    &\multicolumn{3}{c}{$0.088_{-0.012}^{+0.015}$} & \multicolumn{3}{c}{$0.351_{-0.026}^{+0.014}$} \\
extra light $\ell_4$ [in \textit{TESS}-band] & \multicolumn{3}{c}{...} & \multicolumn{3}{c}{$0.270_{-0.046}^{+0.019}$} \\
$(M_V)_\mathrm{tot}$  &\multicolumn{3}{c}{$3.59_{-0.08}^{+0.11}$} & \multicolumn{3}{c}{$1.73_{-0.03}^{+0.07}$} \\
distance [pc]           &\multicolumn{3}{c}{$2443_{-109}^{+119}$} & \multicolumn{3}{c}{$1374_{-14}^{+16}$} \\  
\hline
\end{tabular}
\end{table*}

\begin{table*}
 \centering
\caption{The same as in Table~\ref{tab: syntheticfit_KIC69640435653126} but for KICs\,5731312 and 8023317.}
 \label{tab: syntheticfit_KIC57313128023317}
\begin{tabular}{@{}lllllll}
\hline
 & \multicolumn{3}{c}{KIC\,5731312} & \multicolumn{3}{c}{KIC\,8023317} \\
\hline
\multicolumn{7}{c}{orbital elements} \\
\hline
   & \multicolumn{2}{c}{Aa--Ab} & A--B  & \multicolumn{2}{c}{Aa--Ab} & A--B  \\
  \hline
  $t_0$ [BJD - 2400000]& \multicolumn{3}{c}{$54953.0$} & \multicolumn{3}{c}{$54953.0$} \\
  $P$ [days] & \multicolumn{2}{c}{$7.946521_{-0.000010}^{+0.000011}$} & $917.00_{-0.92}^{+0.98}$ & \multicolumn{2}{c}{$16.57544_{-0.00020}^{+0.00021}$} & $605.4_{-2.2}^{+2.3}$  \\
  $a$ [R$_\odot$] & \multicolumn{2}{c}{$18.20_{-0.09}^{+0.10}$} & $448.0_{-2.1}^{+2.5}$ & \multicolumn{2}{c}{$32.63_{-0.42}^{+0.45}$} & $368.1_{-4.4}^{+5.5}$ \\
  $e$ & \multicolumn{2}{c}{$0.45866_{-0.00033}^{+0.00048}$} & $0.5745_{-0.0079}^{+0.0077}$ & \multicolumn{2}{c}{$0.27889_{-0.00068}^{+0.00041}$} & $0.252_{-0.013}^{+0.013}$ \\
  $\omega$ [deg]& \multicolumn{2}{c}{$206.82_{-0.12}^{+0.15}$} & $36.06_{-1.59}^{+1.38}$ & \multicolumn{2}{c}{$152.90_{-0.15}^{+0.32}$} & $159.14_{-2.66}^{+2.91}$ \\ 
  $i$ [deg] & \multicolumn{2}{c}{$87.378_{-0.107}^{+0.153}$} & $67.649_{-0.354}^{+0.487}$ & \multicolumn{2}{c}{$87.019_{-0.046}^{+0.045}$} & $71.836_{-0.267}^{+0.369}$ \\
  $\mathcal{T}_0^\mathrm{inf}$ [BJD - 2400000]& \multicolumn{2}{c}{$54968.0942_{-0.0002}^{+0.0002}$} & ... & \multicolumn{2}{c}{$54979.7311_{-0.0003}^{+0.0003}$} & $-$ \\
  $\tau$ [BJD - 2400000]& \multicolumn{2}{c}{$54967.5686_{-0.0017}^{+0.0023}$} & $54830.70_{-1.97}^{+1.70}$ & \multicolumn{2}{c}{$54975.7719_{-0.0061}^{+0.0135}$} & $55023.12_{-4.22}^{+4.82}$ \\
  $\Omega$ [deg] & \multicolumn{2}{c}{$0.0$} & $35.18_{-0.43}^{+0.41}$ & \multicolumn{2}{c}{$0.0$} & $-55.25_{-0.81}^{+0.86}$ \\
  $i_\mathrm{mut}$ [deg] & \multicolumn{3}{c}{$39.41_{-0.31}^{+0.32}$} & \multicolumn{3}{c}{$55.71_{-0.82}^{+0.79}$} \\
  $\omega^\mathrm{dyn}$ [deg]& \multicolumn{2}{c}{$263.9_{-0.6}^{+0.5}$} & $281.1_{-1.7}^{+1.5}$ & \multicolumn{2}{c}{$79.4_{-0.3}^{+0.4}$} & $255.9_{-2.6}^{+2.7}$ \\
  $i^\mathrm{dyn}$ [deg] & \multicolumn{2}{c}{$27.20_{-0.34}^{+0.33}$} & $12.22_{-0.37}^{+0.37}$ & \multicolumn{2}{c}{$33.38_{-0.88}^{+0.85}$} & $22.33_{-0.53}^{+0.52}$ \\
  $\Omega^\mathrm{dyn}$ [deg] & \multicolumn{2}{c}{$118.9_{-0.6}^{+0.6}$} & $298.9_{-0.6}^{+0.6}$ & \multicolumn{2}{c}{$257.9_{-0.4}^{+0.5}$} & $77.9_{-0.4}^{+0.5}$ \\
  $i_\mathrm{inv}$ [deg] & \multicolumn{3}{c}{$73.22_{-0.24}^{+0.23}$} & \multicolumn{3}{c}{$78.59_{-0.21}^{+0.28}$} \\
  $\Omega_\mathrm{inv}$ [deg] & \multicolumn{3}{c}{$23.62_{-0.39}^{+0.39}$} & \multicolumn{3}{c}{$-32.62_{-0.84}^{+0.90}$} \\
  \hline
  mass ratio $[q=m_\mathrm{sec}/m_\mathrm{pri}]$ & \multicolumn{2}{c}{$0.665_{-0.042}^{+0.031}$} & $0.120_{-0.004}^{+0.004}$ & \multicolumn{2}{c}{$0.303_{-0.013}^{+0.010}$} & $0.079_{-0.003}^{+0.003}$ \\
  $K_\mathrm{pri}$ [km\,s$^{-1}$] & \multicolumn{2}{c}{$52.05_{-2.28}^{+1.63}$} & $2.99_{-0.08}^{+0.08}$  & \multicolumn{2}{c}{$23.99_{-0.50}^{+0.48}$} & $2.25_{-0.07}^{+0.07}$ \\ 
  $K_\mathrm{sec}$ [km\,s$^{-1}$] & \multicolumn{2}{c}{$78.48_{-1.40}^{+1.50}$} & $24.96_{-0.21}^{+0.20}$ & \multicolumn{2}{c}{$79.45_{-1.56}^{+2.15}$} & $28.51_{-0.46}^{+0.34}$ \\ 
  \hline
  \multicolumn{7}{c}{Apsidal motion related parameters} \\
  \hline
 $P_\mathrm{apse}$ [year] & \multicolumn{2}{c}{$-4787_{-306}^{+280}$} & $18193_{-1238}^{+1368}$ & \multicolumn{2}{c}{$-653_{-36}^{+31}$} & $-6567_{-929}^{+676}$ \\ 
 $P_\mathrm{apse}^\mathrm{dyn}$ [year] & \multicolumn{2}{c}{$1298_{-33}^{+34}$} & $873_{-18}^{+18}$ & \multicolumn{2}{c}{$-1707_{-342}^{+229}$} & $1103_{-54}^{+56}$ \\ 
$P_\mathrm{node}^\mathrm{dyn}$ [year] & \multicolumn{3}{c}{$899_{-18}^{+18}$}  & \multicolumn{3}{c}{$884_{-23}^{+24}$} \\
$\Delta\omega_\mathrm{3b}$ [arcsec/cycle] & \multicolumn{2}{c}{$-7.1_{-0.4}^{+0.3}$} & $179_{-13}^{+13}$ & \multicolumn{2}{c}{$-91_{-5}^{+5}$} & $-327_{-38}^{+41}$ \\ 
$\Delta\omega_\mathrm{GR}$ [arcsec/cycle] & \multicolumn{2}{c}{$0.736_{-0.007}^{+0.008}$} & $0.0395_{-0.0005}^{+0.0005}$ & \multicolumn{2}{c}{$0.465_{-0.012}^{+0.013}$} & $0.044_{-0.001}^{+0.001}$ \\ 
$\Delta\omega_\mathrm{tide}$ [arcsec/cycle] & \multicolumn{2}{c}{$0.47_{-0.02}^{+0.02}$} & ... & \multicolumn{2}{c}{$0.516_{-0.035}^{+0.034}$} & ... \\ 
  \hline  
\multicolumn{7}{c}{stellar parameters} \\
\hline
   & Aa & Ab &  B & Aa & Ab &  B \\
  \hline
 \multicolumn{7}{c}{Relative quantities} \\
  \hline
 fractional radius [$R/a$]  & $0.0399_{-0.0005}^{+0.0007}$ & $0.0277_{-0.0015}^{+0.0011}$  & $0.000409_{-0.000009}^{+0.000009}$ & $0.0600_{-0.0007}^{+0.0007}$ & $0.0121_{-0.0002}^{+0.0002}$  & $0.000474_{-0.000015}^{+0.000017}$ \\
 temperature relative to $(T_\mathrm{eff})_\mathrm{Aa}$ & $1$ & $0.7272_{-0.0225}^{+0.0189}$ & $0.5733_{-0.0053}^{+0.0058}$ & $1$ & $0.5402_{-0.0061}^{+0.0056}$ & $0.4393_{-0.0063}^{+0.0056}$ \\
 fractional flux [in \textit{Kepler}-band] & $0.8986_{-0.0202}^{+0.0207}$ & $0.0984_{-0.0208}^{+0.0200}$ & $0.0031_{-0.0003}^{+0.0003}$ & $0.9978_{-0.0002}^{+0.0001}$ & $0.0021_{-0.0001}^{+0.0002}$ & $0.0001_{-0.00001}^{+0.00001}$ \\
 fractional flux [in \textit{TESS}-band] & $0.8774_{-0.0220}^{+0.0233}$ & $0.1194_{-0.0233}^{+0.0218}$ & $0.0033_{-0.0003}^{+0.0003}$ & $0.8479_{-0.0177}^{+0.0256}$ & $0.0023_{-0.0002}^{+0.0002}$ & $0.0001_{-0.00001}^{+0.00001}$ \\
 fractional flux [in \textit{r'}-band] & $0.9086_{-0.0193}^{+0.0197}$ & $0.0892_{-0.0198}^{+0.0192}$ & $0.0022_{-0.0002}^{+0.0002}$ & ... & ... & ... \\
 \hline
 \multicolumn{7}{c}{Physical Quantities} \\
  \hline 
 $m$ [M$_\odot$] & $0.773_{-0.013}^{+0.008}$ & $0.510_{-0.026}^{+0.022}$ & $0.153_{-0.005}^{+0.004}$ & $1.300_{-0.058}^{+0.071}$ & $0.392_{-0.009}^{+0.008}$ & $0.134_{-0.006}^{+0.006}$ \\
 $R$ [R$_\odot$] & $0.728_{-0.009}^{+0.009}$ & $0.505_{-0.029}^{+0.023}$ & $0.184_{-0.004}^{+0.004}$ & $1.963_{-0.046}^{+0.029}$ & $0.394_{-0.010}^{+0.008}$ & $0.174_{-0.007}^{+0.008}$ \\
 $T_\mathrm{eff}$ [K]& $5079_{-56}^{+47}$ & $3693_{-91}^{+68}$ & $2911_{-26}^{+28}$ & $5869_{-73}^{+66}$ & $3166_{-42}^{+45}$ & $2575_{-36}^{+35}$ \\
 $L_\mathrm{bol}$ [L$_\odot$] & $0.316_{-0.018}^{+0.020}$ & $0.042_{-0.008}^{+0.008}$ & $0.0022_{-0.0002}^{+0.0002}$ & $4.084_{-0.215}^{+0.196}$ & $0.014_{-0.001}^{+0.001}$ & $0.0012_{-0.0001}^{+0.0001}$ \\
 $M_\mathrm{bol}$ & $6.02_{-0.07}^{+0.06}$ & $8.20_{-0.18}^{+0.23}$ & $11.43_{-0.08}^{+0.08}$ & $3.24_{-0.05}^{+0.06}$ & $9.41_{-0.06}^{+0.06}$ & $12.07_{-0.12}^{+0.11}$ \\
 $M_V           $ & $6.27_{-0.08}^{+0.09}$ & $9.56_{-0.26}^{+0.36}$ & $14.68_{-0.19}^{+0.19}$ & $3.27_{-0.06}^{+0.07}$ & $11.91_{-0.20}^{+0.18}$ & $16.72_{-0.17}^{+0.15}$ \\
 $\log g$ [dex] & $4.599_{-0.006}^{+0.007}$ & $4.739_{-0.021}^{+0.027}$ & $5.095_{-0.007}^{+0.007}$ & $3.967_{-0.011}^{+0.013}$ & $4.839_{-0.009}^{+0.013}$ & $5.077_{-0.020}^{+0.021}$ \\
 \hline
\multicolumn{7}{c}{Global system parameters} \\
  \hline
$\log$(age) [dex] &\multicolumn{3}{c}{$9.775_{-0.046}^{+0.043}$} &\multicolumn{3}{c}{$9.639_{-0.064}^{+0.036}$} \\
$[M/H]$  [dex]    &\multicolumn{3}{c}{$-0.122_{-0.040}^{+0.027}$} &\multicolumn{3}{c}{$0.330_{-0.150}^{+0.064}$} \\
$E(B-V)$ [mag]    &\multicolumn{3}{c}{$0.038_{-0.011}^{+0.009}$} &\multicolumn{3}{c}{$0.080_{-0.019}^{+0.026}$} \\
extra light $\ell_4$ [in \textit{TESS}-band] & \multicolumn{3}{c}{...} & \multicolumn{3}{c}{$0.150_{-0.026}^{+0.018}$} \\
$(M_V)_\mathrm{tot}$  &\multicolumn{3}{c}{$6.22_{-0.07}^{+0.08}$} &\multicolumn{3}{c}{$3.27_{-0.06}^{+0.07}$} \\
distance [pc]         &\multicolumn{3}{c}{$353_{-7}^{+6}$} &\multicolumn{3}{c}{$814_{-22}^{+13}$} \\  
\hline
\end{tabular}

\end{table*}

The median values of the orbital and physical parameters, as well as some derived quantities, of the four triple systems, computed from the MCMC posteriors and their $1\sigma$ uncertainties are tabulated in Tables~\ref{tab: syntheticfit_KIC69640435653126} and \ref{tab: syntheticfit_KIC57313128023317}. Furthermore, the observed vs.~model lightcurves are plotted in Figs.~\ref{fig:keplerlcs}--\ref{fig:K57313128023317lcfits}, while the observed vs.~model ETV curves are shown in Fig.~\ref{fig:ETVswithfit}.

\section{Discussion}
\label{sec:discussion}

In this section, first we briefly describe the photodynamical results for each individual system, and then we study the short and longer term dynamical evolutions of our triples. For the investigation of the short- and long-term dynamical evolutions we integrate the future motion of the triple star systems with two different numerical integrators. For the short-term ($\leq1$\,Myr-long) integration we use the very same integrator which is built into {\sc Lightcurvefactory}, and described in \citet{borkovitsetal19a,borkovitsetal19b}.  Here we note that this integrator works directly with the equations of the orbital motions, including point-mass, tidal and relativistic contributions, and the Eulerian equations of uni-axial stellar rotations. Its main advantage is that it returns at every integration step the spatial coordinates and velocities of each star and, therefore, can be used directly to generate the system lightcurve.  It also gives the osculating orbital elements at each instant and, hence, the observational consequences of the motions can be easily studied.  The disadvantages are, however, that neither stellar evolution nor tidal dissipation are built in and, therefore, it cannot be used for realistic simulations on much longer timescales. Moreover, another disadvantage is that it is much slower than the method we use for the investigations of the long-term (up to several Gyrs) dynamical evolution of the systems. For this latter approach we use the triple stellar evolution code \textit{TRES} \citep{Too16} which incorporates secular dynamical evolution including the quadrupole \& octupole level interactions, stellar evolution including stellar winds \& stellar interaction processes such as mass transfer, common-envelope evolution, mergers, tidal forces, and gravitational wave emission.\footnote{Note, we have also numerically integrated the dynamical evolution of the systems with \textit{TRES} on the shorter timescales and obtained similar (dynamical) results to those found with the built in integrator of {\sc Lightcurvefactory} for all four systems considered in this paper.}

Finally, we discuss the implications of our findings on the system parameters for the formation processes of multiple stellar systems in general.

\subsection{KIC 69640403}
\label{sec:5.1}

This is the only triply eclipsing triple among our sample, and the system is found to be nearly flat. In contrast to the previous findings of \citet{borkovitsetal15,borkovitsetal16}, here we find the mutual inclination of the inner and outer orbital planes to be $i_\mathrm{mut}=4\fdg14\pm0\fdg09$. This is also the only system in the present study for which the outer stellar component was found to be the most massive one.  This tertiary is a slightly evolved twin of our Sun ($m_\mathrm{B}=1.00\pm0.08\,\mathrm{M}_\odot$; $R_\mathrm{B}=1.98\pm0.09\,\mathrm{R}_\odot$; $T_\mathrm{eff,B}=5500\pm50\,\mathrm{K}$), and it emits more than 96\% of the total flux of the system (at least in the \textit{Kepler} photometric band). The inner binary consists of two very similar red dwarf stars ($m_\mathrm{Aa}=0.59\pm0.04\,\mathrm{M}_\odot$; $R_\mathrm{Aa}=0.59\pm0.03\,\mathrm{R}_\odot$; $T_\mathrm{eff,Aa}=4130\pm50\,\mathrm{K}$ and $m_\mathrm{Ab}=0.57\pm0.04\,\mathrm{M}_\odot$; $R_\mathrm{Ab}=0.57\pm0.03\,\mathrm{R}_\odot$; $T_\mathrm{eff,Ab}=4040\pm40\,\mathrm{K}$, respectively). This triple was found to be the oldest ($\tau=9_{-2}^{+1}$\,Gyr) and most distant ($d=2450\pm120$\,pc) among the set of studied systems.  The distance is greater by $\approx2\sigma$ than the Gaia EDR3 distance, but such an inconsistency, in the case of a triple system with an outer period near 2/3 of a year, might be expected. Both the inner and outer mass ratios are relatively close to unity ($q_\mathrm{in}=0.97\pm0.01$ and $q_\mathrm{out}=0.87\pm0.01$) which, together with the flatness and compact nature of the system, imply that this system might have formed through a sequential disk instability mechanism \citep{tokovininmoe20,tokovinin21} which may lead to compact triple systems with both the inner and outer mass ratios close to unity \citep[double twins,][]{tokovinin18}.

\begin{figure*}
\begin{center}
\includegraphics[width=0.50 \textwidth]{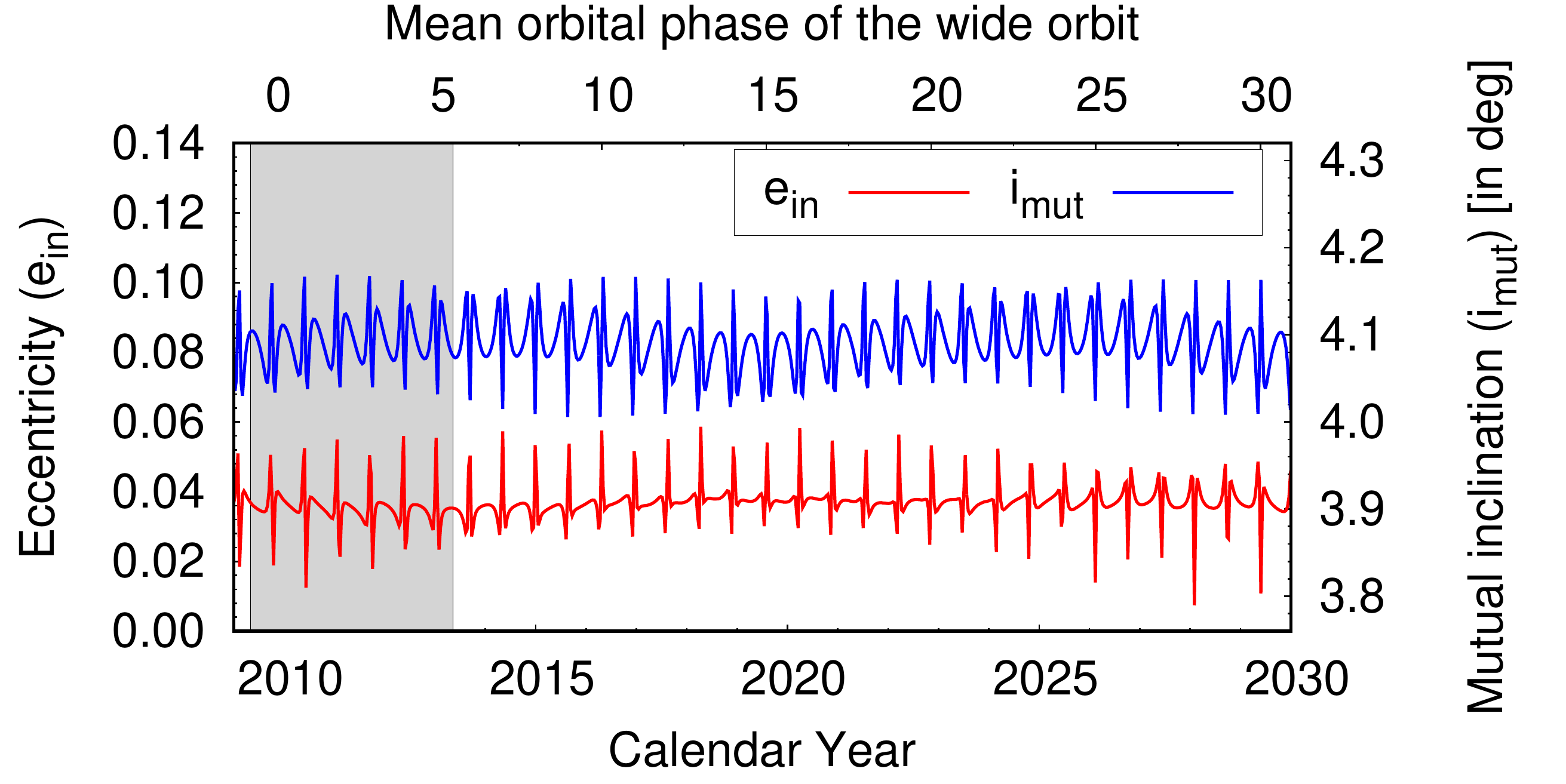}\includegraphics[width=0.50 \textwidth]{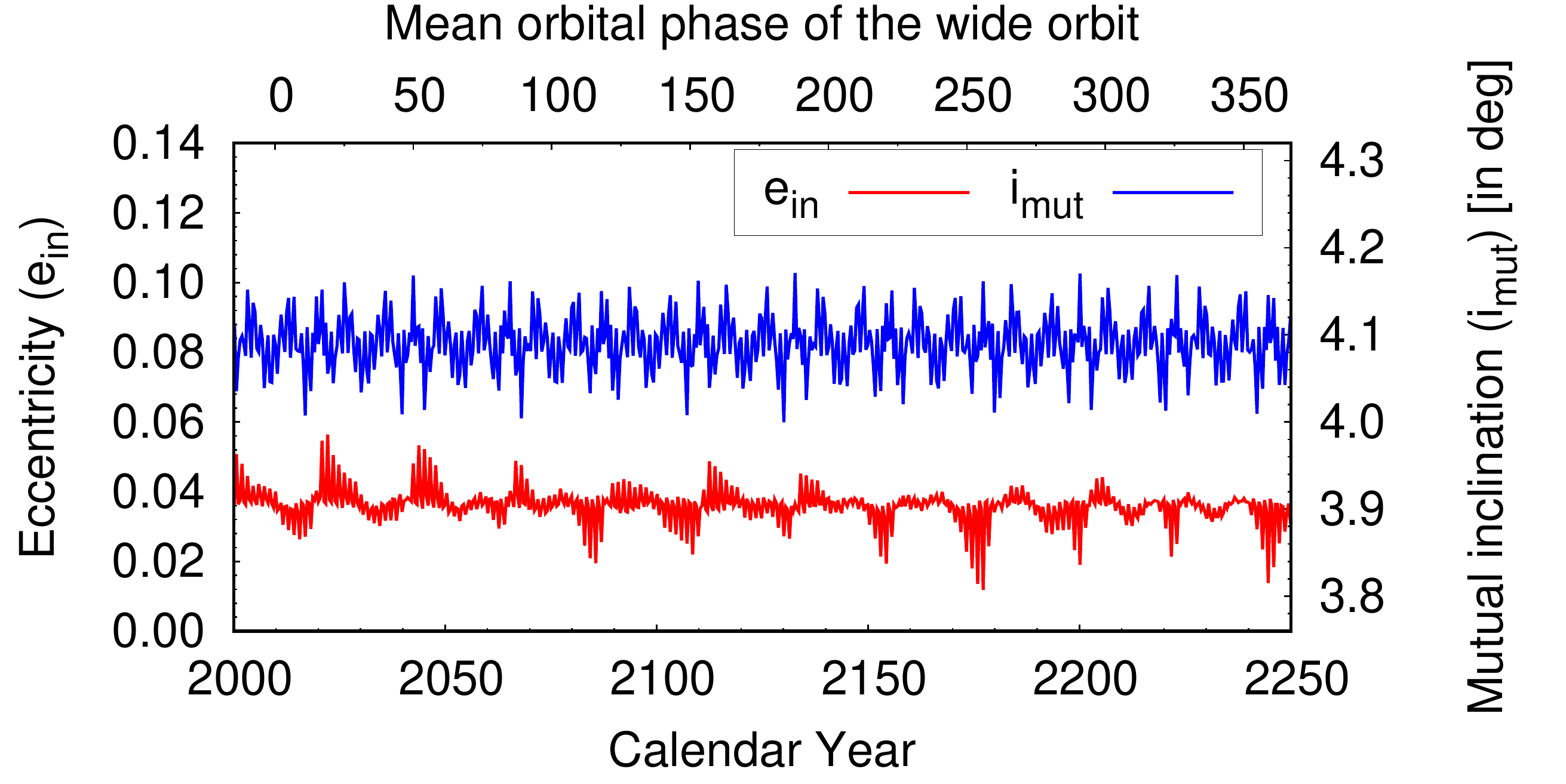}
\includegraphics[width=0.50 \textwidth]{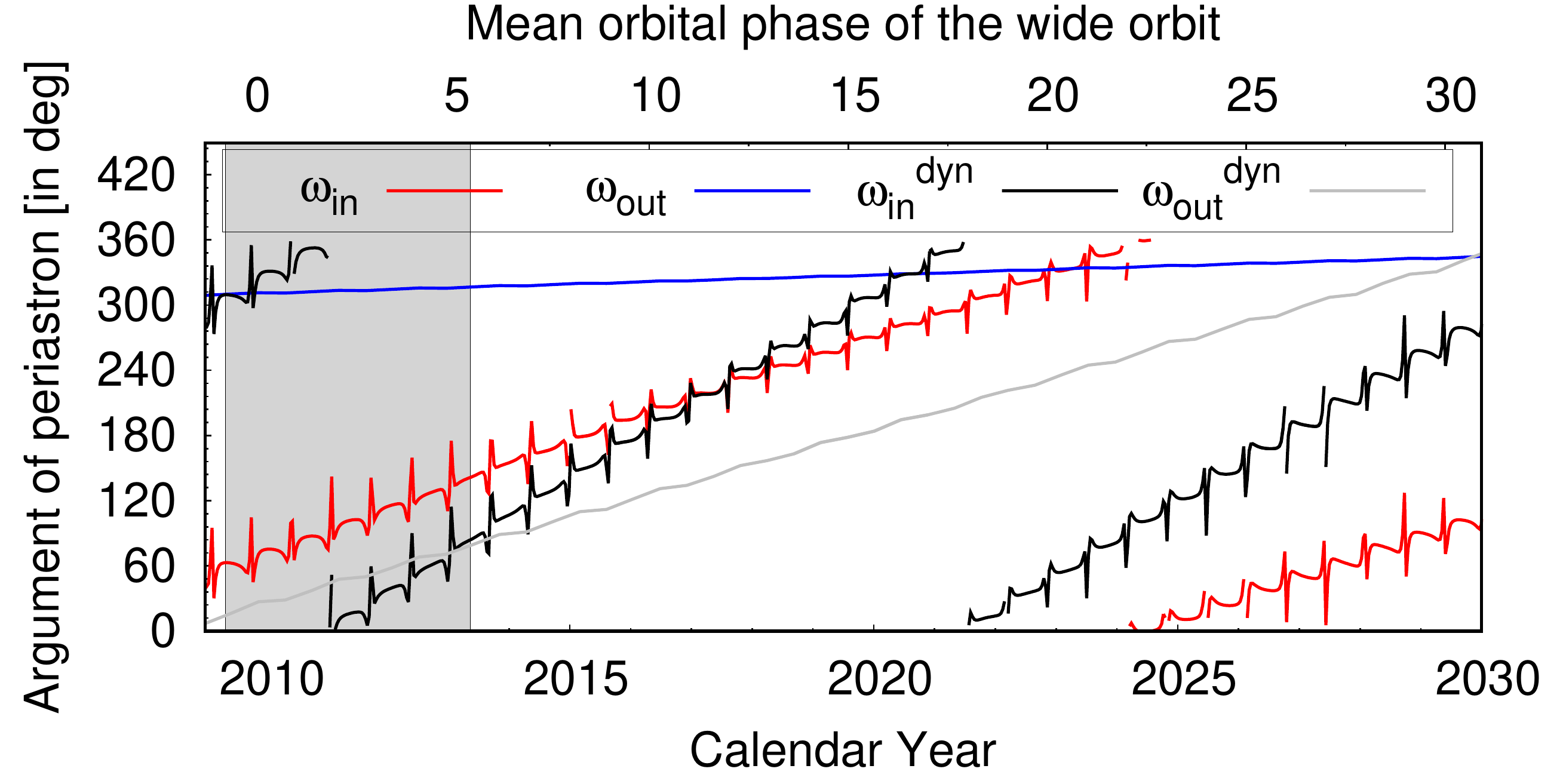}\includegraphics[width=0.50 \textwidth]{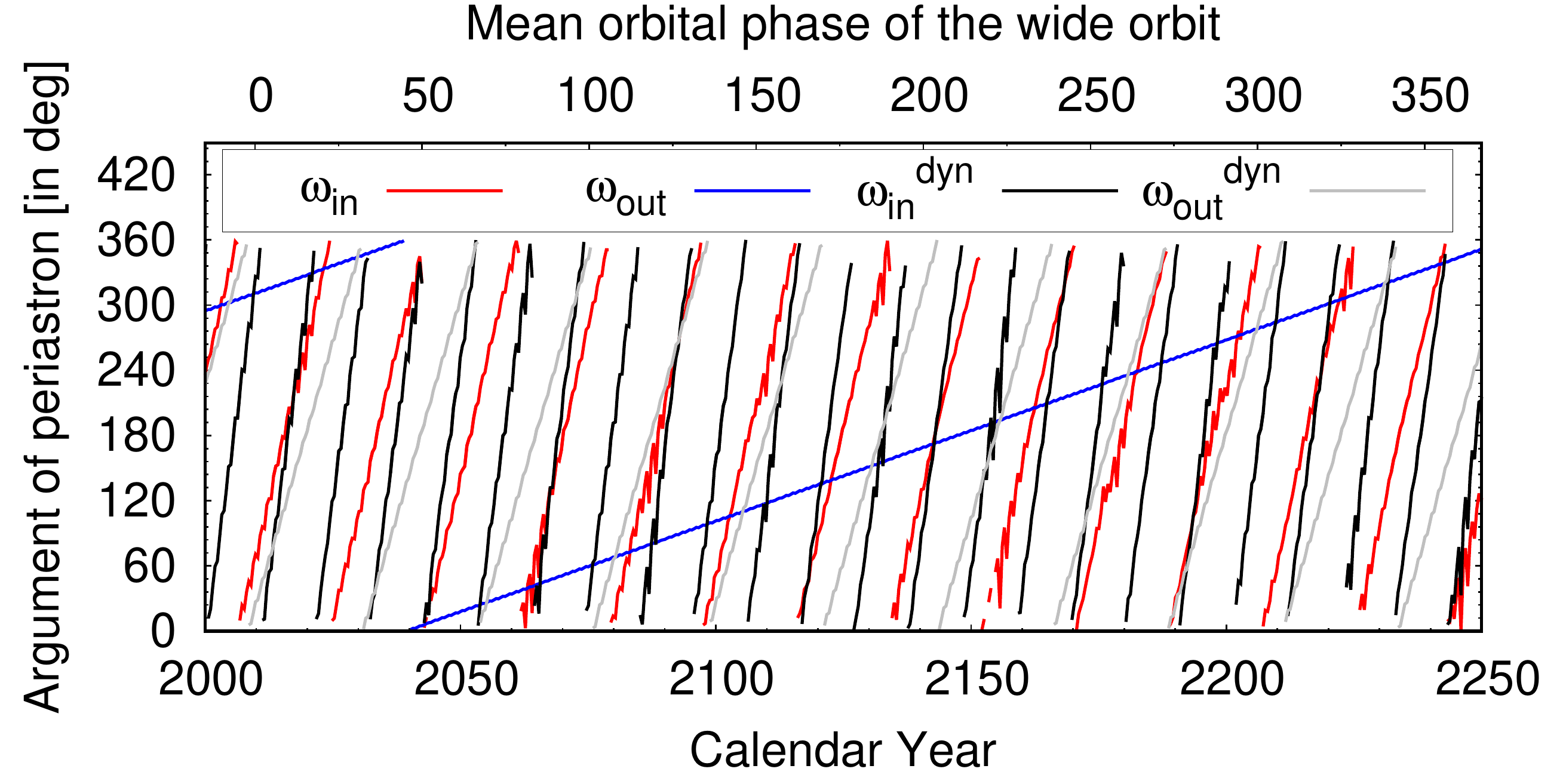}
\includegraphics[width=0.50 \textwidth]{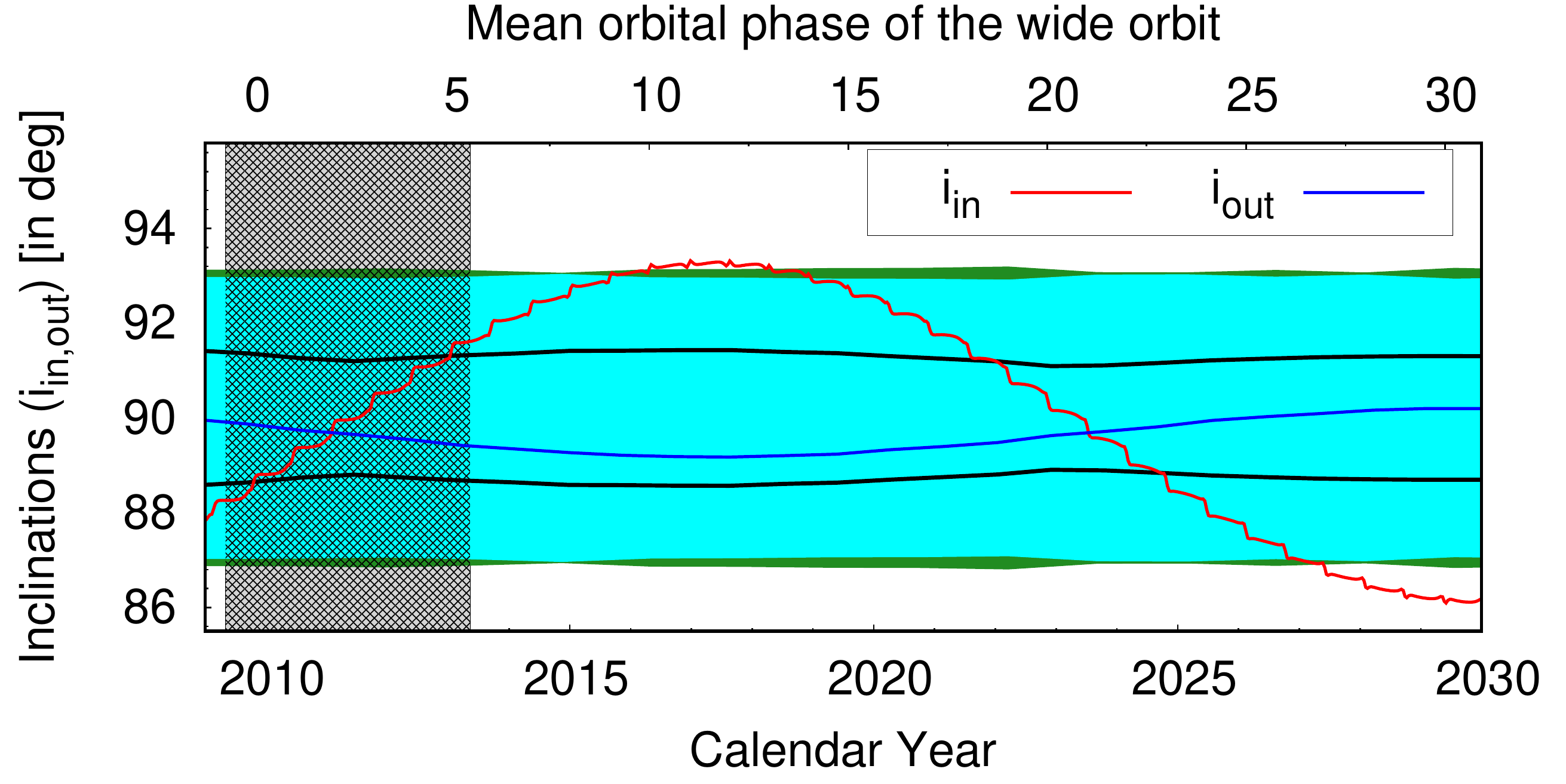}\includegraphics[width=0.50 \textwidth]{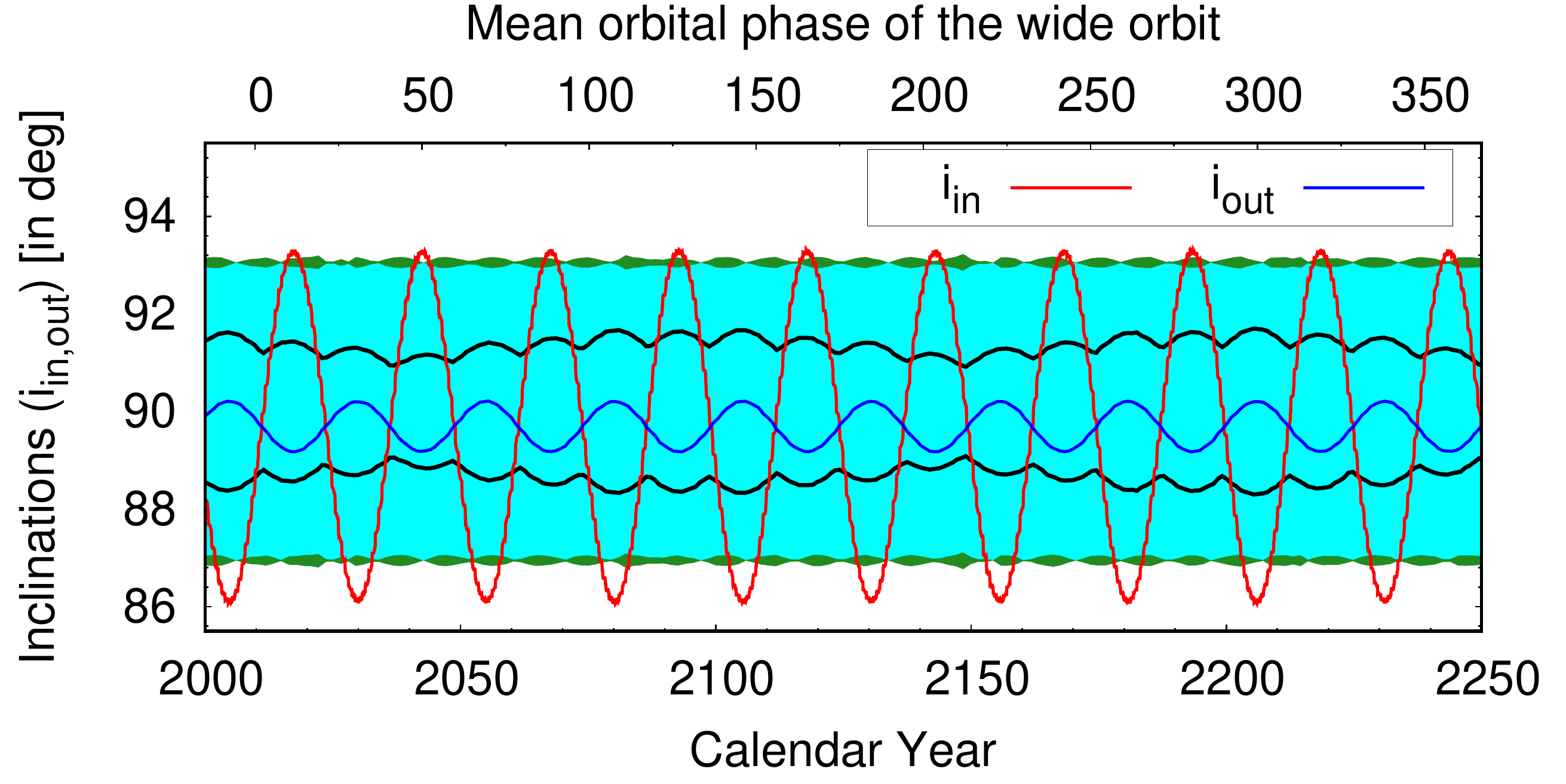}
\caption{Variations of some of the osculating orbital elements in the triple system KIC 6964043. The left panels represent short timescale variations from 2009 to 2030, while the right panels show 250 years. The vertical shaded area in the left panels denote the time domain of the original \textit{Kepler} mission. Note, that the medium-period perturbation effects, especially the bumps around the periastron passages of the third component, are also nicely visible on the left panels. {\em Upper panels:} Variations of the inner eccentricity ($e_\mathrm{in}$) and the mutual inclination ($i_\mathrm{mut}$). 
 {\em Middle panels:} The variations of the observable and dynamical arguments of periastrons of the inner and outer orbits ($\omega_\mathrm{in,out}$ and $\omega_\mathrm{in,out}^\mathrm{dyn}$, respectively). 
{\em Lower panels:} Variations of the observable inclinations of the inner and outer orbital planes ($i_\mathrm{in}$, $i_\mathrm{out}$, respectively). The green area represents the domain where at least one of the inner EB's eclipses should be observed, while the cyan area stands for the inclination domain where both eclipses can be observed. Moreover, the horizontal black lines stand for the upper and lower outer inclination ($i_\mathrm{out}$) limits of the domain where outer, third-body eclipses might occur. 
See text for further details. }
\label{fig:K6964043orbelements_numint} 
\end{center}
\end{figure*}  

Regarding the orbital parameters, while the inner orbit is only slightly eccentric ($e_\mathrm{in}=0.0037\pm0.0005$), the eccentricity of the outer orbit is moderately high ($e_\mathrm{out}=0.4781\pm0.0006$).  As a consequence, the periastron distance of the outer orbit is $a_\mathrm{out}(1-e_\mathrm{out})\approx5.1\,a_\mathrm{in}$, which places the system quite close to the dynamical stability limit; however, the system is definitely within the dynamically stable region. This is clearly illustrated in the upper panels of Fig.~\ref{fig:K6964043orbelements_numint} which show the variations of the inner eccentricity and the mutual inclination over the first 250 years of our one million year-long numerical integration. This short-term stability is mainly guaranteed by the nearly perfect flatness of the system, which leads to the suppression of any long-term or secular perturbations in the semi-major axes, as well as to the almost complete disappearance of the quadrupole order long-term perturbations in the eccentricities and the mutual inclination. Moreover, due to the nearly equal masses of the inner binary, the octupole-order terms also remain very small. 

In the middle and bottom panels of Fig.~\ref{fig:K6964043orbelements_numint} we show decades-long (left) and centuries-long (right) variations of some further orbital elements which have readily detectable observational consequences. In the middle panels the time evolution of the observational and dynamical arguments of periastrons ($\omega_\mathrm{in,out}$ and $\omega^\mathrm{dyn}_\mathrm{in,out}$, respectively) are shown.  Similar to the other three triples, the apsidal advances of both the inner and outer orbits are clearly dominated by third-body perturbations over the relativistic and tidal effects (see the corresponding rows in Table~\ref{tab: syntheticfit_KIC69640435653126}).  Due to the flatness of the system, again the apsidal motion is practically linear in time, apart from the small amplitude $P_\mathrm{out}$-medium period timescale variations. The theoretical\footnote{For the calculations we used the formulae described in \citet{borkovitsetal15}, Appendix C.} inner and outer apsidal motion periods in the dynamical frame are $P_\mathrm{apse}^\mathrm{dyn}=13$ and 199\,years, respectively. Naturally, in the observational frame a much longer apsidal motion period will be observed due to the counterbalancing effect of the nodal regression.\footnote{It can be readily shown that the observable and dynamical apsidal motion rates relate each other as $\Delta\omega=\Delta\omega^\mathrm{dyn}+\Delta\Omega^\mathrm{dyn}\cos i_\mathrm{dyn}-\Delta\Omega\cos i$ \citep[see, e.g.][]{borkovitsetal07,borkovitsetal15}. For eclipsing binaries $\cos i\ll1$ hence, the last term is negligible (though it was taken into account for calculating the theoretical observable apsidal motion periods -- $P_\mathrm{apse}$, tabulated in Tables~\ref{tab: syntheticfit_KIC69640435653126} and \ref{tab: syntheticfit_KIC57313128023317}) and therefore, for the negative sign of $\Delta\Omega^\mathrm{dyn}$, one can infer, that the observable apsidal advance rate should be lower than the dynamical one.} 
Thus, the apsidal motion of the inner binary in the observational frame of reference, which can be detected directly through the anticorrelated primary and secondary ETV curves (Fig.~\ref{fig:ETVswithfit}), has a theoretical period of $P_\mathrm{apse}=28.59\pm0.02$\,yr, in accord with both the numerical integrations and the observed ETV curves.

The bottom panels of Fig.~\ref{fig:K6964043orbelements_numint} display the inclination angle variations of both the inner and outer orbits. The precession period is about 25\,years. (The theoretical value is $P_\mathrm{node}=25.0\pm0.1$\,yr.) In these panels we also denote with cyan (or green) regions of the inclination-angle domain where both (or at least one) eclipses of the inner binary can be observed. The borders of these regions are calculated from the equation 
\begin{equation}
|\cos i_\mathrm{in}|=\frac{R_\mathrm{Aa}+R_\mathrm{Ab}}{a_\mathrm{in}}\frac{1\pm|e_\mathrm{in}\sin\omega_\mathrm{in}|}{1-e_\mathrm{in}^2},
\label{Eq:innereclipsecond}
\end{equation}
where, on the right hand side, the positive sign gives the inclination-angle limit for a single eclipse (i.e., the green area), while the negative sign gives the limit where both eclipses should occur (the cyan area).  As one can see, during the first part of the \textit{Kepler}-observations the inner inclination became closer and closer to $90\degr$, and the inner orbit was seen exactly edge-on from the Earth around 2011.5.  Therefore, during the second part of the \textit{Kepler} measurements the orbital plane was viewed under a continuously decreasing angle.  This is in nice agreement with the first increasing, then constant (total eclipses) and, finally, decreasing eclipse depths (Fig.~\ref{fig:keplerlcs}) that were seen. Moreover, one can also see in the lower left panel of Fig.~\ref{fig:K6964043orbelements_numint}, that during the Sector 14 and 15 \textit{TESS}-observations ($\approx2019.6$), the inner binary was at the very edge of the eclipsing regime. Strictly speaking, the inner eclipses were present, however, their amplitudes were two orders of magnitude smaller than the scatter of the \textit{TESS} lightcurve of this very faint ($T=15.0$\,mag) source.  And, indeed, no third-body eclipse was seen during the two sectors of \textit{TESS} observations, therefore, the eclipsing nature of this system remained hidden from \textit{TESS}. In contrast to this, during the scheduled Sector 54 and 55 observations ($\approx2022.6$) the inner binary will again be very close to its next edge-on position ($i_\mathrm{in}\approx90\fdg7$) offering much better conditions to observe the shallow regular eclipses.

In the same panels we also denote, with thin black lines, the inclination limits of the outer inclination domain for the possibility of third-body eclipses. These are defined as follows:
\begin{eqnarray}
|\cos i_\mathrm{out}|&=&\frac{R_\mathrm{B}+R_\mathrm{Ab}+\frac{a_\mathrm{in}}{1+q_\mathrm{out}}\left(1+\tan^2i_\mathrm{in}\cos^2\Delta\Omega\right)^{-1/2}}{a_\mathrm{out}} \nonumber \\
&&\times\frac{1+|e_\mathrm{out}\sin\omega_\mathrm{out}|}{1-e_\mathrm{out}^2}.
\label{Eq:outereclipsecond}
\end{eqnarray}
(We will discuss this expression in Appendix~\ref{app:inclinationlimits}.) As one can see, the system continuously remains in the domain of the third-body eclipses.

In addition we assess the long-term future evolution of KIC 6964043. On longer timescales of a few 100 Myr (Fig.\,\ref{fig:6964043_evol}), KIC 6964043 displays evolution of the eccentricity of the outer orbit due to the octupole term. The outer eccentricity slowly increases from the initial value of $\approx 0.48$ to $\approx 0.50$ at 765 Myr. At this point, the system formally crosses into the dynamically unstable regime. Given the current proximity to the stability limit, it is not surprising that the system's hierarchy may break down. As the secular theory that \textit{TRES} is based on is no longer valid at this point, we stop the simulation.  Based on \cite{toonenetal22} we expect the system to remain in the dynamically unstable regime as an unstable but bound triple for on average $10^4$ crossing times, about 5 kyr. The most likely outcome of such an interaction is the dissolution of the system into a binary and single star, or the collision of two of the stars reducing the system from a triple into a binary. A summary of the evolution in shown in Fig.\,\ref{fig:cartoon1}.

\begin{figure}
\includegraphics[width=1.01\columnwidth]{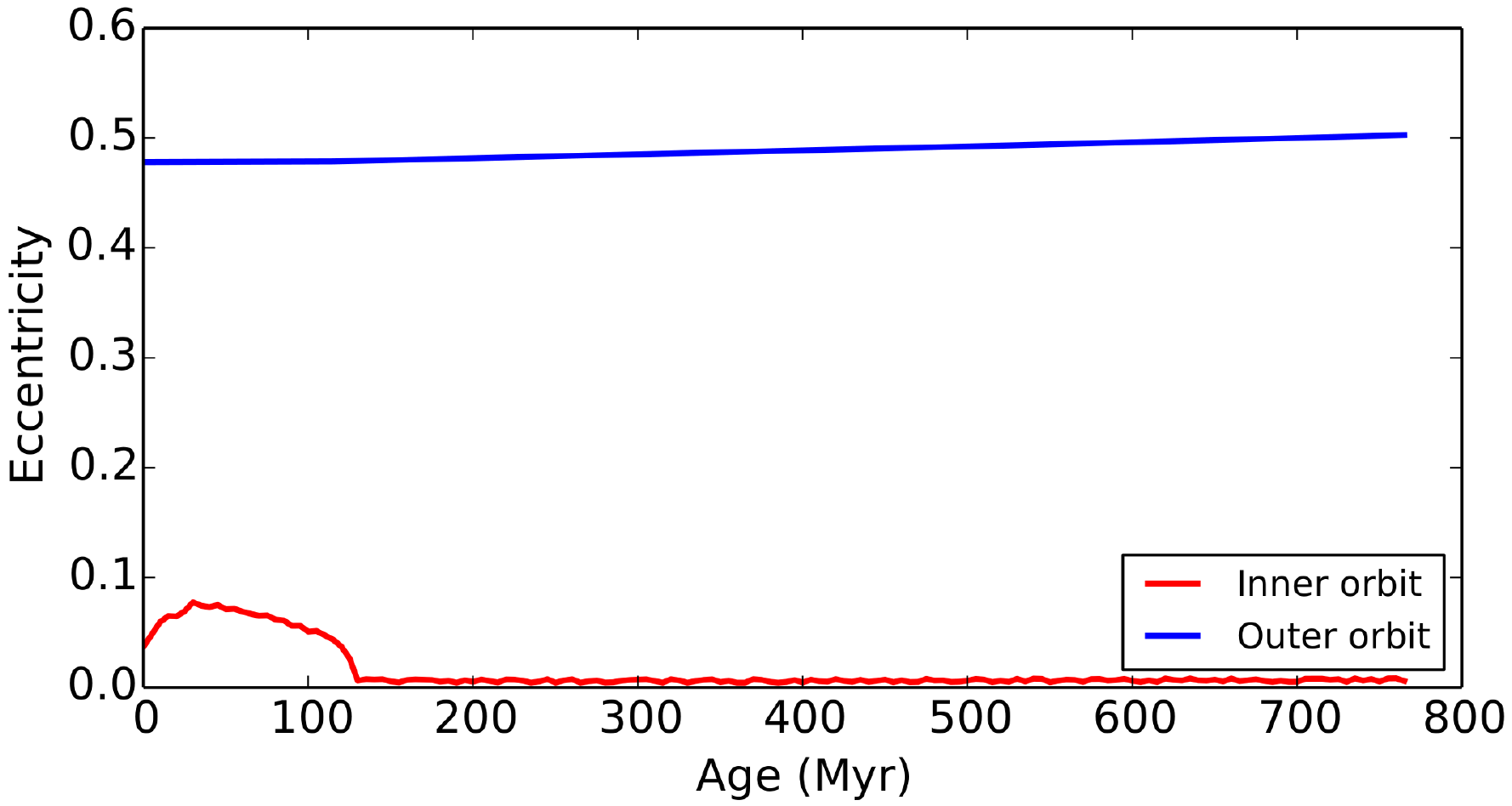}  
 \caption{Evolution of the system eccentricities in KIC 6964043.}
\label{fig:6964043_evol}
\end{figure}  

\begin{figure}
\includegraphics[width=0.99\columnwidth]{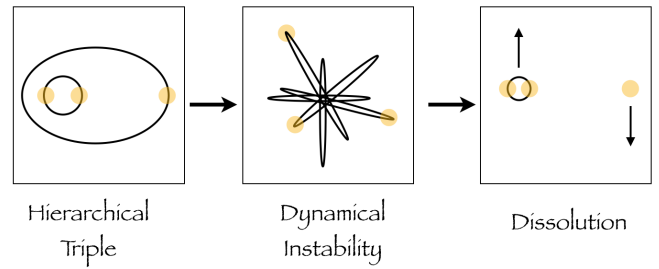}
 \caption{Schematic evolution of the triple star system KIC 6964043.}
\label{fig:cartoon1}
\end{figure}  

\subsection{KIC 5653126}
\label{sec:5.2}

This triple contains the most massive stars in our sample, and it has the second highest inner and outer mass ratios ($q_\mathrm{in}=0.736\pm0.006$; $q_\mathrm{out}=0.418\pm0.004$).  The most massive component is the early F-type primary of the inner pair ($m_\mathrm{Aa}=1.77\pm0.03\,\mathrm{M}_\odot$; $R_\mathrm{Aa}=2.20\pm0.02\,\mathrm{R}_\odot$; $T_\mathrm{eff,Aa}=7060\pm80\,\mathrm{K}$), while its binary companion and the tertiary star are very similar late F-type stars ($m_\mathrm{Ab}=1.30\pm0.02\,\mathrm{M}_\odot$; $R_\mathrm{Ab}=1.31\pm0.02\,\mathrm{R}_\odot$; $T_\mathrm{eff,Ab}=6260\pm60\,\mathrm{K}$ and $m_\mathrm{B}=1.29\pm0.03\,\mathrm{M}_\odot$; $R_\mathrm{B}=1.28\pm0.04\,\mathrm{R}_\odot$; $T_\mathrm{eff,B}=6230\pm60\,\mathrm{K}$). The age of the system is found to be $\tau=1.0\pm0.1$\,Gyr.  Our analysis yields a distance of $d=1375\pm15$\,pc, which agrees with the trigonometric Gaia EDR3 distance to within $3\sigma$.

Turning to the orbital parameters, both the inner and outer orbits are moderately eccentric ($e_\mathrm{in}=0.294\pm0.002$ and $e_\mathrm{out}=0.179\pm0.001$). The mutual inclination of the orbital planes was found to be $i_\mathrm{mut}=12\fdg3\pm0\fdg2$, but at that epoch both orbits had very similar observable inclination angles. At the beginning of the \textit{Kepler} observations, the inclination angles were $i_\mathrm{in}=87\fdg13\pm0\fdg02$ and $i_\mathrm{out}=87\fdg6\pm0\fdg5$ for the inner and outer orbital planes, respectively.  Note that, despite the absence of third-body eclipses, which would strongly constrain the outer inclination angle as well as the mutual inclination, we were able to find these quantities with remarkable accuracies of some tenths of a degree (which is even better than the accuracies obtained for the same parameters in the case of some \textit{TESS} observed triply eclipsing triples -- \citealp[see, e.g.,][]{borkovitsetal22}).  This is a consequence of the fact that the variation of the inner inclination angle and, hence, the depths and durations of the inner eclipses are extremely sensitive to these parameters. (Note, for the former system, KIC~6964043, where both the eclipse depth variations and third body eclipses were detected continuously over a 3-year-long timescale, we were able to obtain even higher accuracies, some hundredths of a degree, for these parameters.)

At this point, we comment on the presence or absence of third-body eclipses. In contrast to the triply eclipsing triple KIC~69640433, in the case of this, and the other two systems we will discuss, in the absence of third-body eclipses, the lightcurves do not carry any information on the radii and effective temperatures of the tertiary components, because their only direct effects on the lightcurves are the dilution of the regular eclipses by their extra flux contamination. Thus, the third star's radius and temperature are largely dependent on the \texttt{PARSEC} isochrones.  The latter, however, are constrained by the third star's mass and mass ratios, which can be inferred quite accurately from the photodynamical analysis.  This is especially true of the effects of the third-body perturbations which drive apsidal motion, leading to ETVs, and orbital plane precession (whose consequences were discussed in the preceding paragraph).  Thus, the fact that the dynamically inferred third star masses led to fully consistent MDN solutions, in our view, again verifies the use of \texttt{PARSEC} isochrones as proxies either for the masses (in the absence of RV data in the case of regular EBs) or radii and other fundamental stellar parameters (in the absence of third-body eclipses).

Turning now to the above mentioned perturbations, in Fig.~\ref{fig:K5653126orbelements_numint} we display the variations of some orbital elements over 30 and $\approx1300$ year-long intervals. In contrast to KIC~6964043, besides the characteristic medium ($P_\mathrm{out}$) timescale perturbations (upper left panel), now some longer timescale periodic variations can be identified both in the inner eccentricity and the mutual inclination.  The amplitudes of these clearly exceed those of the medium period effects, though the relative variations in both orbital elements remain at the few percents level.  As one can see, $e_\mathrm{in}$ and $i_\mathrm{mut}$ vary in an anticorrelated manner, and the characteristic period of these cycles is about half of the dynamical apsidal motion period. These findings are in good agreement with the theories of the quadrupole-order long-period perturbations in the low mutual inclination regime \citep[see, e.g.,][]{mazehshaham79,soderhjelm82,borkovitsetal07}. Moreover, another even longer period variation in the eccentricity can also be noticed, which we attribute to the octupole-order perturbations \citep[e.g.][]{naozetal13}. These variations, however, again remain small, and our longer, one million year numerical integration displays the same picture.

The variations of the observationally more relevant orbital elements, again, are shown in the middle (apsidal motion) and bottom (orbital precession) panels. The behaviour of the different arguments of periastron are similar to those of KIC 6964043, but the periods are somewhat longer, in accordance with the longer characteristic timescale of $P_\mathrm{out}^2/P_\mathrm{in}$. The calculated (theoretical) observable apsidal motion periods are $P_\mathrm{apse,in}=309\pm4$\,yr and $P_\mathrm{apse,out}=1204\pm4$\,yr for the inner and outer orbits, respectively.  Besides the longer timescales, another minor difference in the behaviour of the rotations of the lines of the apsides relative to the previous system is that the dynamical apsidal motion no longer remains strictly linear, but a low amplitude cyclic variation (with a period similar to that of the eccentricity variation period) is also superposed on the linear trend. For such a low mutual inclination, however, the amplitude of this extra cyclic term, from an observational point of view, remains negligible.

In the bottom panels, the variations of the observable inclination angles, and the domains within which eclipses are possible are plotted. The precession period is calculated to be $P_\mathrm{node}=225\pm3$\,yr. During one cycle, the inner binary exhibits eclipses only in two, separated, $\approx40$\,yr-long intervals. The current eclipsing session will terminate at the very end of 2031. What is more interesting, is that there are short intervals when third-body eclipses are also possible. The last such interval of outer eclipses occurred circa the 1970s.  In order to check whether third body eclipses had actually occurred during that period, we back-integrated our best solution (i.e. the set of input parameters which yielded the lowest net $\chi^2$ value) to 1950, and found two third-body events during 12-14 January 1976 and 11-15 September 1978. The same integration has revealed that the next third-body eclipse is expected only in the year 2143.  As one can see in the lower right panel of Fig.~\ref{fig:K5653126orbelements_numint}, these third-body eclipses may occur when the inner binary is the farthest from its eclipsing binary state. At the current time only one such triple star system is known with certainly, and it currently shows third-body eclipses without inner binary eclipses. This is KIC~2835289, a 0.86-day-period ellipsoidal variable, orbited by a third star with a period of $P_\mathrm{out}=755$\,d, from which two third-body eclipses were observed with \textit{Kepler} \citep{conroyetal14,conroyetal15}, and a further one with \textit{TESS}. Moreover, \cite{borkovitsetal20a} found that TIC~167692429, which currently is an EB with remarkable eclipse depth variations, has also produced third-body eclipses in the recent past, {\it before} the beginning of its recent eclipsing session. And, the last of those events was actually identified with great likelihood in the archival SuperWASP observations of that triple.

\begin{figure*}
\begin{center}
\includegraphics[width=0.50 \textwidth]{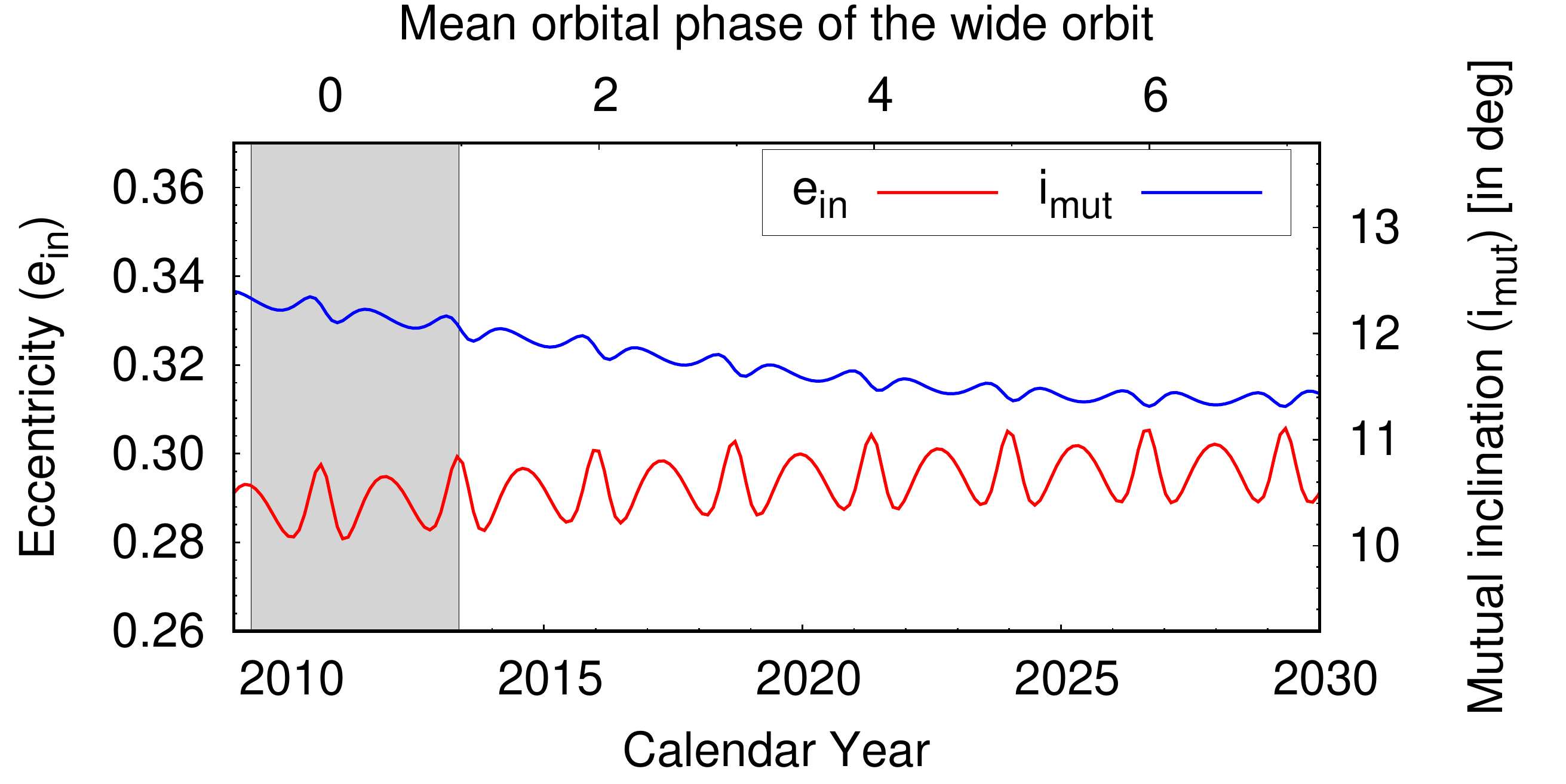}\includegraphics[width=0.50 \textwidth]{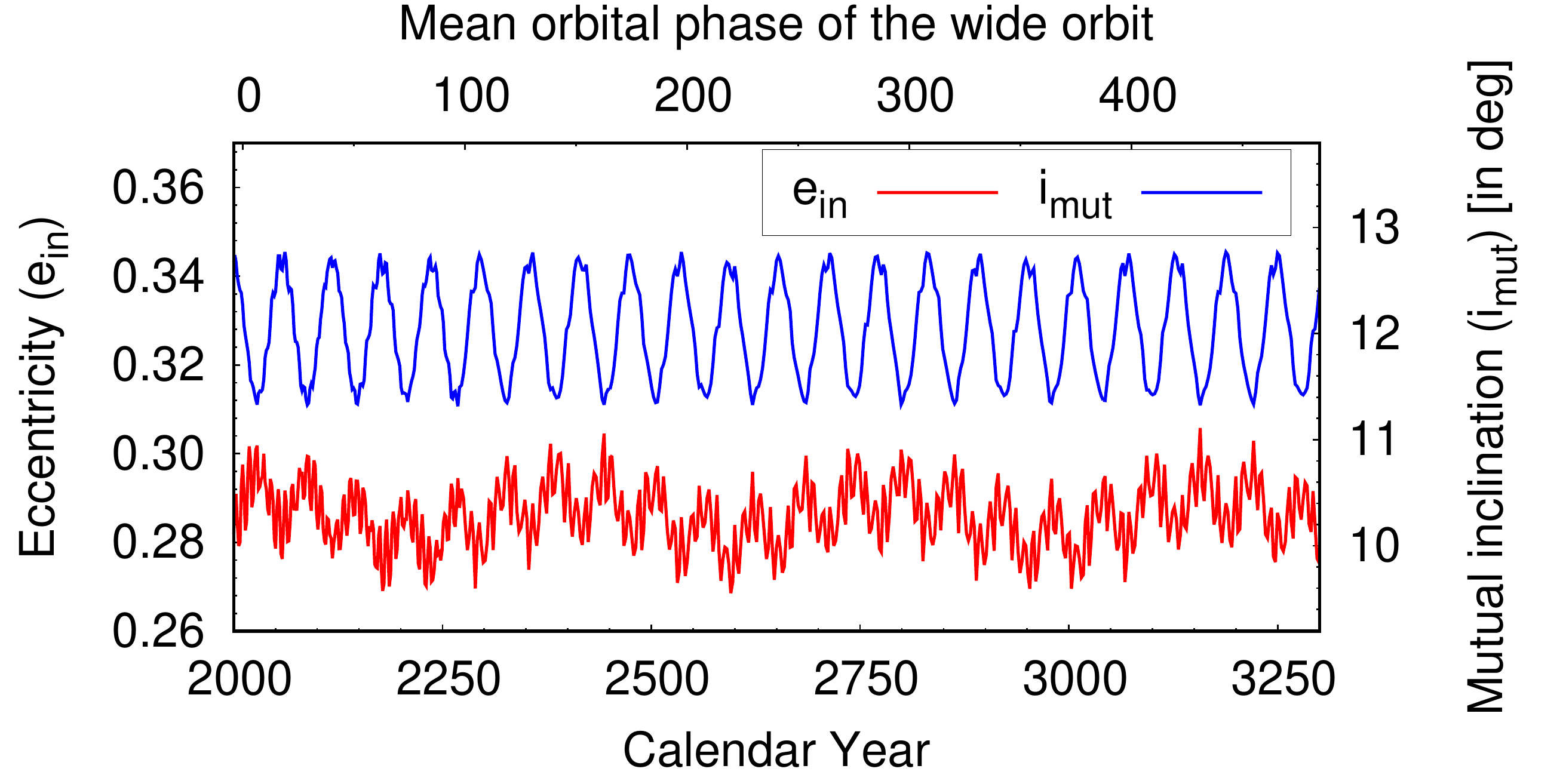}
\includegraphics[width=0.50 \textwidth]{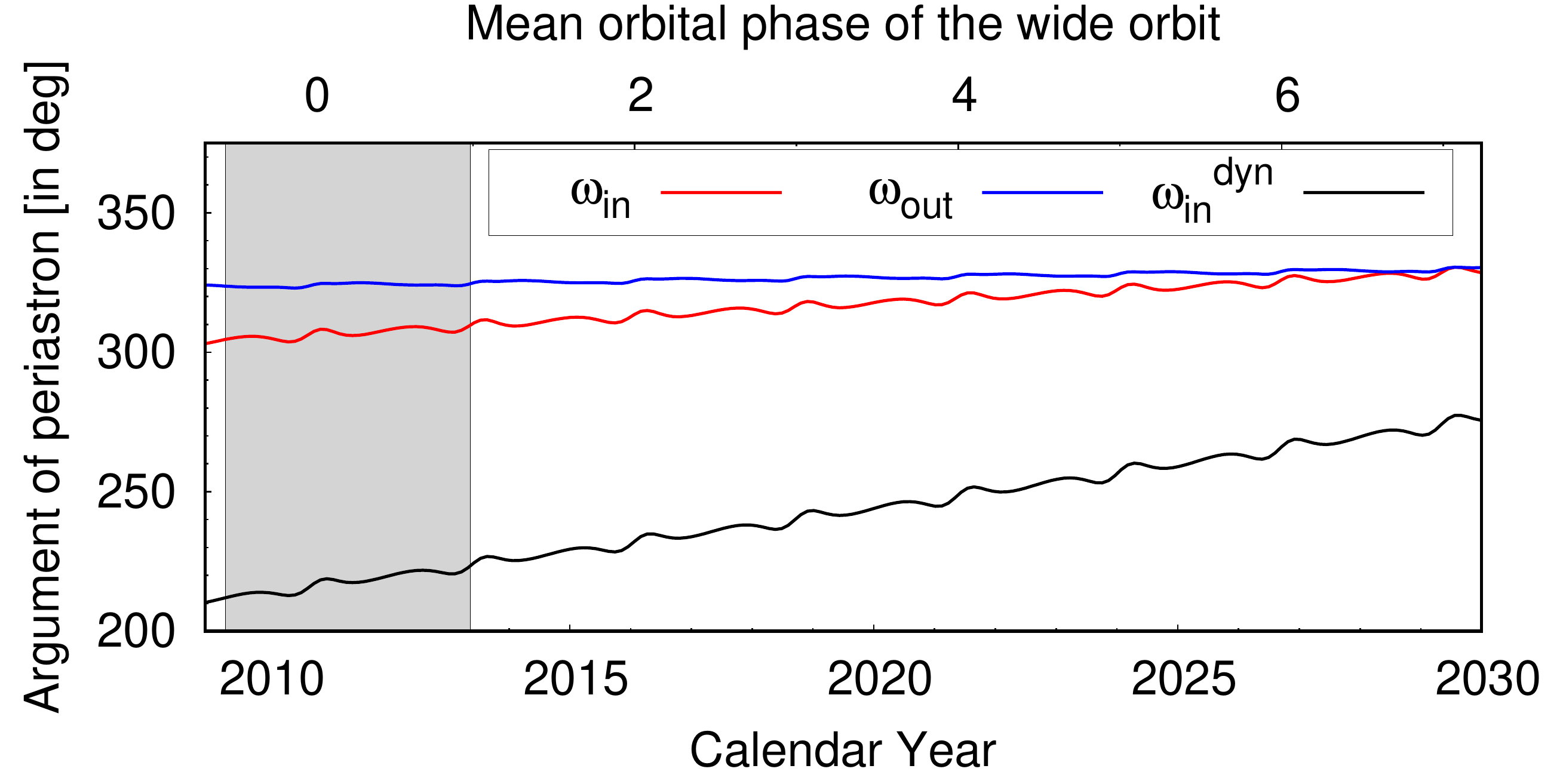}\includegraphics[width=0.50 \textwidth]{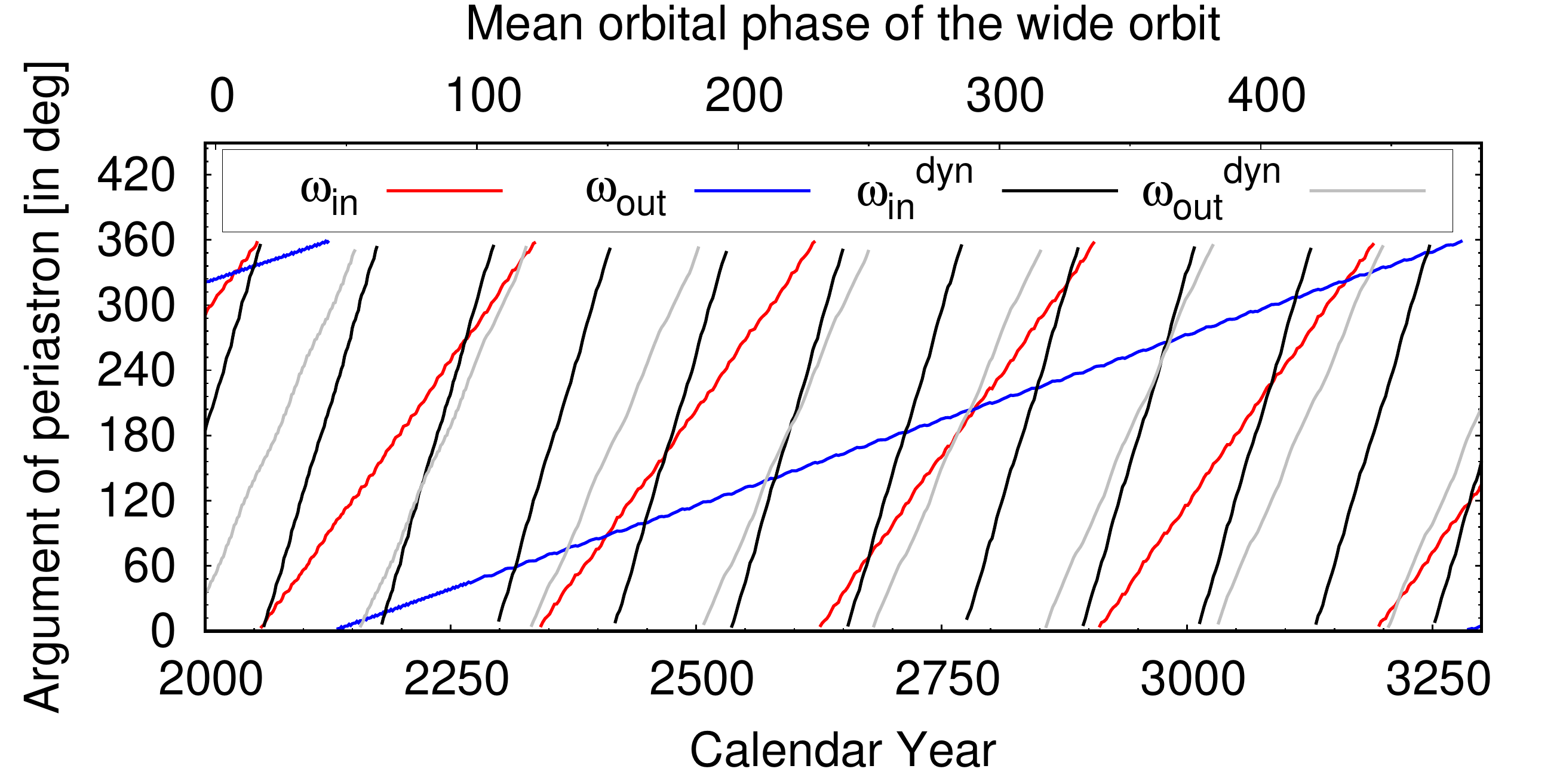}
\includegraphics[width=0.50 \textwidth]{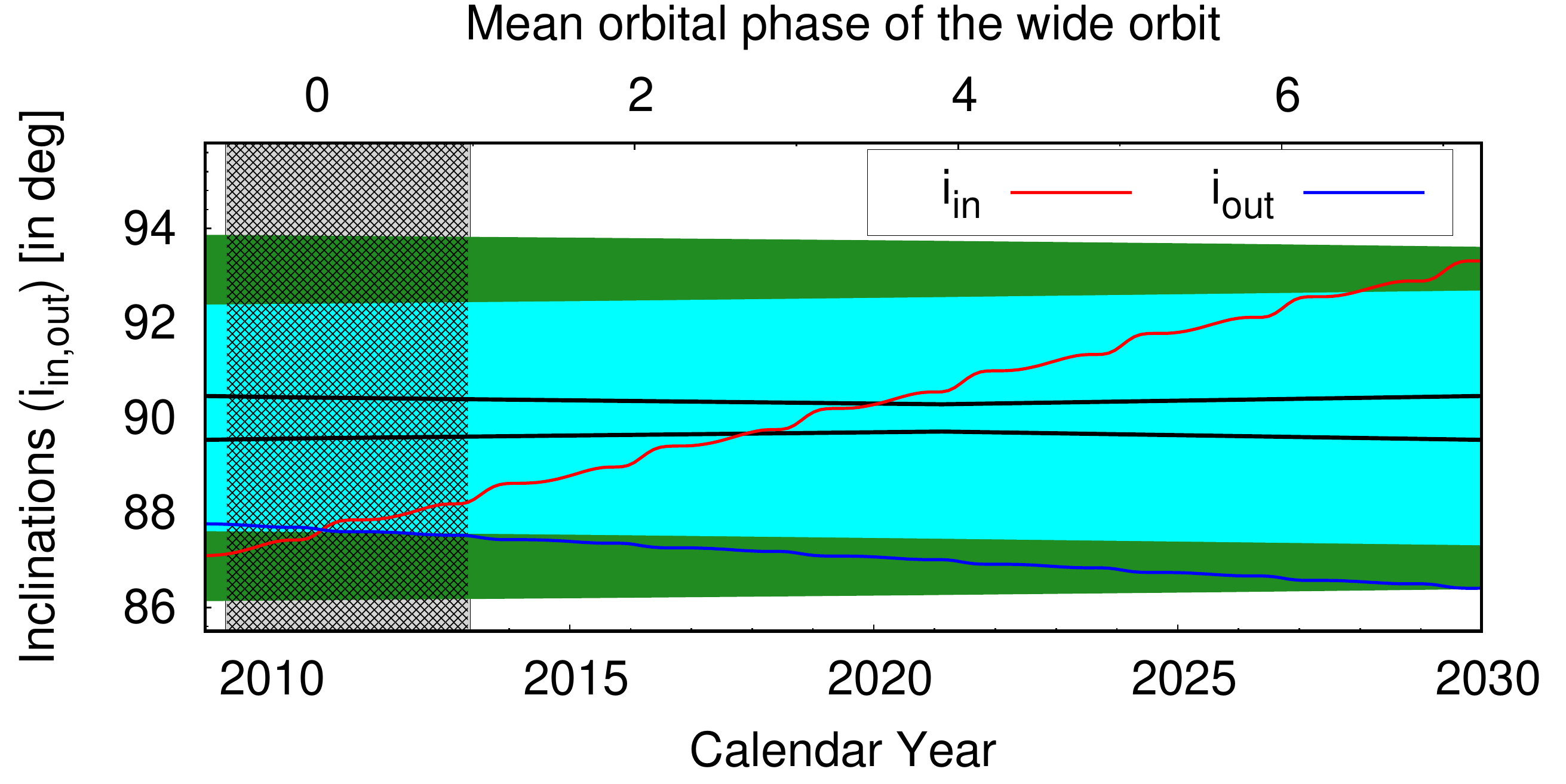}\includegraphics[width=0.50 \textwidth]{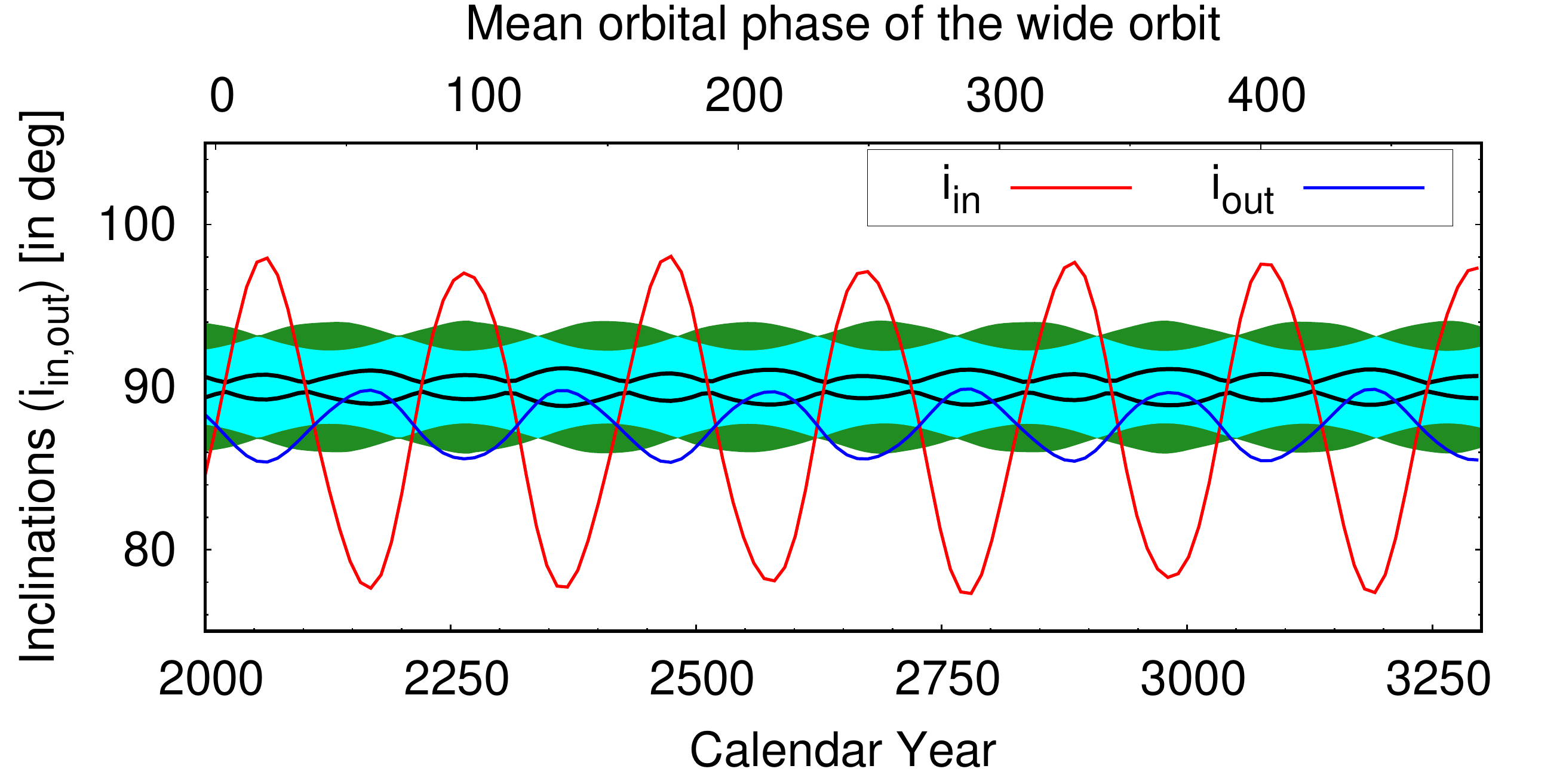}
\caption{Variations of the same orbital elements as in Fig.~\ref{fig:K6964043orbelements_numint} above, but for KIC 5653126. Note, that the right panels here and also for the other two systems represent an interval of 1300 years. 
See text for details.}
\label{fig:K5653126orbelements_numint} 
\end{center}
\end{figure*}  

As with KIC 6964043, we address the long-term future evolution of KIC 5653126 by making use of the \textit{TRES} code.  The long-term future of KIC 5653126 is different that of the previous system as in the case of KIC 5653126 the stars are massive enough to evolve off the MS in a Hubble time (see Fig.~\ref{fig:cartoon2}). As the primary of the inner binary, which is the most massive star in this triple ascends the first giant branch and expands rapidly, the system changes due to three reasons; (1) tidal torques amplify leading to the near circularisation of the inner binary (Fig. \,\ref{fig:5653126_ev}). The inclination freezes out at about 30 degrees, after increasing steadily from the current inclination of about 12 degrees. (2) the primary star loses 0.004\, M$_{\odot}$ of envelope material due to its stellar winds. As a result there is a slight increase in the outer semi-major axis from 675.8\,R$_{\odot}$ to 676.42\,R$_{\odot}$. There is also a widening effect from the wind mass loss on the inner orbit, however the dominant effect is a slight shrinkage of the semi-major axis from 69.77\,R$_{\odot}$ to 69.46\,R$_{\odot}$ due to stellar tides. (3) Lastly, and most importantly, at 1910 Myr, the primary star fills its Roche lobe. 

\begin{figure}
\includegraphics[width=0.99\columnwidth]{ev2_KIC.png}  
\vglue0.2cm \hglue1.1cm
\includegraphics[width=0.75\columnwidth]{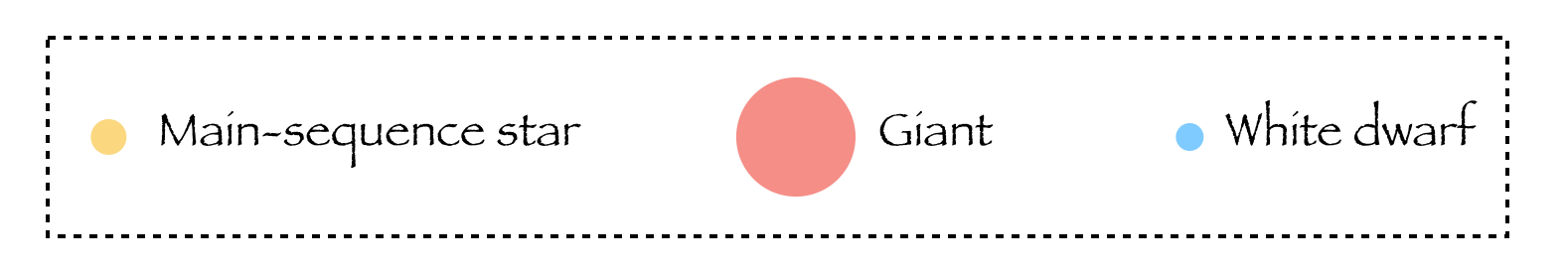}
 \caption{Schematic evolution of the triple star system KIC 5653126. }
\label{fig:cartoon2}
\end{figure}  

\begin{figure}
\includegraphics[width=1.02\columnwidth]{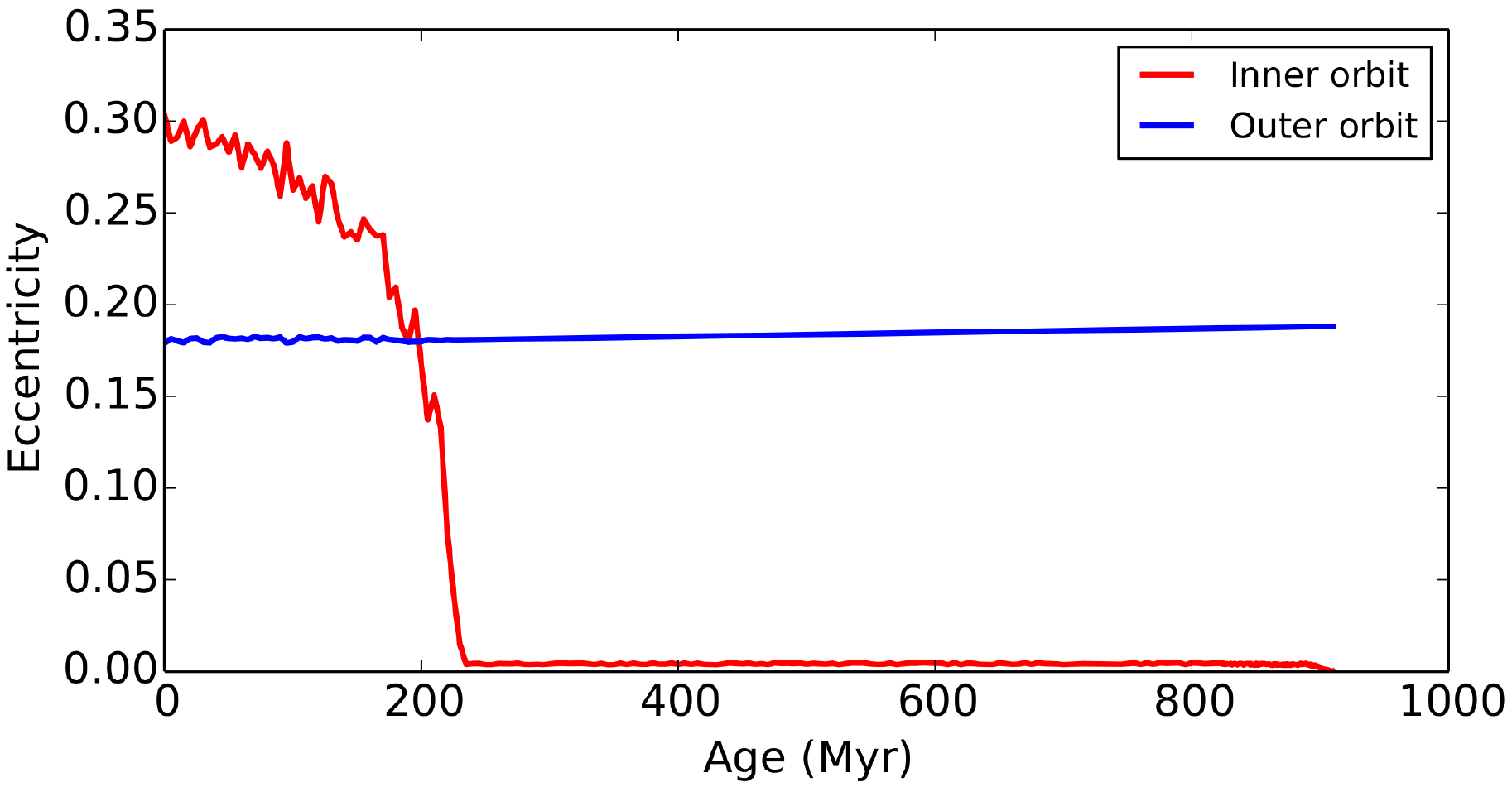}
 \caption{Evolution of the system eccentricities in KIC 5653126.}
\label{fig:5653126_ev}
\end{figure}  

Given the evolutionary state of the donor star and its deep convective envelope, the ensuing mass transfer phase proceeds in an unstable way, quickly leading to a common-envelope (CE) phase (for a review, see \citealt{Iva13}). During this phase the secondary star orbits around the core of the primary star inside its envelope. Friction causes the secondary's orbit to decay leading to the merger of the secondary with the core of the star, or a tight binary if the envelope can be ejected.  Adopting the classical energy-based CE-prescription \citep[i.e. the $\alpha$-model;][]{Pac76, Web84, Liv88, DeK90}, the CE-phase leads to a merger of the two stars in the inner binary. This holds for both the classical values of the CE-efficiency ($\alpha\approx 1$) as well as the observationally derived reduced efficiencies \citep[$\alpha\approx 0.25$;][]{Zor10, Too13, Cam14, Zor22}. The resulting merger product is a giant-like object with a compact helium-rich core of 0.32\,M$_{\odot}$ and a massive hydrogen-rich envelope of 2.75\,M$_{\odot}$ assuming a conservative merger. 

As this object evolves further, it will initiate core helium burning followed by helium-shell burning  as it ascends the asymptotic giant branch (AGB). During this phase, when the merger remnant is still orbited by the initial tertiary star, it fills its Roche lobe again. This second CE-phase does not lead to a merger, as the pre-CE orbit is much wider than for the first CE-phase. After the second CE-phase, the system consists of a 0.65\,M$_{\odot}$ white dwarf (i.e., the old core of the AGB-star) and a 1.29\,M$_{\odot}$ main-sequence companion. The orbit of the post-CE system depends critically on the CE-efficiency i.e. the efficiency with which the orbital energy is converted to unbind the envelope. In both cases of inefficient ($\alpha=1$) and efficient ($\alpha=0.25$) CE-evolution, the system experience an ultimate mass transfer event, with the white dwarf as the accretor star, and the donor star either on the main-sequence or giant branch, respectively. In both cases the system ends its life as a single white dwarf of mass $0.8-0.9$\,M$_{\odot}$. Even though, the systems started out its life as a triple star system, we expect it will end its life as a single star. 

\subsection{KIC 5731312}
\label{sec:5.3}

Now we turn to the two dynamically more interesting, high mutual inclination systems. KIC~5731312 is the least massive triple in our sample. The inner binary is formed by a K and an M-type dwarf ($m_\mathrm{Aa}=0.77\pm0.01\,\mathrm{M}_\odot$; $R_\mathrm{Aa}=0.73\pm0.01\,\mathrm{R}_\odot$; $T_\mathrm{eff,Aa}=5080\pm60\,\mathrm{K}$ and $m_\mathrm{Ab}=0.51\pm0.02\,\mathrm{M}_\odot$; $R_\mathrm{Ab}=0.51\pm0.03\,\mathrm{R}_\odot$; $T_\mathrm{eff,Ab}=3690\pm30\,\mathrm{K}$, respectively), while the distant, third component is a very low mass, cool, late M star ($m_\mathrm{B}=0.15\pm0.01\,\mathrm{M}_\odot$; $R_\mathrm{B}=0.18\pm0.01\,\mathrm{R}_\odot$; $T_\mathrm{eff,B}=2910\pm30\,\mathrm{K}$). Thus, the mass ratios are far from unity ($q_\mathrm{in}=0.665\pm0.040$; $q_\mathrm{out}=0.120\pm0.004$) and, in such a way, they differ substantially from the mass ratios of the flat triples investigated above. This fact, together with the inclined tertiary orbit, certainly imply a different formation scenario. The system is found to be quite old ($\tau=5.9\pm0.6$\,Gyr). Its photodynamically obtained distance ($d=353\pm7$\,pc) is in excellent agreement with the Gaia EDR3 distance.

As mentioned in Sect.~\ref{sec:targets} this triple is by far the least tight one in our sample ($P_\mathrm{out}/P_\mathrm{in}\approx122$). Moreover, the outer mass ratio is also very low. As a consequence, one can observe remarkable, readily detectable ETVs in this system largely due to the high inner and outer eccentricities ($e_\mathrm{in}=0.4587\pm0.0005$ and $e_\mathrm{out}=0.575\pm0.008$). In an opposite, i.e. nearly circular, case the presence of the distant tertiary probably would have remained unnoticed even via an ETV analysis based on the highly accurate timing data that were obtained from \textit{Kepler} observations. While the semi-major axes of the two orbits are found to be $a_\mathrm{in}=18.2\pm0.1\,\mathrm{R}_\odot$ and $a_\mathrm{out}=448\pm2\,\mathrm{R}_\odot$, respectively, at the times of apastron passage of the inner orbit, and periastron passage of the outer orbit, the orbital separations are $r_\mathrm{in}^\mathrm{apo}\approx26.5\,\mathrm{R}_\odot$ and $r_\mathrm{out}^\mathrm{peri}\approx190.6\,\mathrm{R}_\odot$. Hence, the ratio of the equivalent periods becomes $P_\mathrm{out}^\mathrm{peri}/P_\mathrm{in}^\mathrm{out}\approx19.2$, temporarily resulting in a very tight triple system. 

The mutual inclination of the two orbits is found to be $i_\mathrm{mut}=39\fdg4\pm0\fdg3$, i.e., the former findings of \citet{borkovitsetal15,borkovitsetal16}, which were based only on an analytic analysis of the ETV curves, is nicely confirmed. This mutual inclination value is very close to the famous mutual inclination limit ($i_\mathrm{mut}^\mathrm{ZKL}=39\fdg23$, and its retrograde counterpart) of the original, quadrupole, asymptotic ZKL theory, which separates the low- and high mutual inclination regimes of the phase space (see the short summary in the Introduction).  However, as discussed next, the configuration of the present system is far from appropriate to be fit in the framework of the original ZKL theory.  Due to the large outer eccentricity and small inner mass ratio, the octupole-order perturbations may become significant, and hence the evolution of the orbital elements should be modeled within the framework of the eccentric ZKL formalism \citep[see~e.g.][]{naoz16}. 

First we introduce again the short-term characteristics of these perturbations in some of the orbital elements, as well as their direct observational consequences. In order to do this, similar to the previous two triples, we display decades and centuries long variations of a few orbital elements obtained from a 1\,Myr-long numerical integration in the various panels of Fig.~\ref{fig:K5731312orbelements_numint}. Now the eccentricity cycles of the inner binary, and the corresponding variations of the mutual inclination with a period of $P_\mathrm{ZKL}\approx690$\,yr and full amplitudes of $\Delta e_\mathrm{in}\approx0.28$ and $\Delta i_\mathrm{mut}\approx10\degr$ are clearly visible in the upper panels.  The characteristics of these variations do not change significantly during the entire 1-million-year integration, however, a slow sinusoidal variation in the amplitudes of $\approx10\%$ on an $\approx$ ten-thousand year timescale is also apparent, probably due to the octupole terms.

As one can see in the middle left panel, the inner EB exhibits retrograde apsidal motion in the observational frame over the current several decades. This results because a negative contribution arises from the nodal regression which temporarily overpowers the effect of the prograde apsidal advance in the dynamical frame of reference. However, as the middle right panel shows, this is only part of a longer-term oscillatory (i.e., libration) cycle. The dynamical apsidal motion of both the inner and outer orbits remain continuously prograde, however their non-linear nature, i.e., the presence of a significant librating motion in addition to the usual circular apsidal motion is readily visible in the middle right panel.  Their period is about 1300-1400\,yr, i.e., twice of that of the eccentricity cycles (which, again, is in perfect agreement with the theory of the secular perturbations). Note, however, that these values do not agree with the theoretical apsidal motion periods tabulated in Table~\ref{tab: syntheticfit_KIC57313128023317}. The reason is that these latter periods were calculated from the instantaneous apsidal motion rates (also listed in the tables) simply by linear interpolation and, hence, do not give accurate results for highly non-linear apsidal motions.

The variations in the visible inclination angles are shown in the bottom panels of Fig.~\ref{fig:K5731312orbelements_numint}. The nodal precession cycle of this triple has a period of $P_\mathrm{node}\approx1400$\,yr.  During such a cycle the inner pair of stars exhibits eclipses only for two short, 80-100 year-long intervals.  The current eclipsing session began around the 1950's, and the EB was seen edge-on at the beginning of the 1990's. During the prime \textit{Kepler} mission, both kinds of eclipses were observable.  Then, however, the eclipse closer to apastron would no longer have been observable since $\sim2016.5$, which is in accord with the 2019 and 2021 \textit{TESS} measurements. The remaining, solo eclipses will also disappear in about 2029.0, and the eclipses will return only in the XXIX-th century. Finally, as one can see, between circa 2300 and 2600 there might be a chance for third-body eclipses to appear. In order to check this possibility, we numerically generated the system's lightcurve for a section of this interval, and found two very shallow (with depth of $\sim0.3\%$) third-body eclipses at the epoch 2453.63. 

\begin{figure*}
\begin{center}
\includegraphics[width=0.50 \textwidth]{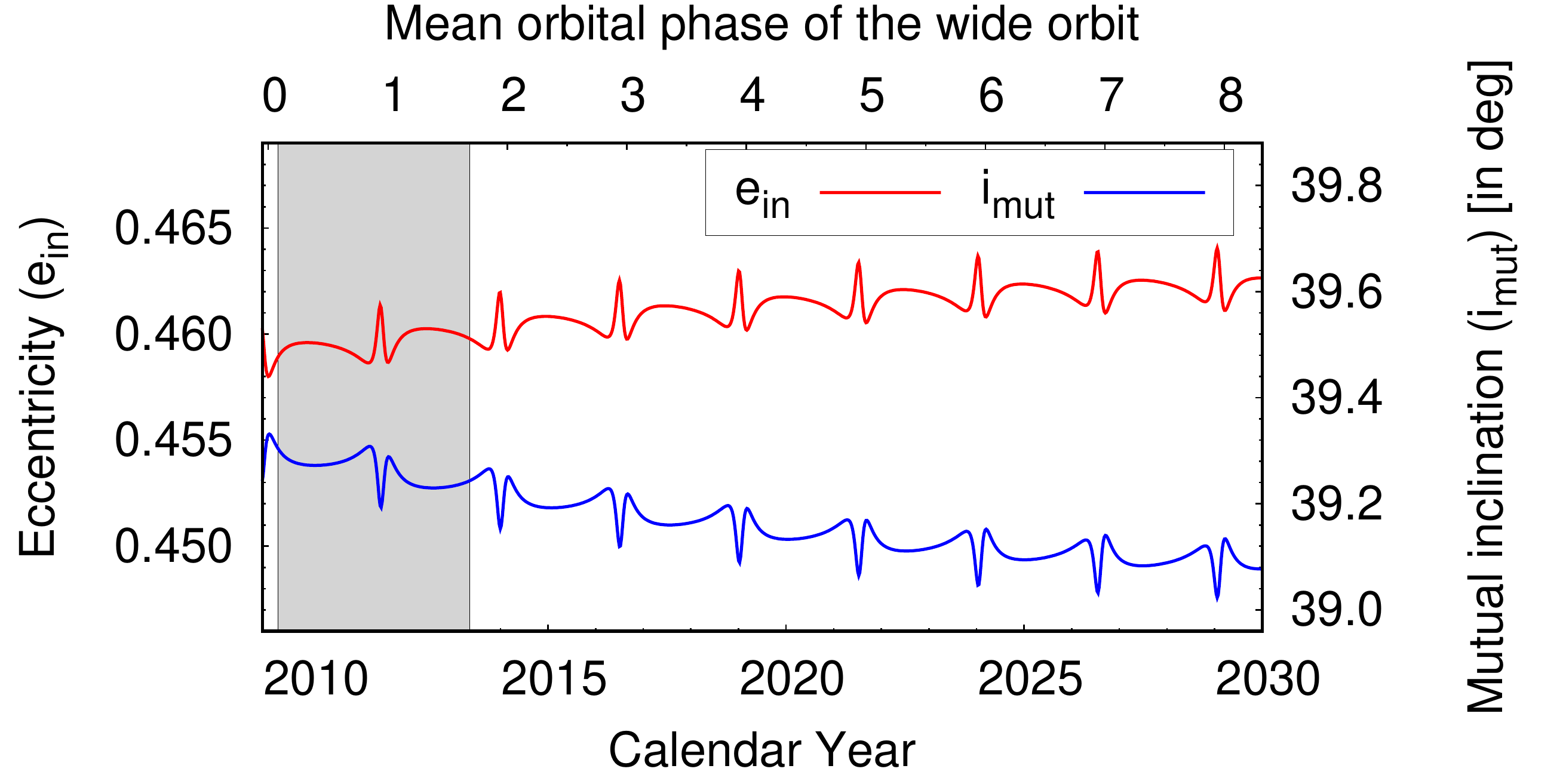}\includegraphics[width=0.50 \textwidth]{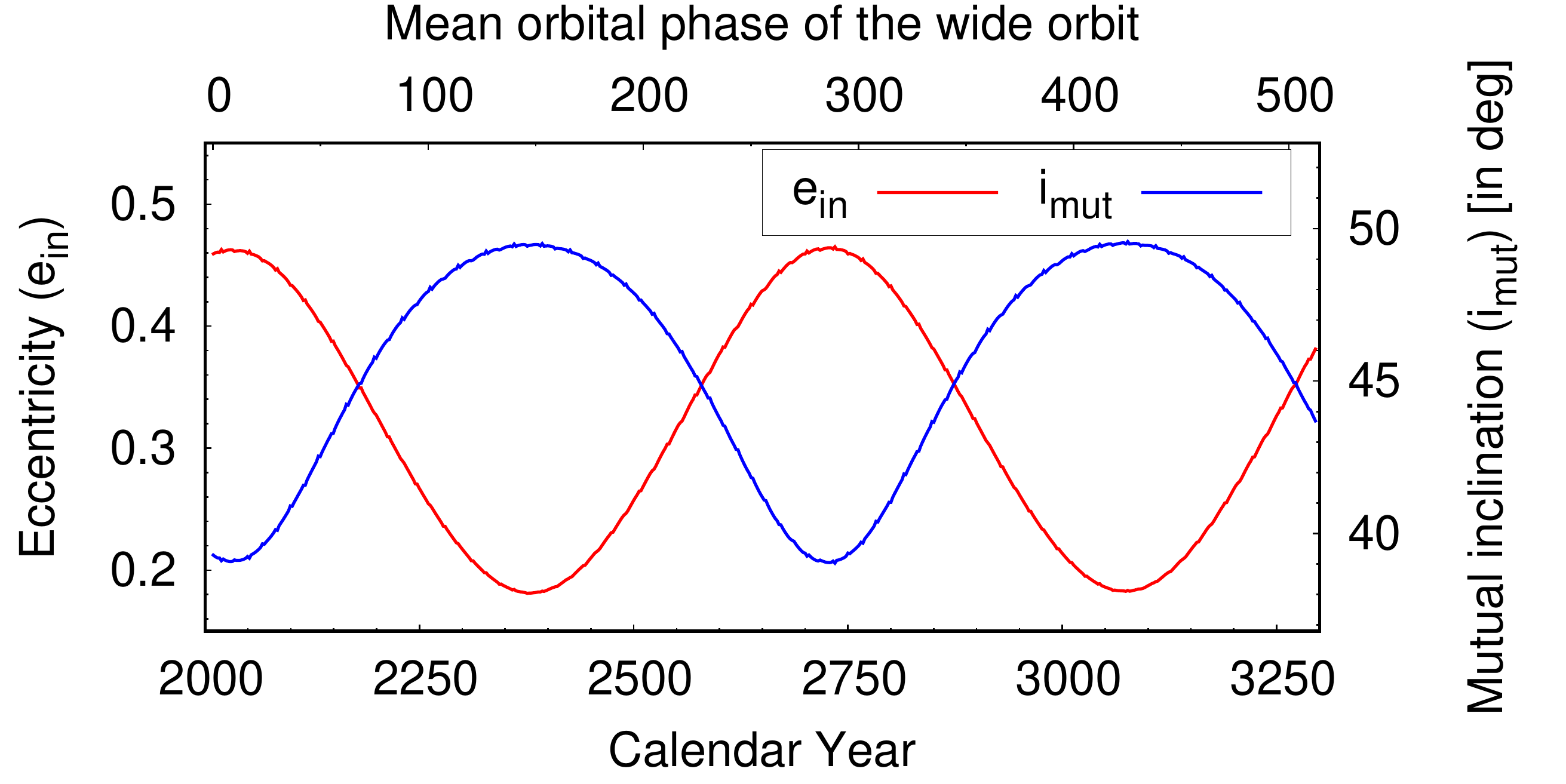}
\includegraphics[width=0.50 \textwidth]{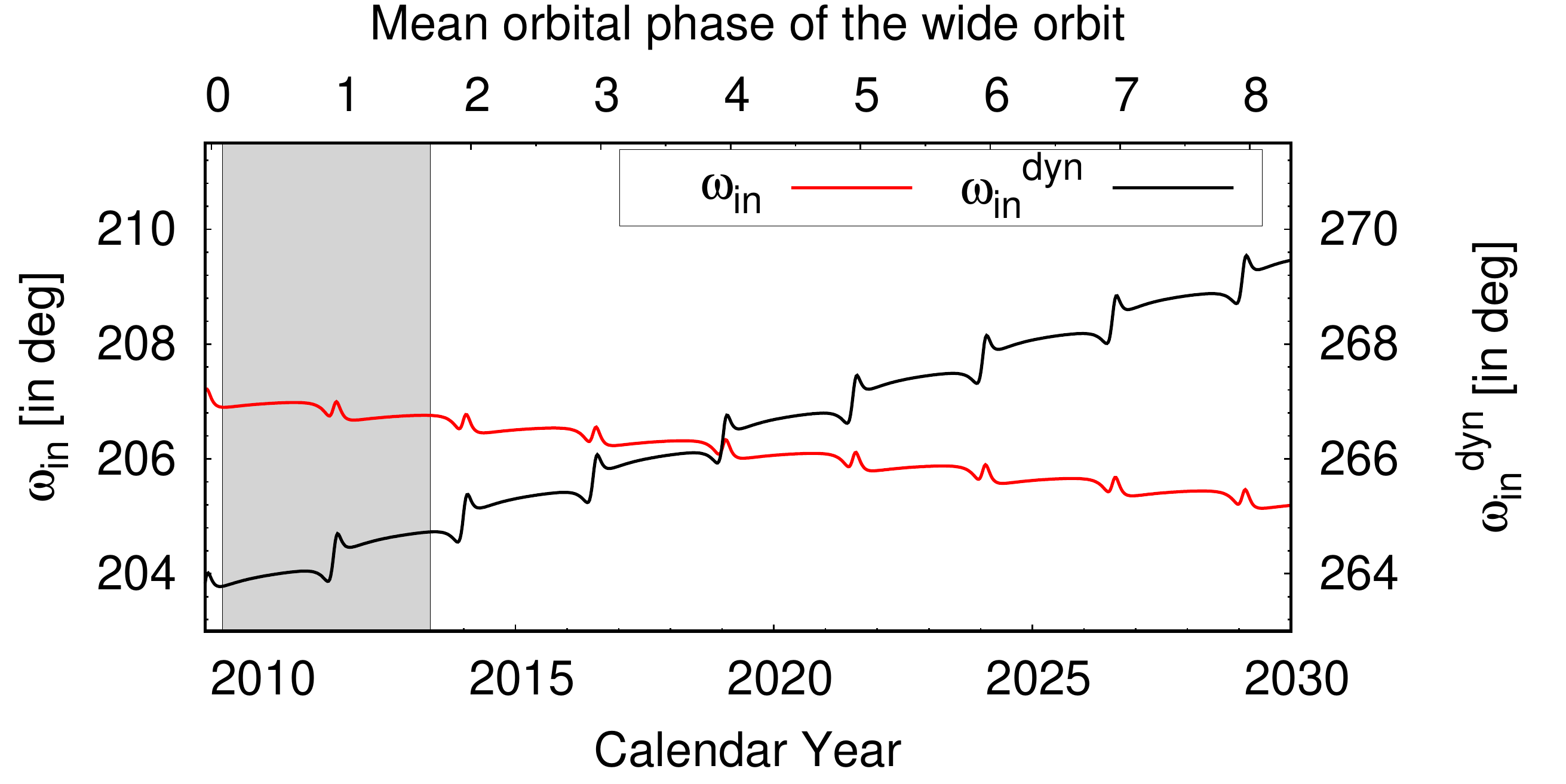}\includegraphics[width=0.50 \textwidth]{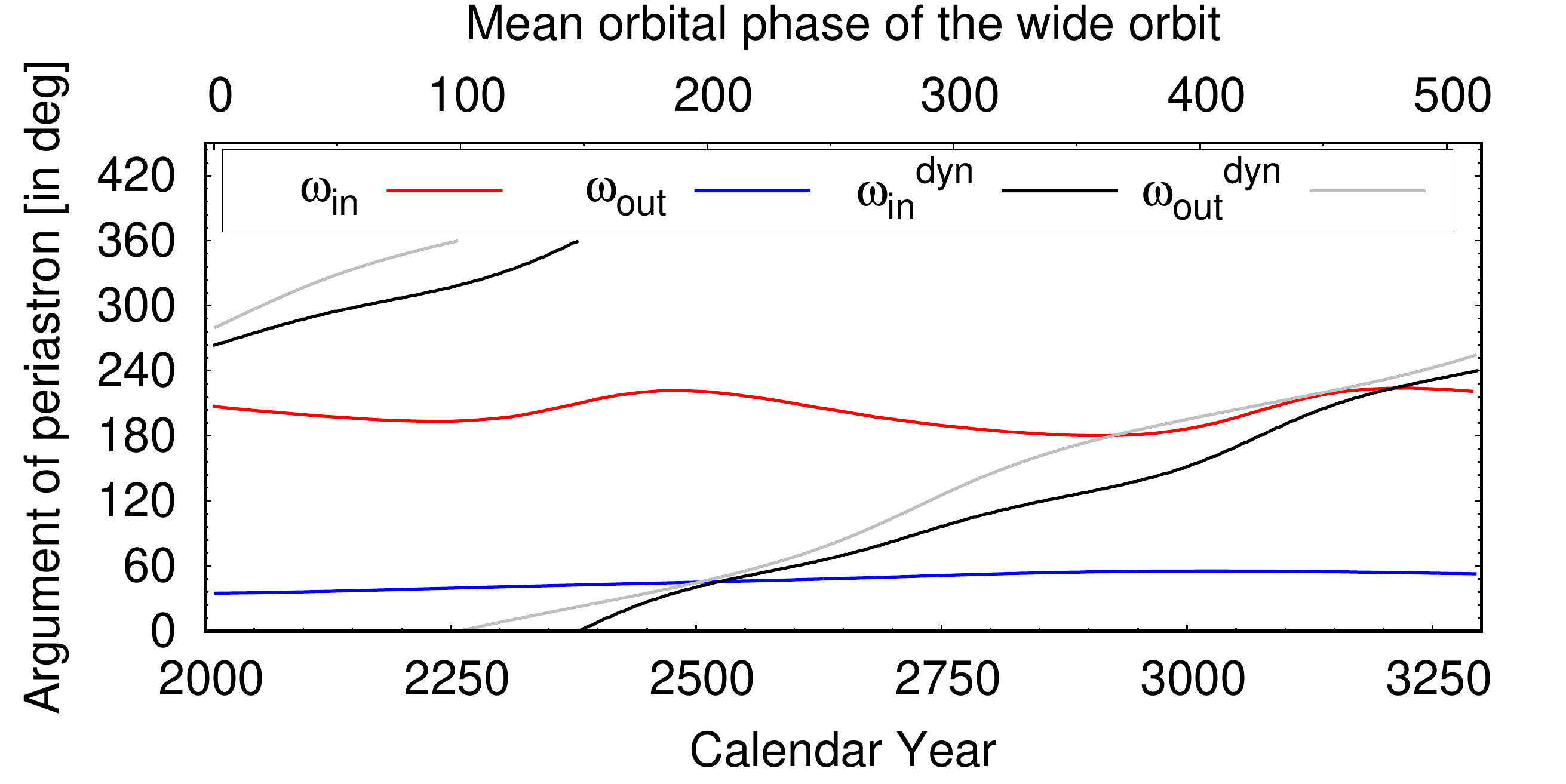}
\includegraphics[width=0.50 \textwidth]{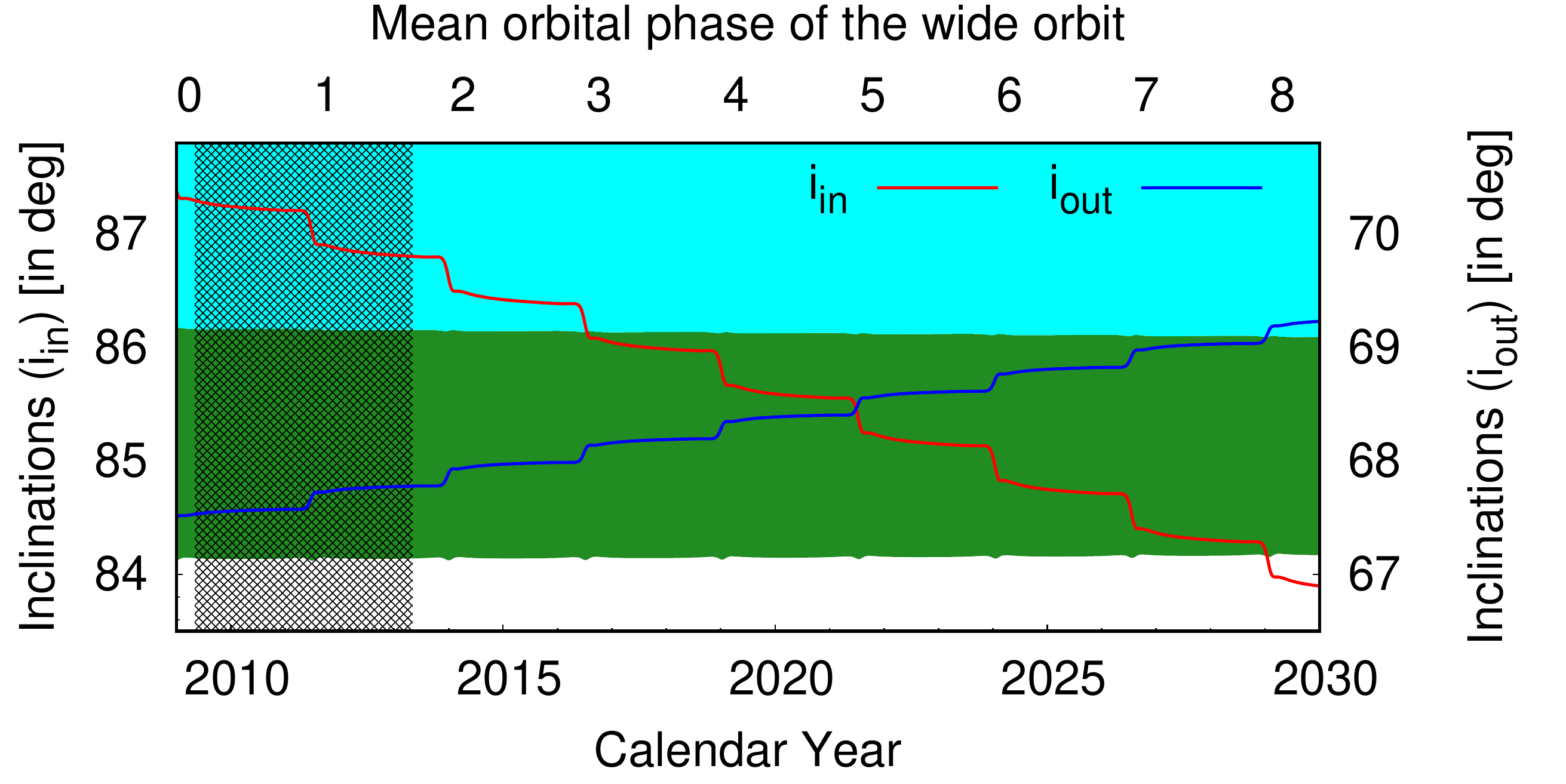}\includegraphics[width=0.50 \textwidth]{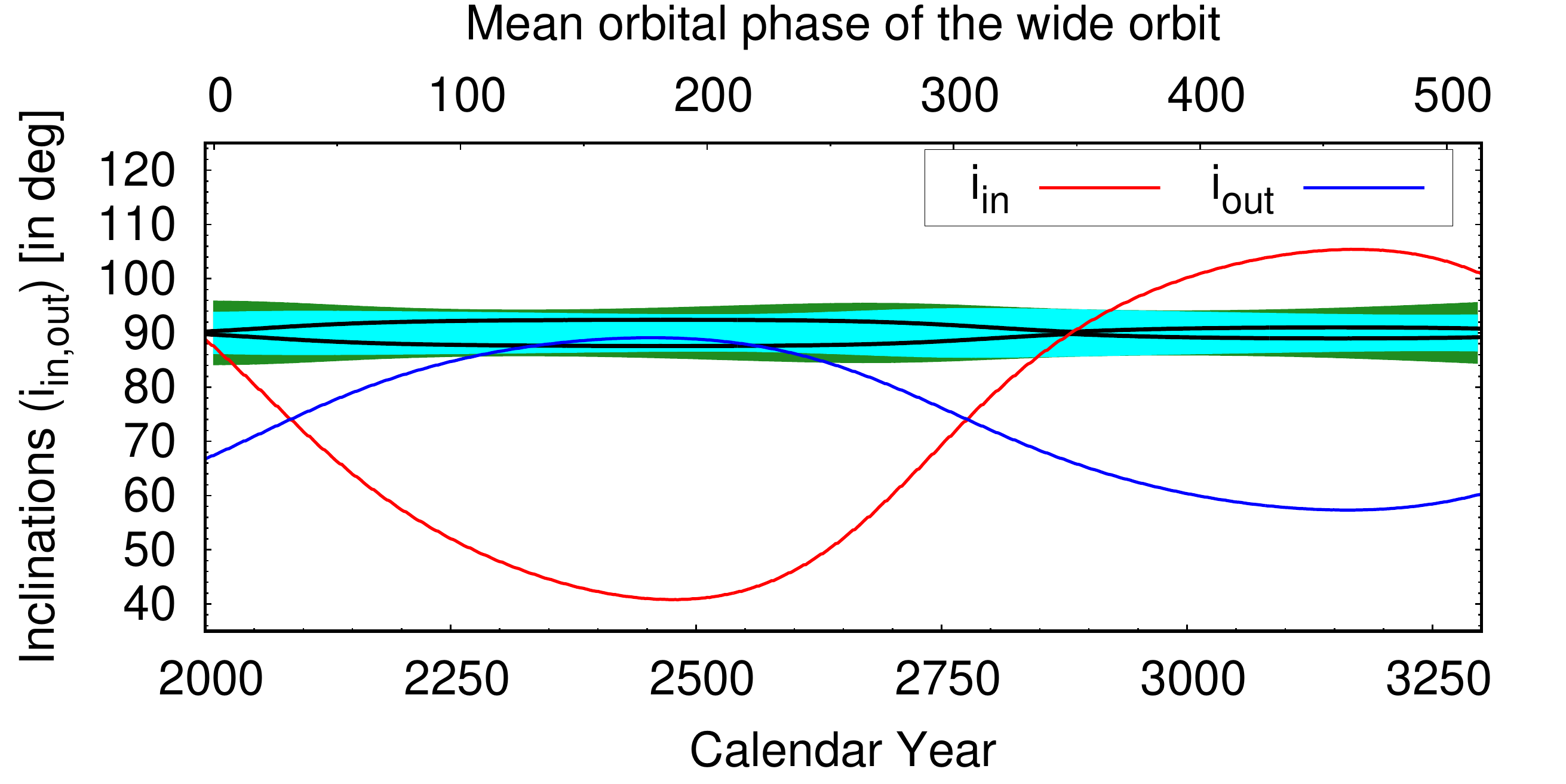}
\caption{Variations of the same orbital elements as in Fig.~\ref{fig:K6964043orbelements_numint} above, but for KIC 5731312.
See text for details.}
\label{fig:K5731312orbelements_numint} 
\end{center}
\end{figure*}  

The long-term future of KIC 5731312 is remeniscent of that of KIC 6964043 (see Fig.~\ref{fig:cartoon1}).  On long timescales, here of the order of more than a few Gyr, we find an extra modulation in the eccentricity of the outer orbit due to the octupole term (Fig. \ref{fig:5731312_ev}). The eccentricity first decreases from the original value of $\approx 0.57$ to 0.55, after which it starts increasing.  When the outer eccentricity reaches a value of $\approx 0.81$ at 3.96 Gyr, the system crosses the stability limit, in analogy of KIC 6964043. Given the stellar masses  of KIC 5731312, the stars are still on the MS when this happens, and will remain so within a Hubble time. The evolution during the dynamically unstable phase is therefore a pure dynamical (gravitational) problem. After many crossing times ($10^4$ crossing time is about 0.75 kyr), the system most likely dissolves into a bound binary and a single star ejecting the low-mass tertiary star from the system. 

\begin{figure}
\includegraphics[width=1.01\columnwidth]{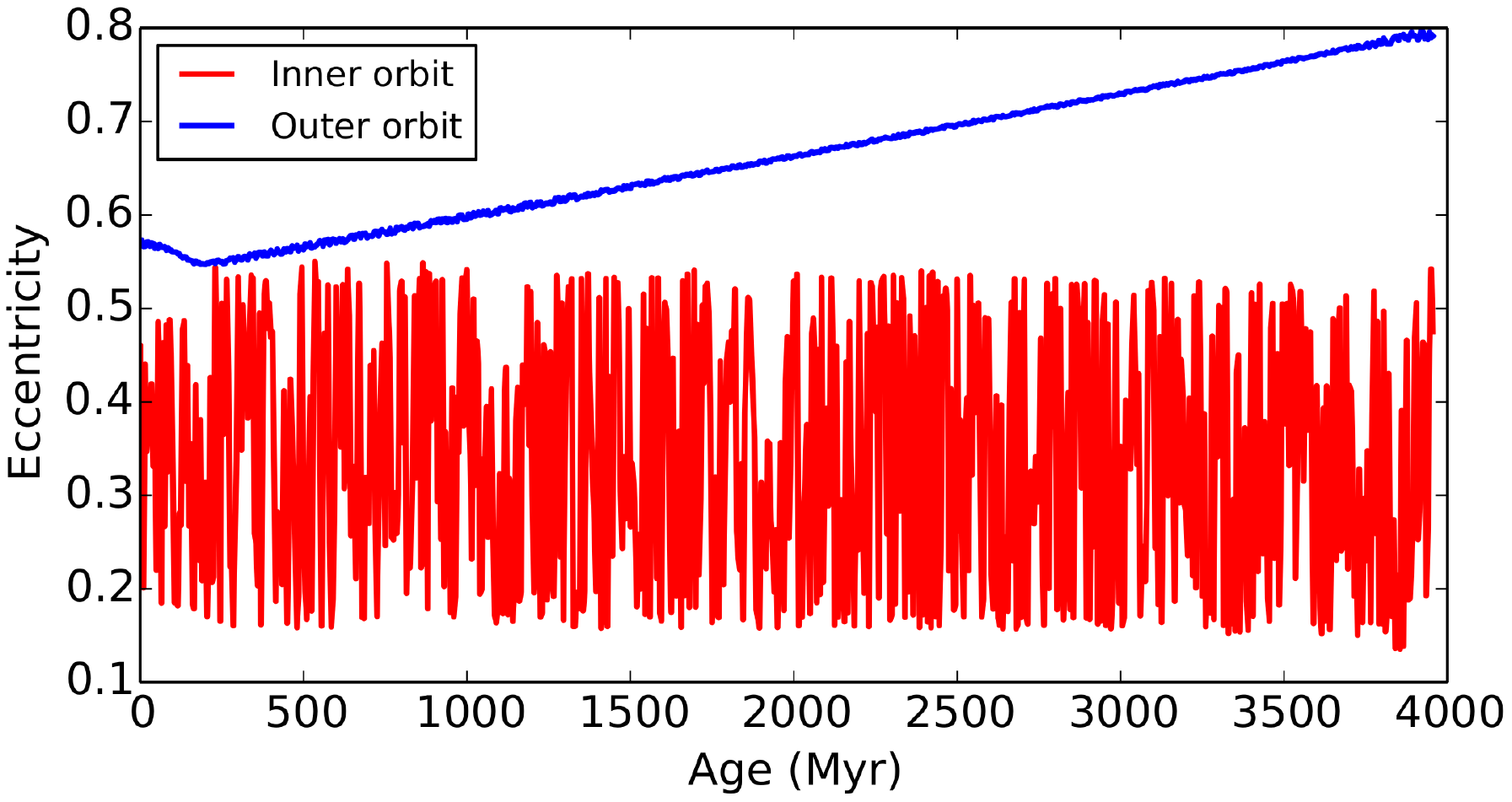} 
 \caption{Evolution of the system eccentricities in KIC 5731312.}
\label{fig:5731312_ev}
\end{figure}  

\subsection{KIC 8023317}
\label{sec:5.4}

This triple has the most extreme inner and outer mass ratios in our sample ($q_\mathrm{in}=0.303\pm0.013$; $q_\mathrm{out}=0.079\pm0.003$). The main stellar component is one of the EB stars, and is a slightly evolved G-type star ($m_\mathrm{Aa}=1.30\pm0.07\,\mathrm{M}_\odot$; $R_\mathrm{Aa}=1.96\pm0.05\,\mathrm{R}_\odot$; $T_\mathrm{eff,Aa}=5870\pm70\,\mathrm{K}$).  Its binary companion is a low mass M-type red dwarf ($m_\mathrm{Ab}=0.39\pm0.01\,\mathrm{M}_\odot$; $R_\mathrm{Ab}=0.39\pm0.01\,\mathrm{R}_\odot$; $T_\mathrm{eff,Ab}=3165\pm45\,\mathrm{K}$).  The tertiary is another even lower mass M star with $m_\mathrm{B}=0.13\pm0.01\,\mathrm{M}_\odot$, $R_\mathrm{B}=0.17\pm0.01\,\mathrm{R}_\odot$, and $T_\mathrm{eff,B}=2575\pm35\,\mathrm{K}$.  The latter star has the lowest mass among all dozen in our full sample. The age of the triple is found to be $\tau=4.3\pm0.5$\,Gyr. The photodynamically obtained distance is $d=814\pm20$\,pc, which is larger than the Gaia EDR3 inferred distance by about 3-$\sigma$. 

The mutual inclination is found to be $i_\mathrm{mut}=55\fdg7\pm0\fdg8$, which is even higher by $\sim6\degr$ than the former result of \citet{borkovitsetal15,borkovitsetal16}. Thus, according to our knowledge, this system has the second highest mutual inclination after Algol amongst all triple star systems with accurately known mutual inclinations.

The ZKL cycles in the inner eccentricity and the mutual inclination are, again, clearly visible (see upper panels of Fig.~\ref{fig:K8023317orbelements_numint}) and look quite similar to those same features of the previous system, KIC 5731312.  One quantitative difference, however, is that the ZKL cycle period is shorter, being $P_\mathrm{ZKL}\approx360$\,yr, in accordance with the shorter $P_\mathrm{out}^2/P_\mathrm{in}$ timescale. The ranges of the variations of both parameters are very close to that of KIC~5731312, but the phase is almost opposite. While KIC~5731312 currently is close to its maximum eccentricity and, hence, minimum mutual inclination phase, KIC~8023317 is close to its smallest inner eccentricity, and largest mutual inclination.

There is, however, an important qualitative difference in the behavior of the apsidal lines of the inner orbits between the present and the previous system. While in the case of KIC~5731312, the dynamical apsidal line circulates, in the case of KIC~8023317 it librates around $\omega_\mathrm{in}^\mathrm{dyn}=90\degr$, with a half-amplitude of $\approx27\degr$.  The period of this libration is the same as that of the inner eccentricity cycles. Though prior numerical \citep[e.g,][]{fordetal00} and theoretical works \citep{naoz16} have shown that the periods of apsidal librations and circulations may alter each other, in the present situation the libration remained unchanged throughout our 1\,Myr-long numerical integration. Regarding the observable apsidal motion of the inner pair, it is again clearly retrograde, as can be seen nicely in the middle panels of Fig.~\ref{fig:K8023317orbelements_numint}.

Finally, we consider the variations of the inclination angles in the bottom panels of Fig.~\ref{fig:K8023317orbelements_numint}. The precession period is about $P_\mathrm{node}\approx650$\,yr.  The current eclipsing session of the inner pair started around 2000.0, and central eclipses will occur around 2023.7. The eclipses will then completely disappear around 2052.7, and will return only near the beginning of 2088.  Moreover, similar to the other three triples studied in this work, third-body eclipses are also possible during given sections of a precession cycle. In contrast to the previous systems, however, in the case of KIC~8023317, the outer inclination crosses both the lower and upper borders of the possible third-body eclipse regions (and hence, the outer orbit may also be seen edge-on two times during a precession cycle). The next session of possible third-body eclipses will begin around 2354 and last until 2404.

\begin{figure*}
\begin{center}
\includegraphics[width=0.50 \textwidth]{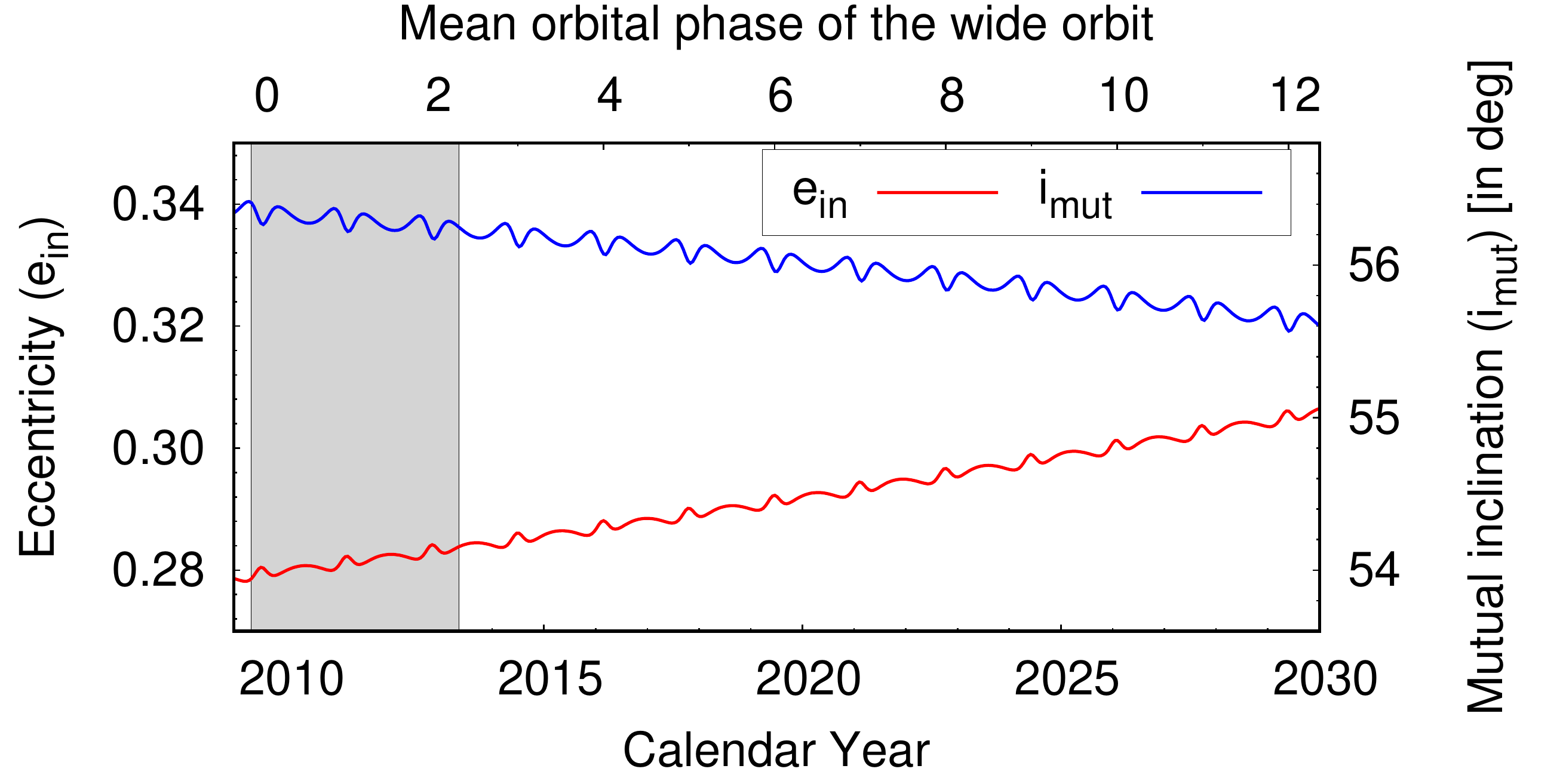}\includegraphics[width=0.50 \textwidth]{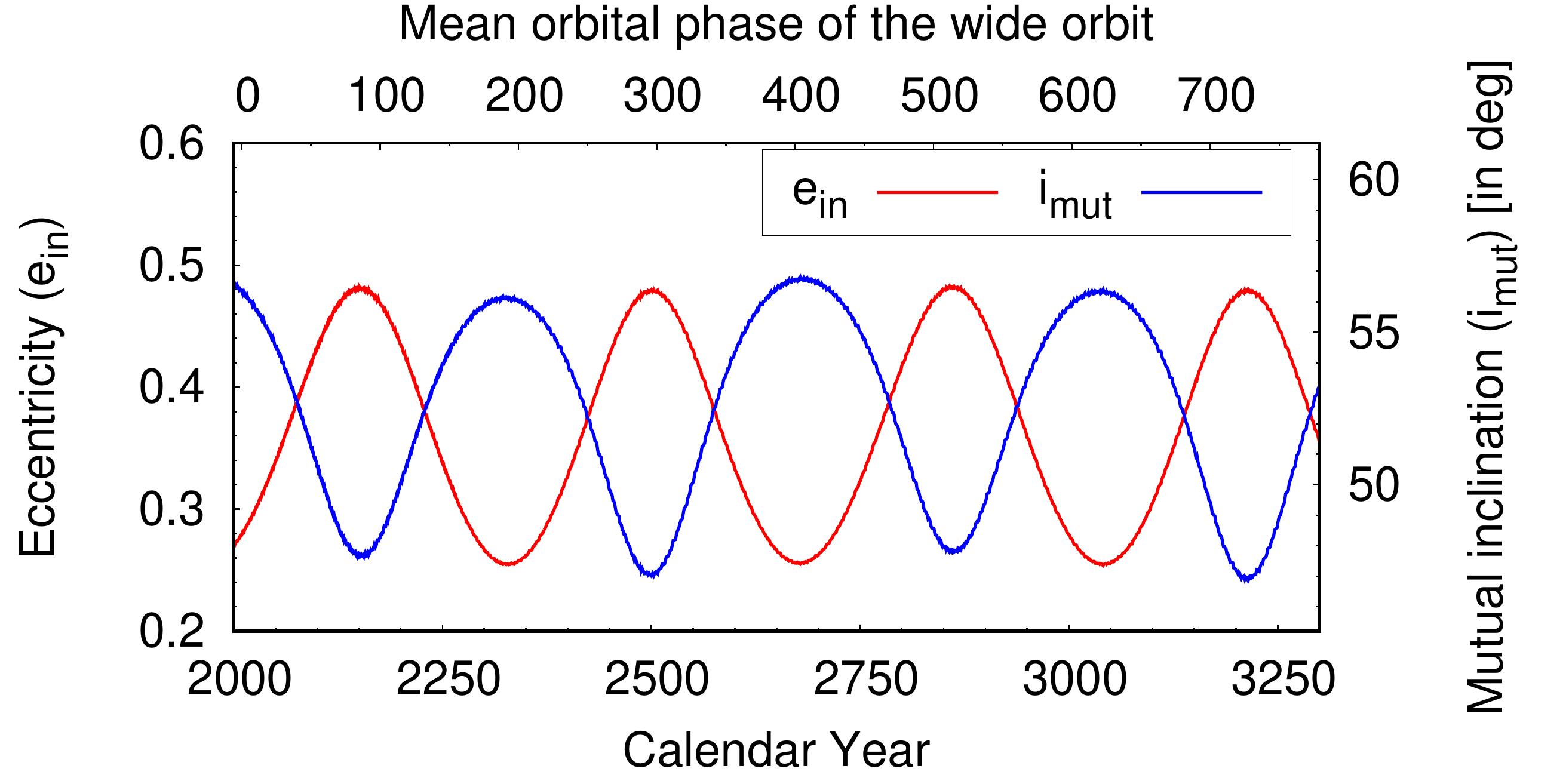}
\includegraphics[width=0.50 \textwidth]{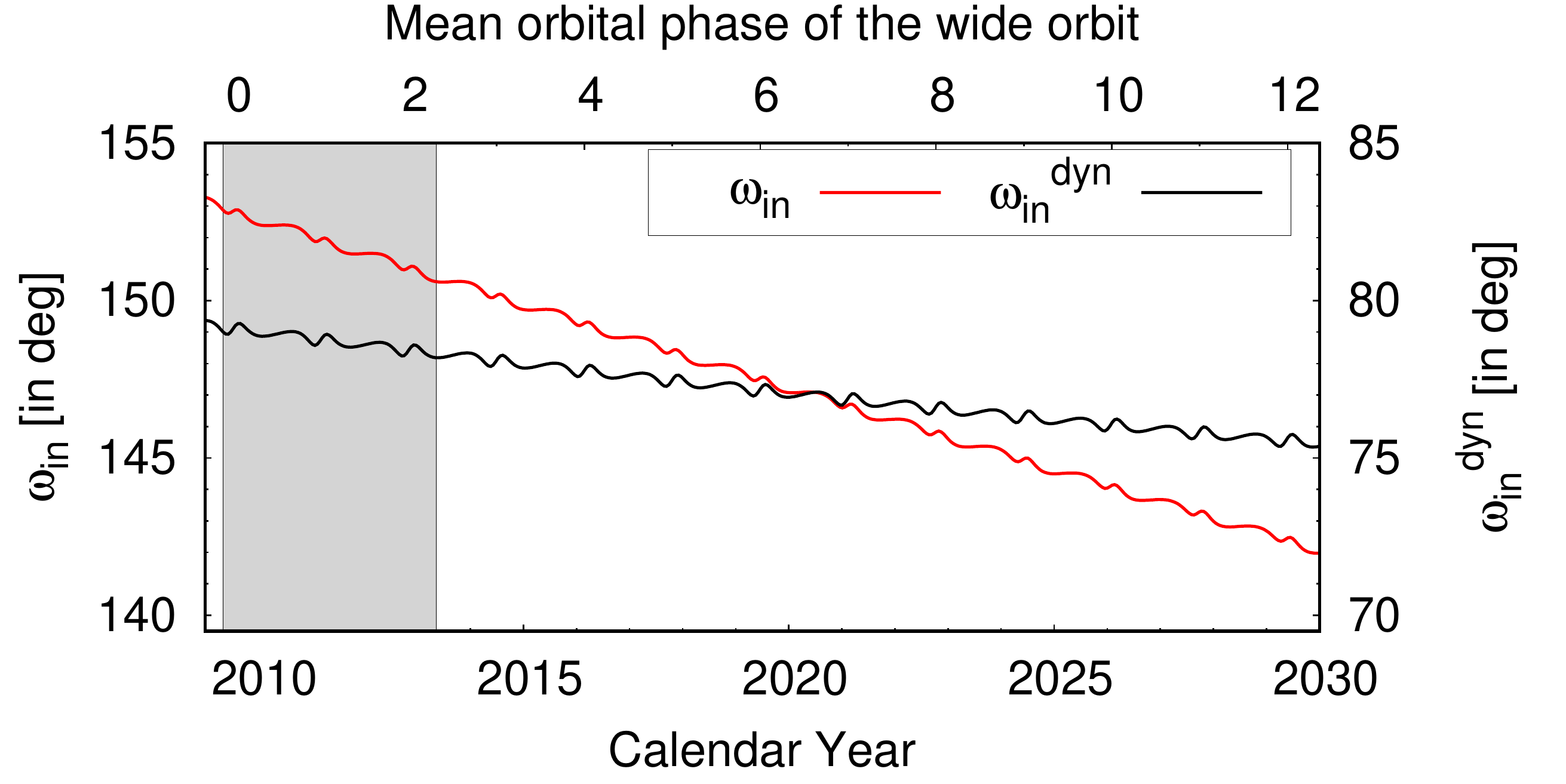}\includegraphics[width=0.50 \textwidth]{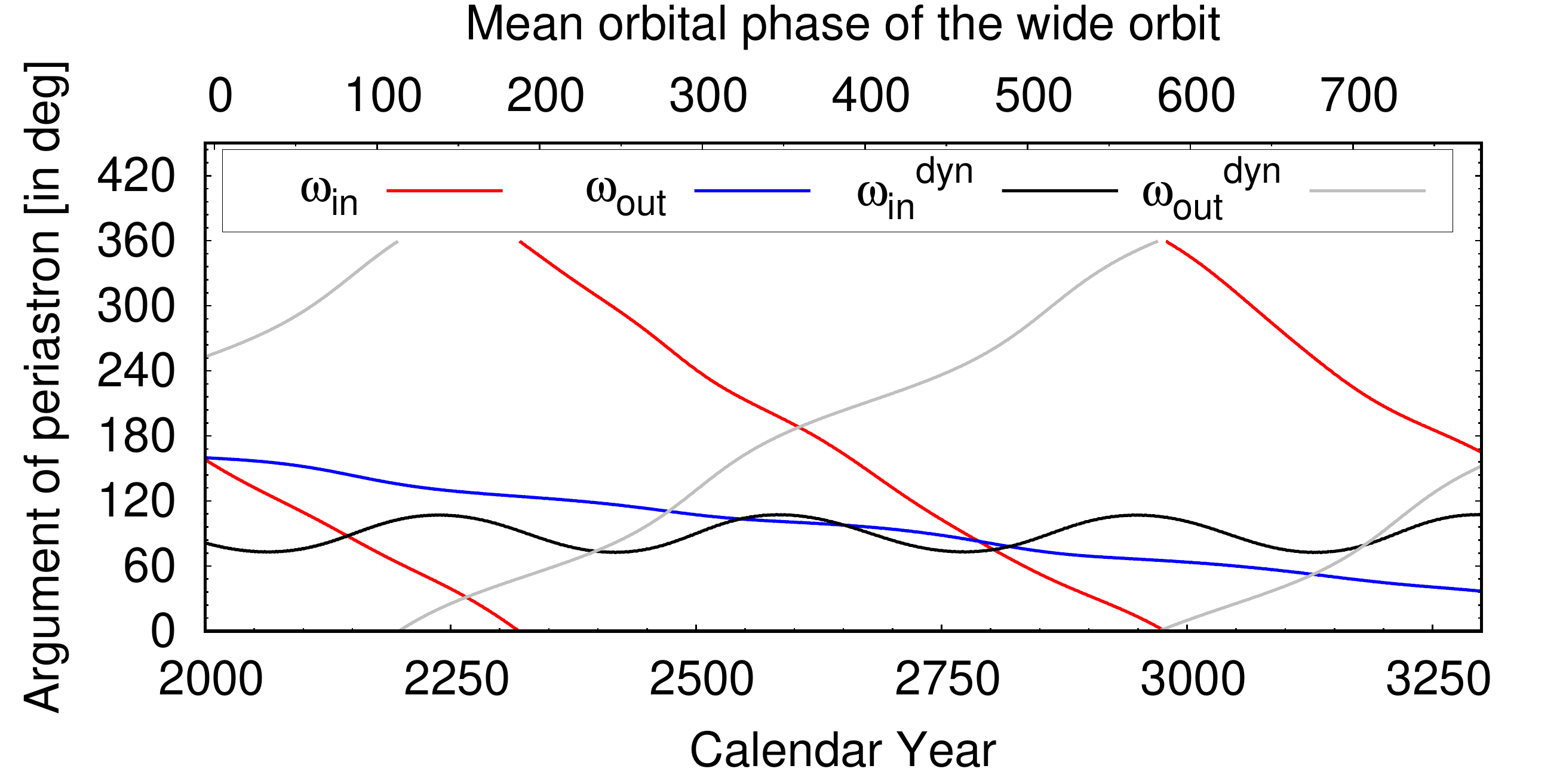}
\includegraphics[width=0.50 \textwidth]{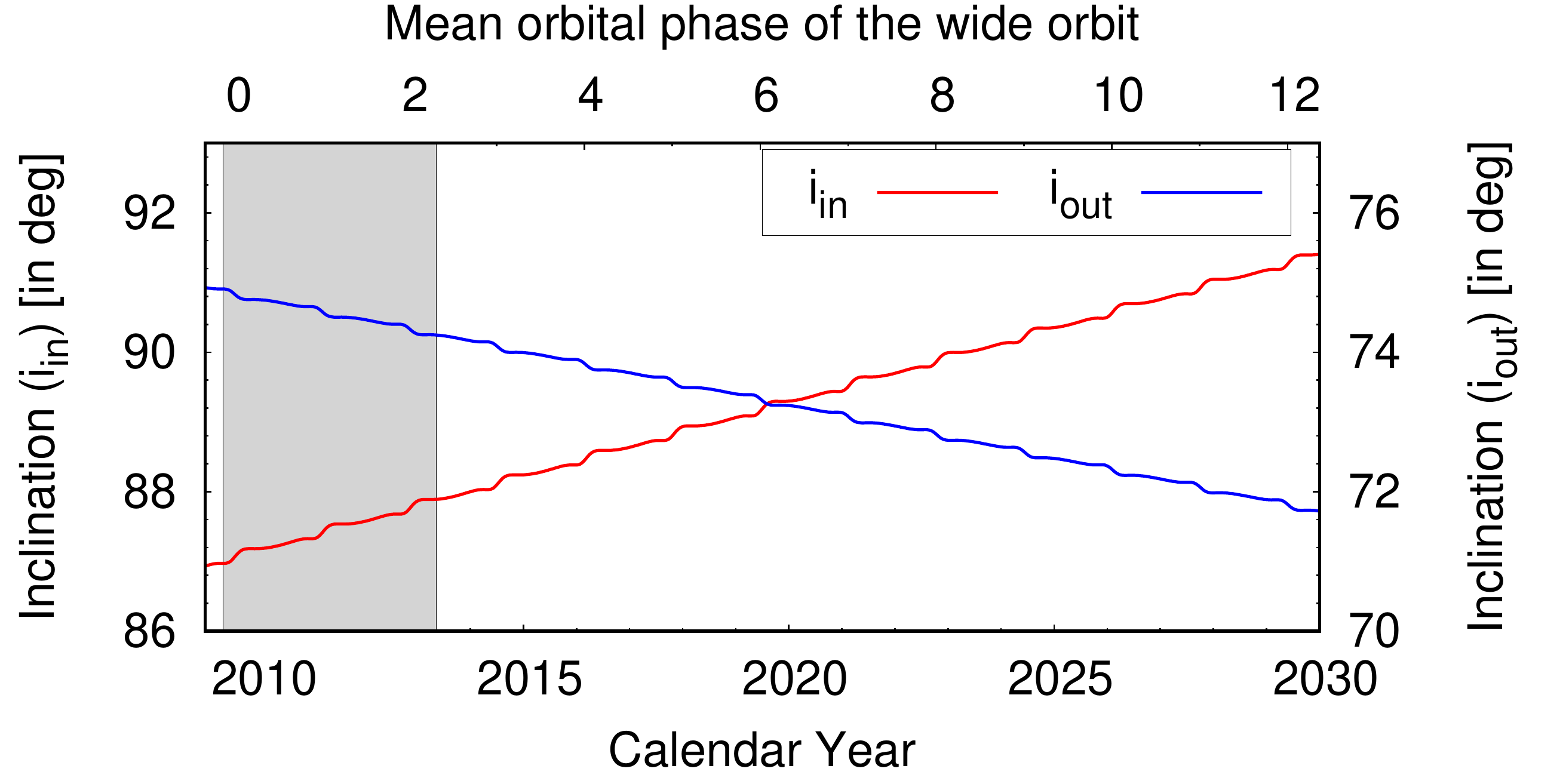}\includegraphics[width=0.50 \textwidth]{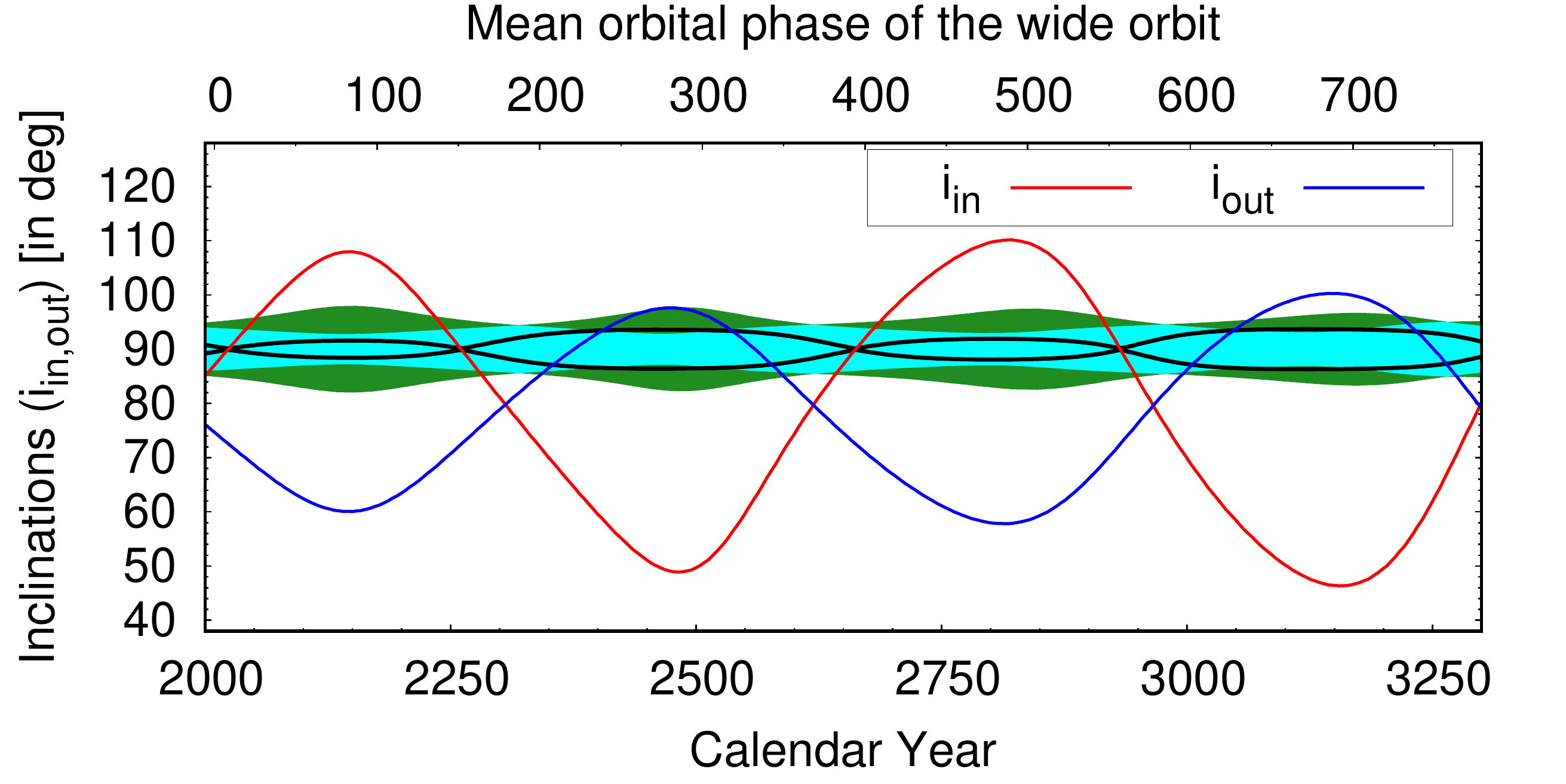}
\caption{Variations of the same orbital elements as in Fig.~\ref{fig:K6964043orbelements_numint} above, but for KIC 8023317.
See text for details.}
\label{fig:K8023317orbelements_numint} 
\end{center}
\end{figure*}  

The long-term evolution of KIC 8023317 is similar to that of KIC 5653126 (see Fig.~\ref{fig:cartoon2}). The primary star is massive enough to evolve off the main-sequence at 4.7 Gyr and fill its Roche lobe as it ascends the first giant branch at 5.3 Gyr. At this point the primary star has lost a small amount of mass due to its stellar wind, i.e., 0.004\,M$_{\odot}$, such that the outer semimajor axis has increased by 0.2\%. The inner orbit did not widen, in fact the inner semimajor axis was reduced from the original 32.6\,R$_{\odot}$ to 24.5\,R$_{\odot}$ due to tidal friction in combination with ZKL cycles \citep{kiselevaetal98}. 
Furthermore, the inner orbit has circularized  by the time of the onset of the mass transfer phase and the inclination freezes out at about 41 degrees.  With the extreme mass ratio of the inner binary, the mass transfer leads to a common-envelope phase and a subsequent merger of the two stars in the inner binary. The merger remnant is a giant-like object with a helium core of 0.24\,M$_{\odot}$ and a relatively thick envelope of at most 1.44\,M$_{\odot}$. In analogue with KIC 5653126, the merger remnant will fill its Roche lobe as an AGB star, initiate a second CE-phase. A compact binary is formed with a 0.52\,M$_{\odot}$ white dwarf component and a 0.13\,M$_{\odot}$ main-sequence companion. In the case of an efficient CE-phase, the post-CE semimajor axis is about 8\,R$_{\odot}$. In the inefficient case it is $\sim$1\,R$_{\odot}$, which is compact enough such that magnetic braking and gravitational wave emission can reduce the orbit to form a cataclysmic variable in a Hubble time. 

\subsection{Implications of the results for different multiple star formation processes}
\label{sec:Max}

As indicated above, there appears to be an anti-correlation between the mutual inclination and tertiary mass. Our two highly misaligned triples, KIC~5731312 with $i_{\rm mut}$ = 39$^{\circ}$ and KIC~8023317 with $i_{\rm mut}$ = 56$^{\circ}$, both contain low-mass M5/6V tertiaries with $m_{\rm B}$ = 0.13\,-\,0.15\,M$_\odot$. Meanwhile, our two relatively aligned triples have more comparable component masses. To determine if this trend is statistically significant, we expand our sample to include all of our previously analyzed triples with reliably measured mutual inclinations and outer mass ratios $q_{\rm out}$ = $m_{\rm B}$/($m_{\rm Aa}$+$m_{\rm Ab}$).  From the \citet{borkovitsetal16} sample of 62 compact {\em Kepler} triples with measured mutual inclinations, we update the four systems presented in this study with the current, more accurate photodynamical results. We also update KIC~7955301 with $i_{\rm mut}$ = 6.2$^{\circ}$ and $q_{\rm out}$ = 0.71 according to our upcoming photodynamical analysis (Gaulme et al., in prep). We exclude KIC~3345675 and KIC~7593110, which have large mutual inclination uncertainties that exceed $\delta i_{\rm mut}$ $>$ 10$^{\circ}$.  We also ignore the only retrograde triple, KIC~7670617 with $i_{\rm mut}$ = 147$^{\circ}$, which likely formed differently than the remaining 59 KIC prograde triples spanning  0$^{\circ}$ $<$ $i_{\rm mut}$ $<$ 56$^{\circ}$. We also add the 12 recently discovered {\em TESS} triples, most of which are nearly coplanar triply eclipsing systems \citep{borkovitsetal20a,borkovitsetal20b,borkovitsetal22,mitnyanetal20,rappaportetal22}. Nonetheless, TIC~042565581 is slightly misaligned with $i_{\rm mut}$ = 5.5$^{\circ}$ and $q_{\rm out}$ = 0.65 while TIC~167692429, (the only one currently non-eclipsing \textit{TESS} triple) is moderately misaligned with $i_{\rm mut}$ = 27$^{\circ}$ and $q_{\rm out}$ = 0.34.

\begin{figure}
\begin{center}
\includegraphics[width=0.50 \textwidth]{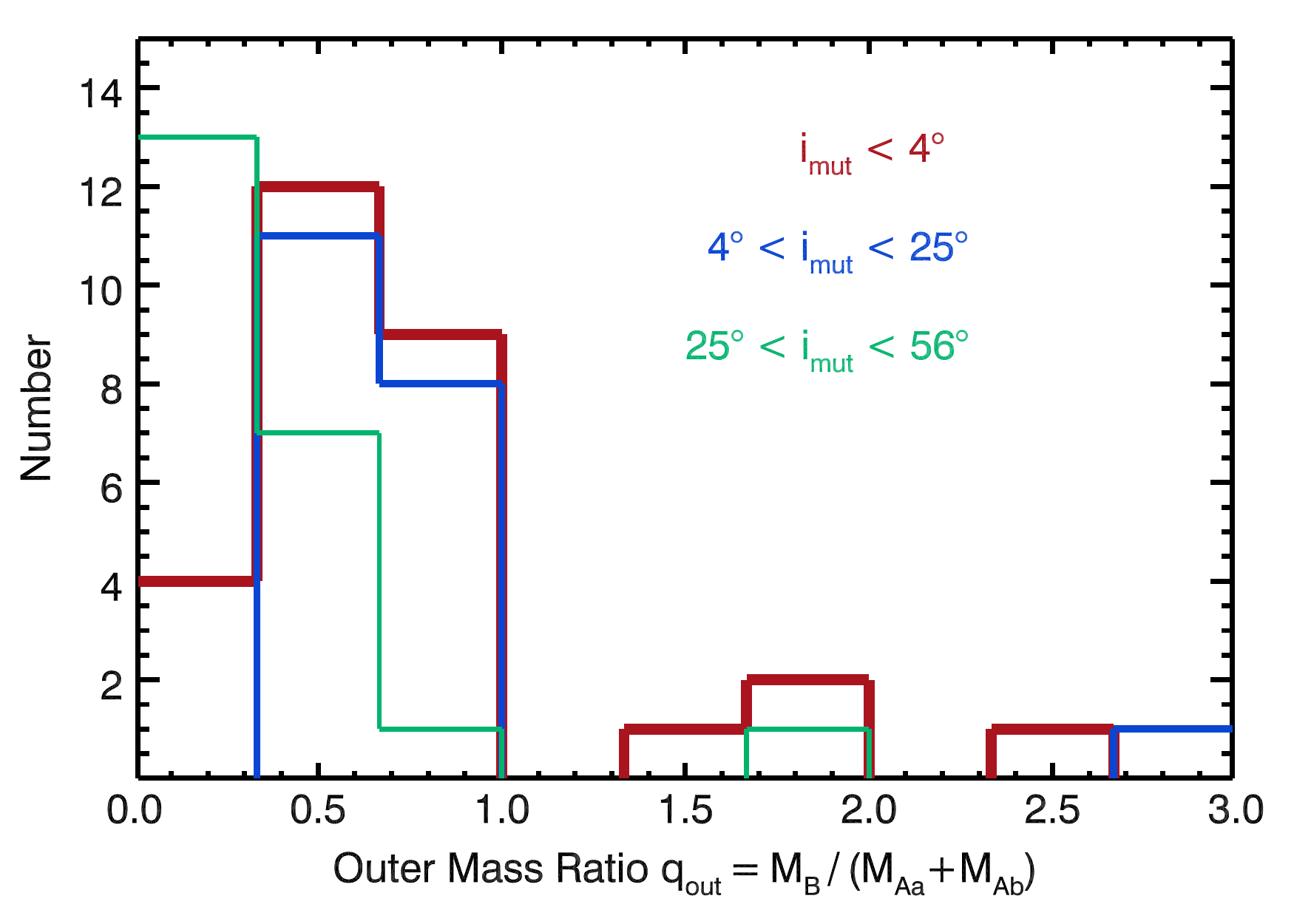}
\caption{The distributions of outer mass ratios for the 29 nearly coplanar triply eclipsing systems with $i_{\rm mut}$ $<$ 4$^{\circ}$ (thick red), the 20 moderately misaligned triples with 4$^{\circ}$ $\le$ $i_{\rm mut}$ $<$ 25$^{\circ}$ (blue), and the 22 highly misaligned triples with 25$^{\circ}$ $\le$ $i_{\rm mut}$ $<$ 56$^{\circ}$ (thin green). More coplanar triples tend to have more massive outer tertiaries, suggesting that compact triples accreted from surrounding protostellar disks that simultaneously aligned the orbits while leading to preferential mass accretion onto the outer tertiaries.}
\label{fig:qout_vs_Imut} 
\end{center}
\end{figure}

A Spearman rank correlation test demonstrates that the mutual inclinations and outer mass ratios of our selected 71 compact triples are anti-correlated with a coefficient of $\rho_{\rm S}$ = $-$0.40 and statistical significance of $p_{\rm S}$ = 0.0006 (3.5$\sigma$). In Fig.~\ref{fig:qout_vs_Imut}, we plot the distributions of $q_{\rm out}$ for three subsamples: (1) the 29 nearly coplanar triply eclipsing systems with $i_{\rm mut}$ $<$ 4$^{\circ}$ that are subject to different selection effects than the non-triply eclipsing systems; (2) the 20 somewhat misaligned triples with 4$^{\circ}$ $\le$ $i_{\rm mut}$ $<$ 25$^{\circ}$; and (3) the 22 highly misaligned triples with 25$^{\circ}$ $\le$ $i_{\rm mut}$ $<$ 56$^{\circ}$. The nearly coplanar triples are concentrated across $q_{\rm out}$ = 0.33\,-\,1.00, but there are a few systems below $q_{\rm out}$ $<$ 0.33 and above $q_{\rm out}$ $>$ 1.00. Similarly, all but one of the 20 somewhat misaligned triples span $q_{\rm out}$ = 0.33\,-\,1.00. The single outlier with $q_{\rm out}$ = 3.0 and $i_{\rm mut}$ = 8$^{\circ}$ is KIC~5897826, which is barely a triply eclipsing system \citep{carteretal11}. Meanwhile, more than half of the highly misaligned triples have small mass ratios $q_{\rm out}$ $<$ 0.33. The single outlier with $i_{\rm mut}$ = 54$^{\circ}$ and $q_{\rm out}$ = 1.8 is KIC~4055092, which has one of the widest tertiaries with $a_{\rm out}$ = 11~AU and therefore likely formed differently from the more compact triples in our sample\footnote{Note, however, that the \citet{borkovitsetal16} ETV solution for KIC~4055092 should be considered only with considerable caution, as it gives an outer period of $P_\mathrm{out}\approx31.6$\,yr which is $\sim8$-times longer than the duration of the analysed \textit{Kepler} dataset.} An Anderson-Darling test demonstrates that the somewhat misaligned and highly misaligned triples have different $q_{\rm out}$ distributions at the $p_{\rm AD}$ = 8$\times$10$^{-6}$ (4.3$\sigma$) significance level.  Moreover, if we limit our sample to the 40 triples with 4$^{\circ}$ $\le$ $i_{\rm mut}$ $<$ 56$^{\circ}$ and $q_{\rm out}$ $\le$ 1.0, then a Spearman rank test yields an even stronger degree of anti-correlation with $\rho_{\rm S}$ = $-$0.66 and $p_{\rm S}$ = 4$\times$10$^{-6}$ (4.5$\sigma$). The anti-correlation between the mutual inclinations and outer mass ratios for this subset of 40 compact triples is clearly visible in Fig.~\ref{fig:scatter_plot}.
\begin{figure}
\begin{center}
\includegraphics[width=0.50 \textwidth]{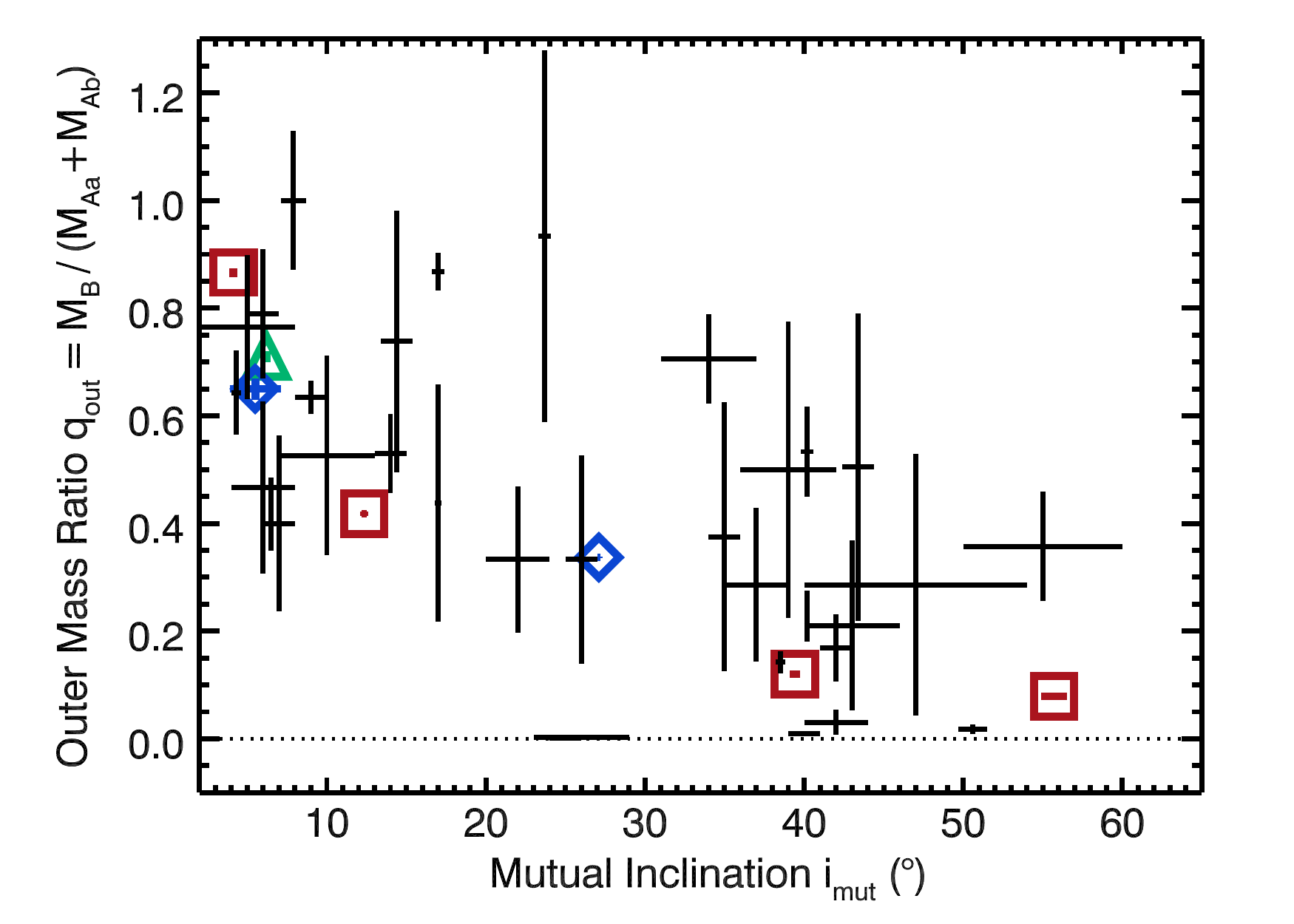}
\caption{Outer mass ratios versus mutual inclinations for the 40 compact triples with 4$^{\circ}$ $\le$ $i_{\rm mut}$ $<$ 56$^{\circ}$ and $q_{\rm out}$ $\le$ 1.0. We highlight the 4 KIC triples presented in this study (red boxes), the two misaligned TIC triples (blue diamonds), and the Gaulme et al. preliminary result for KIC~7955301 (green triangle). The outer mass ratios systematically decrease with mutual inclination at the 4.5$\sigma$ level. A more massive tertiary likely accreted from a massive circumtriple protostellar disk that aligned both the inner and outer orbits with the disk.}
\label{fig:scatter_plot} 
\end{center}
\end{figure}  

Compact triples with $a_{\rm out}$ $\lesssim$ 10~au likely formed either through two successive episodes of disk fragmentation or disk-mediated capture of the outer tertiary \citep{TobinKratter2016, Bate2018, tokovininmoe20, Offner2022}. In both scenarios, the tertiary tends to be initially less massive than the inner binary, but the triple subsequently accretes from the surrounding protostellar disk. For an isolated protobinary that accretes from a circumbinary disk, the low-mass secondary sweeps out a larger orbit and thus accretes most of the infalling material, effectively driving the binary mass ratio $q$ = $M_2$/$M_1$ toward larger values \citep{BateBonnell1997, Farris2014, YoungClarke2015}.  Close binaries exhibit a large excess of twins with $q$ $>$ 0.95 \citep{Tokovinin2000, MoeDiStefano2017, ElBadry2019, tokovininmoe20}, consistent with the predictions of the hydrodynamical simulations of circumbinary disk accretion. Similarly, a low-mass tertiary in a compact triple will accrete most of the mass from the circumtriple disk, thereby increasing the outer mass ratio $q_{\rm out}$ \citep{tokovininmoe20}.

For a protobinary with small to moderate eccentricity, circumbinary disk accretion also tends to align the binary orbit to the plane of the disk \citep{PapaloizouTerquem1995, LubowOgilvie2000, LodatoFacchini2013, Foucart2014, MartinZhuArmitage2020}. Close pre-MS binaries undergo tidal circularization \citep{MeibomMathieu2005, MoeKratter2018, ZanazziWu2021}. Indeed, pre-MS binaries with $P$ $<$ 30~days have small to moderate eccentricities and nearly coplanar circumbinary disks as expected \citet{Czekala2019}. A significant majority of the inner binaries in our sample of compact triples also have small to moderate eccentricities and short periods, and so we expect the inner binary and circumbinary disk to have aligned prior to the formation or disk-capture of the tertiary.  A notable exception includes KIC~5731312 investigated in this study, which has an inner binary with a relatively large eccentricity $e_{\rm in}$ = 0.46 given its short period of $P_{\rm in}$ = 7.95 days. Nonetheless, KIC~5731312 also has a highly inclined tertiary with $i_{\rm mut}$ = 39.4$^{\circ}$, precisely matching the critical von Zeipel-Kozai-Lidov angle (see Section~\ref{sec:5.3}), and suggesting this particular triple dynamically evolved via ZKL cycles coupled to tidal friction \citep{kiselevaetal98,FabryckyTremaine2007, NaozFabrycky2014}.  For the rest of our systems, once the tertiary forms in the outer disk via disk fragmentation (born nearly coplanar) or is captured by the disk, subsequent inclination evolution of the tertiary depends on the mass of the disk \citep{Martin2016}. For massive disks, the tertiary accretes much of the mass and angular momentum from the disk, thereby dissipating into a coplanar configuration. Meanwhile, for low-mass disks with negligible accretion onto the tertiary, \citet{Martin2016} demonstrated that the tertiary can remain at large mutual inclinations or even be pumped to large inclinations via secular oscillations.

The combination of the tertiary accreting most of the mass and angular momentum from the surrounding disk explains the observed anti-correlation between $q_{\rm out}$ and $i_{\rm mut}$. Accretion from a massive disk will substantially increase $q_{\rm out}$ while dissipating the triple toward coplanarity. Moreover, compact triples that formed via two episodes of disk fragmentation should be especially coplanar, perhaps explaining the excess of triply eclipsing systems with $i_{\rm mut}$ $<$ 4$^{\circ}$ relative to moderately misaligned triples, albeit an observational selection bias could also play a role. Meanwhile, a tertiary that is captured by a low-mass disk late in the formation of the system will accrete only marginal additional mass and its inclination with respect to the inner binary will change only slightly. Hence the observed population of low-mass tertiaries tend to have large mutual inclinations.


\section{Conclusions}
\label{sec:conclusions}

In this paper we have carried out complex photodynamical analyses of four compact, tight, hierarchical triple stellar systems.  These systems were observed quasi-continuously  with high photometric precision by the \textit{Kepler} spacecraft during its four-year-long prime mission, and then were more recently revisited with the \textit{TESS} space telescope. All  four triples consist of currently eclipsing, eccentric inner binaries which, due to strong third-body interactions and the non-alignment of the inner and outer orbits, produce rapid and large amplitude eclipse timing and depth variations.  These serve as ideal targets for precise, in-depth dynamical analyses.  Moreover, one of the four targets, KIC~6964043, also exhibits third-body eclipses, thereby allowing further refinements of the system parameters.

On the other hand, unfortunately, due the faintness of the systems no RV measurements are available for any of them.  Thus, we used an analysis of the composite system SED in combination with astrophysical-model dependent \texttt{PARSEC}  isochrones as proxies for the determination of the individual stellar masses and, indirectly, radii. In such a manner, we were able to obtain fundamental stellar parameters with accuracies of about $2-8\%$. This relatively lower accuracy in these physical stellar parameters, however, does not affect our main purpose of mapping the full 3-dimensional configurations of the triples and their dynamics with unprecedentedly high precision.  This results from the fact that the geometry and dynamics of the systems are mainly determined by such relative dimensionless quantities as the mass ratios and the orbital positions of the bodies relative to each other. Thus, we were able to determine the orbital elements, both in the observational and the dynamical frames of references, for all systems and subsystems with precisions of better than $1\%$.  Here we mainly highlight the precise determinations not only of the mutual inclination angles, but also of the current positions of the dynamical lines of apsides, and dynamical nodes. These are direct important key parameters in the equations of secular perturbation theory and, hence, in addition to the mutual inclination of the orbits and the ratios of the masses and separations (periods), they substantially influence the future dynamical evolution of the systems. Thus, knowing these parameters, we were able to model realistically the future dynamical, and corresponding astrophysical, evolution of the four triple systems.

Two of the four triples, KICs~6964043 and 5653126 were found to be practically flat. In these systems our short-term numerical integrations have shown only minor orbital variations in the forthcoming 1 million years, despite the fact that KIC~6964043 is an extremely tight triple system. On the other hand, integrating the secular dynamical evolution of these systems over longer timescales, we find that KIC~6964043 is expected to become dynamically unstable and most likely dissolve within the next 1\,Gyr.  In contrast to this scenario, KIC~5653126 will pass through two common envelope phases and also stellar merger(s) before it will end its life most likely as a single white dwarf.

The other two systems, KICs~5731312 and 8023317 were demonstrated to belong to the high mutual inclination regime, at least in the sense of the original ZKL theory. To the best of our knowledge, KIC 8023317 has the second highest accurately known mutual inclination angle among the known hierarchical triple stellar systems.  (The highest mutual inclination system is Algol, where the outer orbit is nearly perpendicular to the inner one; \citealp{lestradeetal93,csizmadiaetal09,zavalaetal10}.)  According to our short-term numerical integrations, both inclined KIC triple systems exhibit characteristic, quasi-sinusoidal variations in their inner eccentricities and mutual inclinations ($e$-cycles of the ZKL-theorem) on some hundred year-long timescales, with amplitudes of 10--30\% of the entire physically meaningful range of these parameters. Moreover, while the dynamical lines of apsides of the inner pair in KIC~5731312 circulates (though this circulation has superposed on it some librations), KIC~8023317 exhibits clear libration, at least during the forthcoming one million years. On the other hand, the observable apsidal motion in both of these highly mutually inclined systems is retrograde, which make these EBs more exotic. Similar to the previous, almost flat systems, the final fates of these triples are driven again by the masses of their components. Our secular dynamical evolution studies predict that the low-mass triple KIC~5731312 will also become unstable on a Gyr timescale, and most probably, the less massive tertiary will be ejected from the system. Finally, the evolution of the more massive triple KIC~8023317 is expected to resemble that of KIC~5653126 with the difference in this case being that the most probable final remnant may be a cataclysimic variable instead of a single white dwarf.

Due to the non-alignment of the inner and outer orbits, all four targets exhibit rapid orbital precession and hence, eclipse depth variations. As a consequence, none of the four inner binaries remain eclipsing during their complete precession cycle.  In this regard the nearly flat system, KIC~6964043 is the least affected, as its inner pair will display eclipses over the vast majority of its precession cycle. In the case of the other three systems, they will exhibit regular, inner eclipses only during short portions of their $\sim$$200-1400$\,yr-long precession cycles. Naturally, this implies that there may be numerous similar tight compact systems which are currently in non-eclipsing phases, but which can be discovered as EBs in the near or far future.

What is more interesting is that our photodynamical results predict that all four systems may well exhibit third-body eclipses during some shorter or longer phases of their complete precession cycles.  In this regard\, KIC~6964043 is exceptional in the sense that it continuously displays third-body eclipses.  We further note that this system is similar to the vast majority of the presently known triply eclipsing triple systems, which were found to be practically flat and, hence, they continuously exhibit regular inner as well as third-body eclipses.  
One might therefore think that finding third-body eclipses in EBs is just an observational selection effect, since EBs are naturally more intensively observed and followed than non-eclipsing stars. By this reasoning, there would be a much smaller chance to discover serendipitously third-body eclipses in the case of a target that exhibits neither regular eclipses nor some other kind of characteristic light variations which would make it worthy of frequent observations.  It turns out that finding third-body eclipses in EBs is indeed a selection effect, but not simply because EBs tend to be better studied.  In Rappaport et al.~(2022) we discussed the fact that substantially more triply eclipsing triples were found in a visual survey of $\sim$$10^6$ preselected EBs than in $\sim$$10^7$ random stars, all taken from the {\it TESS} full frame images.  The selection effects involved include the facts that: (i) the more compact triple systems that we are studying (i.e., with $P_{\rm out} \lesssim 1000$ days) are expected to be nearly flat (see Sect.~\ref{sec:Max}); (ii) the shorter the outer period in a triple system (i.e., the more compact), the wider the outer inclination domain where third-body eclipses may occur; and (iii) once an eclipsing binary has been found, it is already somewhat preferentially aligned so as to increase the eclipse pobability of an outer triple star.  Therefore, in such a manner, the enhancement of the third-body eclipsers amongst the EBs (forming {\it triply} eclipsing triples) may be an indirect consequence of the formation mechanism(s) of most compact triples, i.e., one that produces practically flat systems.

Our results show that there might be a fair number of episodically appearing third-body eclipser systems amongst inclined compact triple systems.\footnote{This situation partially resembles the findings of \citet{martintriaud15} who claimed that most of the inclined circumbinary planets orbiting EBs should produce transits for shorter or longer times during their orbital precession cycles.}  The cases of KIC~2835289 and TIC~167692429 were already mentioned above (Sect.~\ref{sec:5.2}). We note, furthermore, that we have also found a few more solo third-body eclipse-like events in the \textit{TESS} observations of other currently non-eclipsing targets, and one such event was especially noteworthy for being detected in the lightcurve of a former eclipsing binary, V699 Cygni\footnote{Shortly before the submission of this manuscript the discovery of a likely third-body eclipse in the former, but no-longer eclipsing binary V699 Cygni was also reported independently by \citet{zascheetal22}}.  Further investigations of these events are in progress.

Finally we discuss briefly the possible connection of our findings with the different formation scenarios of multiple systems. What is quite evident at first sight, is that the mass ratios are markedly different for the flat and the inclined systems, which may imply different formation scenarios. 

\section*{Acknowledgments}


TB acknowledges the support of the PRODEX Experiment Agreement No. 4000137122 between the ELTE E\"otv\"os Lor\'and University and the European Space Agency (ESA-D/SCI-LE-2021-0025), and the support of the city of Szombathely.

ST acknowledges support from the Netherlands Research Council NWO (VENI 639.041.645 and VIDI 203.061 grants).

The operation of the BRC80 robotic telescope of Baja Astronomical Observatory has been supported by the project ``Transient Astrophysical Objects'' GINOP 2.3.2-15-2016-00033 of the National Research, Development and Innovation Office (NKFIH), Hungary, funded by the European Union.


This paper includes data collected by the \textit{TESS} mission. Funding for the \textit{TESS} mission is provided by the NASA Science Mission directorate. Some of the data presented in this paper were obtained from the Mikulski Archive for Space Telescopes (MAST). STScI is operated by the Association of Universities for Research in Astronomy, Inc., under NASA contract NAS5-26555. Support for MAST for non-HST data is provided by the NASA Office of Space Science via grant NNX09AF08G and by other grants and contracts.



This work has made use  of data  from the European  Space Agency (ESA)  mission {\it Gaia}\footnote{\url{https://www.cosmos.esa.int/gaia}},  processed  by  the {\it   Gaia}   Data   Processing   and  Analysis   Consortium   (DPAC)\footnote{\url{https://www.cosmos.esa.int/web/gaia/dpac/consortium}}.  Funding for the DPAC  has been provided  by national  institutions, in  particular the institutions participating in the {\it Gaia} Multilateral Agreement.

This publication makes use of data products from the Wide-field Infrared Survey Explorer, which is a joint project of the University of California, Los Angeles, and the Jet Propulsion Laboratory/California Institute of Technology, funded by the National Aeronautics and Space Administration. 

This publication makes use of data products from the Two Micron All Sky Survey, which is a joint project of the University of Massachusetts and the Infrared Processing and Analysis Center/California Institute of Technology, funded by the National Aeronautics and Space Administration and the National Science Foundation.

We  used the  Simbad  service  operated by  the  Centre des  Donn\'ees Stellaires (Strasbourg,  France) and the ESO  Science Archive Facility services (data  obtained under request number 396301).   

This research made use of Lightkurve, a Python package for Kepler and TESS data analysis \citep{lightkurve}.

This research made use of Astropy,\footnote{\url{http://www.astropy.org}} a community-developed core Python package for Astronomy \citep{astropy:2013, astropy:2018}.

This research made use of astroquery: An Astronomical Web-querying Package in Python \citep{astroquery}.

\section*{Data availability}

The \textit{TESS} data underlying this article were accessed from MAST (Barbara A. Mikulski Archive for Space Telescopes) Portal (\url{https://mast.stsci.edu/portal/Mashup/Clients/Mast/Portal.html}). The ASAS-SN archival photometric data were accessed from \url{https://asas-sn.osu.edu/}. The ATLAS archival photometric data were accessed from \url{https://fallingstar-data.com/forcedphot/queue/}. A part of the data were derived from sources in public domain as given in the respective footnotes. The derived data generated in this research and the code used for the photodynamical analysis will be shared on reasonable request to the corresponding author.




\appendix


\section{Determination of the limits of the inclination domain where third-body eclipses might occur}
\label{app:inclinationlimits}

In order to find reasonable estimations for the inclination limits of the occurrence of third-body eclipses, we follow the same formalism as was used formerly in Appendix~A of \citet{borkovitsetal13}. There, we have shown that using the formalism of Jacobian position vectors, the spatial distances of the centers of the three stars can be calculated as 

\begin{eqnarray}
\vec{d}_\mathrm{AaAb} & =& \vec\rho_1 \\
\vec{d}_\mathrm{AaB}&=&\vec\rho_2+\frac{q_1}{1+q_1}\vec\rho_1 \\
\vec{d}_\mathrm{AbB}&=&\vec\rho_2-\frac{1}{1+q_1}\vec\rho_1,
\end{eqnarray}
respectively.\footnote{In order to save space in this Appendix, we hereafter use indices $_{1,2}$, instead of $_\mathrm{in,out}$ to distinguish parameters referring to the inner and outer orbits, respectively.} Let's use an observational frame of reference in which the origin is the center of mass of the inner binary, `X-Y' is the tangential plane of the sky, and the `Z' axis is directed away from the observer. Then the Cartesian coordinates of the first two Jacobians can be written as
\begin{eqnarray}
x&=&\rho[\cos u\cos\Omega-\sin u\sin\Omega\cos i], \\
y&=&\rho[\cos u\sin\Omega+\sin u\cos\Omega\cos i], \\
z&=&\rho\sin u\sin i,
\end{eqnarray}
where $u$ denotes the longitude of second/third body (to which the first/second Jacobian points) along the given orbit, measured from the ascending node. (In the eccentric case $u=v+\omega$, where $v$ stands for the true anomaly.) Then, the sky-projected distances can readily be obtained as
\begin{eqnarray}
d_\mathrm{AaAb}^{xy}&=&\rho_1\sqrt{1-\sin^2i_1\sin^2u_1}, \label{Eq:inner_projected_distance}\\
d_\mathrm{AbB}^{xy}&=&\rho_2\left[1-\sin^2i_2\sin^2u_2\right. \nonumber \\
&&-2\frac{1}{1-q_1}\left(\lambda-\sin i_1\sin u_1\sin i_2\sin u_2\right) \nonumber \\
&&\left.+\left(\frac{1}{1+q_1}\frac{\rho_2}{\rho_1}\right)^2\left(1-\sin^2i_1\sin^2u_1\right)\right]^{1/2},
\label{Eq:outer_projected_distance}
\end{eqnarray}
where $\lambda$ denotes the direction cosine between the two Jacobian vectors.  Note, the expression referring to distance $d_\mathrm{AaB}^{xy}$ is not shown, however, one can obtain it very easily by replacing $-1/(1+q_1)$ with $q_1/(1+q_1)$.)  It is clear that eclipses may occur when the sky-projected distances of any of the two bodies become less than the sum of their radii.  For regular two body eclipses then Eq.~(\ref{Eq:inner_projected_distance}) leads readily to the usual condition of $R_1+R_2\geq a|\cos i|$, which is, however, strictly valid only for circular orbits.  It also directly follows that at times of mid-eclipse, $u_1=\pm\pi/2$. In eccentric, not exactly edge-on cases, however, the smallest distance may occur somewhat earlier or later than the conjunction occurs. The departure is, however, in almost all cases, small, and therefore, negligible \citep{gimenezgarcia83}. Thus, when we calculated and utilized the condition given by Eq.~(\ref{Eq:innereclipsecond}) for the inner eclipses, we were working with the conjunction points (i.e., $u_1=\pm\pi/2$).

The case of third-body eclipses is much more complex, and no robust, exact condition(s) can be given.  Thus, we adopted the following assumptions. Again, one can expect third-body eclipses around the conjunctions of the outer orbit.  Hence, we arbitrarily set $u_2=\pm\pi/2$.  At these times, the projection of the second Jacobian vector onto the plane of the sky becomes 
\begin{equation}
\vec{\rho}_2^{\,xy}=\rho_2\cos i_2(\mp\sin\Omega_2;\pm\cos\Omega_2;0).
\end{equation}
Then, we look for that value of $u_1$, at which the sky-projected vector of the first Jacobian ($\vec\rho_1^{\,xy}$) is parallel to the former $\vec\rho_2^{\,xy}$, which may allow us to find the minimum sky-projected distance between the two bodies. The two vectors become parallel when their cross product yields zero. From this condition, and assuming that $\cos i_2\neq0$, one can easily calculate that
\begin{equation}
\sin^2u_1=\frac{\cos^2\Delta\Omega}{1-\sin^2i_1\sin^2\Delta\Omega}.
\end{equation}
Thus, one can readily find that the length of the sky-projected vector $\vec\rho_1^{\,xy}$ becomes
\begin{eqnarray}
\rho_1^{xy}&=&\rho_1\sqrt{1-\sin^2i_1\sin^2u_1} \nonumber \\
&=&\rho_1\sqrt{\frac{\cos^2i_1}{1-\sin^2i_1\sin^2\Delta\Omega}}.
\end{eqnarray}
Then, as this vector is parallel to the sky-projected vector of the second Jacobian, one can write directly that
\begin{eqnarray}
d_\mathrm{AbB}^{xy}&=&\rho_2|\cos i_2|\pm\frac{1}{1+q_1}\rho_1|\cos i_1|\sqrt{\frac{1}{1-\sin^2i_1\sin^2\Delta\Omega}} \nonumber \\
&=&\rho_2|\cos i_2|\pm\frac{1}{1+q_1}\rho_1\sqrt{\frac{\cos^2i_1}{\cos^2i_1+\sin^2i_1\cos^2\Delta\Omega}} \nonumber \\ 
&=&\rho_2|\cos i_2|\pm\frac{1}{1+q_1}\rho_1\sqrt{\frac{1}{1+\tan^2i_1\cos^2\Delta\Omega}}.
\end{eqnarray}
From this last form of the expression one can easily obtain Eq.~(\ref{Eq:outereclipsecond}), where we also made a further simplification, replacing $\rho_1$ with $a_1$ on the right hand side of the equation.

\section{Tables of determined times of minima for all the four systems}
\label{app:ToMs}

In this appendix, we tabulate the individual mid-minima times of the primary and secondary eclipses, including \textit{Kepler}, \textit{TESS}, and ground-based observed ones, for the inner EBs of the four triples considered in this study (Tables B1-B4).

The complete Appendix~\ref{app:ToMs} is available supplementary material to the journal, and also in the present arXiv version.

\begin{table*}
\caption{Binary Mid Eclipse Times for KIC~6964043}
 \label{tab:K69640431ToM}
{\tiny 
\begin{tabular}{@{}lrllrllrllrl}
\hline
BJD & Cycle  & std. dev. & BJD & Cycle  & std. dev. & BJD & Cycle  & std. dev. & BJD & Cycle  & std. dev. \\ 
$-2\,400\,000$ & no. &   \multicolumn{1}{c}{$(d)$} & $-2\,400\,000$ & no. &   \multicolumn{1}{c}{$(d)$} & $-2\,400\,000$ & no. &   \multicolumn{1}{c}{$(d)$} & $-2\,400\,000$ & no. &   \multicolumn{1}{c}{$(d)$} \\ 
\hline
55190.173588 &    0.0 & 0.000688 & 55495.729755 &   28.5 & 0.000505 & 55796.037501 &   56.5 & 0.000497 & 56112.482348 &   86.0 & 0.000672 \\
55195.460530 &    0.5 & 0.000538 & 55501.102486 &   29.0 & 0.000536 & 55801.369553 &   57.0 & 0.000446 & 56117.968120 &   86.5 & 0.001017 \\
55200.886326 &    1.0 & 0.000801 & 55506.447625 &   29.5 & 0.000968 & 55806.764662 &   57.5 & 0.000393 & 56133.913187 &   88.0 & 0.000587 \\
55206.172452 &    1.5 & 0.000961 & 55511.820059 &   30.0 & 0.000861 & 55812.101279 &   58.0 & 0.000409 & 56139.407798 &   88.5 & 0.000847 \\
55211.601150 &    2.0 & 0.000762 & 55517.160827 &   30.5 & 0.000711 & 55822.865967 &   59.0 & 0.001426 & 56144.627374 &   89.0 & 0.000677 \\
55222.317438 &    3.0 & 0.001062 & 55522.536941 &   31.0 & 0.000633 & 55828.367561 &   59.5 & 0.000570 & 56150.125700 &   89.5 & 0.000971 \\
55227.602391 &    3.5 & 0.000799 & 55527.877431 &   31.5 & 0.000478 & 55839.081236 &   60.5 & 0.000504 & 56155.340121 &   90.0 & 0.000732 \\
55238.318601 &    4.5 & 0.000860 & 55533.256203 &   32.0 & 0.000498 & 55844.365770 &   61.0 & 0.000421 & 56160.839199 &   90.5 & 0.000876 \\
55243.743685 &    5.0 & 0.000709 & 55538.593485 &   32.5 & 0.000475 & 55849.818469 &   61.5 & 0.000388 & 56166.054369 &   91.0 & 0.000773 \\
55249.038230 &    5.5 & 0.000625 & 55549.311845 &   33.5 & 0.000531 & 55855.108304 &   62.0 & 0.000468 & 56171.555981 &   91.5 & 0.000516 \\
55254.459530 &    6.0 & 0.001183 & 55570.764524 &   35.5 & 0.000438 & 55860.550658 &   62.5 & 0.000667 & 56176.771913 &   92.0 & 0.000721 \\
55259.747089 &    6.5 & 0.000986 & 55576.159985 &   36.0 & 0.000452 & 55876.550659 &   64.0 & 0.000525 & 56182.270350 &   92.5 & 0.000706 \\
55265.170662 &    7.0 & 0.000728 & 55581.552540 &   36.5 & 0.000568 & 55881.997269 &   64.5 & 0.000952 & 56187.482710 &   93.0 & 0.000766 \\
55270.466723 &    7.5 & 0.000901 & 55586.935079 &   37.0 & 0.000443 & 55887.267179 &   65.0 & 0.000511 & 56192.985929 &   93.5 & 0.000837 \\
55281.174261 &    8.5 & 0.001023 & 55592.366837 &   37.5 & 0.000638 & 55892.716930 &   65.5 & 0.000433 & 56198.196368 &   94.0 & 0.000747 \\
55286.606988 &    9.0 & 0.001019 & 55597.676834 &   38.0 & 0.000492 & 55897.980067 &   66.0 & 0.000485 & 56203.702472 &   94.5 & 0.001134 \\
55291.891233 &    9.5 & 0.001037 & 55603.078800 &   38.5 & 0.000459 & 55908.695388 &   67.0 & 0.000676 & 56208.911051 &   95.0 & 0.000690 \\
55297.325645 &   10.0 & 0.000912 & 55608.431983 &   39.0 & 0.000517 & 55914.146946 &   67.5 & 0.000578 & 56214.415334 &   95.5 & 0.001477 \\
55302.608118 &   10.5 & 0.000573 & 55613.819266 &   39.5 & 0.000514 & 55919.411396 &   68.0 & 0.000457 & 56219.627442 &   96.0 & 0.000992 \\
55313.329315 &   11.5 & 0.000951 & 55619.165008 &   40.0 & 0.000423 & 55924.865240 &   68.5 & 0.000612 & 56225.132630 &   96.5 & 0.000603 \\
55318.767221 &   12.0 & 0.000690 & 55624.552763 &   40.5 & 0.000641 & 55930.120478 &   69.0 & 0.000488 & 56230.344551 &   97.0 & 0.000501 \\
55324.047906 &   12.5 & 0.000922 & 55629.889603 &   41.0 & 0.000592 & 55935.578562 &   69.5 & 0.000839 & 56235.848362 &   97.5 & 0.000688 \\
55329.494681 &   13.0 & 0.000680 & 55635.278777 &   41.5 & 0.000496 & 55940.836280 &   70.0 & 0.000587 & 56241.060220 &   98.0 & 0.000805 \\
55334.787302 &   13.5 & 0.000661 & 55646.079263 &   42.5 & 0.001026 & 55946.298236 &   70.5 & 0.000539 & 56251.772460 &   99.0 & 0.001024 \\
55345.606170 &   14.5 & 0.000999 & 55651.325473 &   43.0 & 0.000501 & 55957.010179 &   71.5 & 0.000538 & 56257.279768 &   99.5 & 0.000939 \\
55351.013111 &   15.0 & 0.000645 & 55656.720776 &   43.5 & 0.000500 & 55962.264805 &   72.0 & 0.000495 & 56262.494709 &  100.0 & 0.000807 \\
55356.358888 &   15.5 & 0.000594 & 55662.040660 &   44.0 & 0.000405 & 55967.726462 &   72.5 & 0.000475 & 56267.995214 &  100.5 & 0.000896 \\
55361.760234 &   16.0 & 0.000642 & 55667.440537 &   44.5 & 0.000550 & 55972.979333 &   73.0 & 0.000497 & 56273.223127 &  101.0 & 0.000942 \\
55367.087933 &   16.5 & 0.000552 & 55672.754895 &   45.0 & 0.000611 & 55978.439783 &   73.5 & 0.000692 & 56278.718490 &  101.5 & 0.000656 \\
55372.506219 &   17.0 & 0.000588 & 55683.469003 &   46.0 & 0.000394 & 55983.693744 &   74.0 & 0.000532 & 56283.949424 &  102.0 & 0.000835 \\
55377.825983 &   17.5 & 0.000392 & 55688.871939 &   46.5 & 0.000706 & 55989.154588 &   74.5 & 0.000553 & 56289.446880 &  102.5 & 0.000905 \\
55383.234593 &   18.0 & 0.000584 & 55694.183544 &   47.0 & 0.000348 & 55999.871173 &   75.5 & 0.000480 & 56294.698965 &  103.0 & 0.002269 \\
55388.557970 &   18.5 & 0.000563 & 55699.587626 &   47.5 & 0.000510 & 56005.126585 &   76.0 & 0.000682 & 56300.244529 &  103.5 & 0.000617 \\
55393.953470 &   19.0 & 0.000537 & 55704.898973 &   48.0 & 0.000607 & 56010.587115 &   76.5 & 0.000490 & 56305.467755 &  104.0 & 0.001790 \\
55399.283659 &   19.5 & 0.000537 & 55710.302091 &   48.5 & 0.000752 & 56015.844256 &   77.0 & 0.000756 & 56321.764224 &  105.5 & 0.001067 \\
55404.672129 &   20.0 & 0.000553 & 55715.614288 &   49.0 & 0.000633 & 56021.303286 &   77.5 & 0.000600 & 56326.965378 &  106.0 & 0.001313 \\
55415.386026 &   21.0 & 0.000488 & 55721.022439 &   49.5 & 0.000926 & 56026.563994 &   78.0 & 0.000575 & 56337.699957 &  107.0 & 0.001383 \\
55420.722963 &   21.5 & 0.000613 & 55726.327539 &   50.0 & 0.000414 & 56032.020670 &   78.5 & 0.001174 & 56348.428551 &  108.0 & 0.001215 \\
55426.100080 &   22.0 & 0.000472 & 55731.734106 &   50.5 & 0.000528 & 56037.290799 &   79.0 & 0.000592 & 56353.951312 &  108.5 & 0.001430 \\
55436.813775 &   23.0 & 0.000491 & 55737.045969 &   51.0 & 0.000394 & 56042.742439 &   79.5 & 0.000545 & 56364.669588 &  109.5 & 0.000918 \\
55442.155904 &   23.5 & 0.000851 & 55742.449747 &   51.5 & 0.000561 & 56053.481243 &   80.5 & 0.000471 & 56375.390259 &  110.5 & 0.000913 \\
55447.527523 &   24.0 & 0.001023 & 55747.763620 &   52.0 & 0.000479 & 56058.772697 &   81.0 & 0.000639 & 56380.578115 &  111.0 & 0.000901 \\
55452.872094 &   24.5 & 0.000811 & 55753.167304 &   52.5 & 0.000346 & 56064.313580 &   81.5 & 0.000577 & 56386.110902 &  111.5 & 0.001041 \\
55458.244107 &   25.0 & 0.000598 & 55758.479819 &   53.0 & 0.000386 & 56069.534880 &   82.0 & 0.000663 & 56396.827476 &  112.5 & 0.001104 \\
55463.585359 &   25.5 & 0.000440 & 55763.884715 &   53.5 & 0.000520 & 56075.056447 &   82.5 & 0.000945 & 56402.001462 &  113.0 & 0.000929 \\
55468.956843 &   26.0 & 0.000574 & 55769.198047 &   54.0 & 0.000442 & 56080.283777 &   83.0 & 0.000710 & 56407.547455 &  113.5 & 0.003209 \\
55474.299841 &   26.5 & 0.000818 & 55774.600402 &   54.5 & 0.000437 & 56085.786810 &   83.5 & 0.000593 & 56412.719525 &  114.0 & 0.000981 \\
55479.670071 &   27.0 & 0.000719 & 55779.917753 &   55.0 & 0.000488 & 56091.030676 &   84.0 & 0.001283 & 56423.431712 &  115.0 & 0.001415 \\
55485.016731 &   27.5 & 0.000656 & 55785.315934 &   55.5 & 0.000453 & 56096.519692 &   84.5 & 0.000957 &&& \\ 
55490.387630 &   28.0 & 0.000607 & 55790.642026 &   56.0 & 0.000557 & 56107.246770 &   85.5 & 0.000691 &&& \\
\hline
\end{tabular}}

{\it Notes.} Integer and half-integer cycle numbers refer to primary and secondary eclipses, respectively. 
\end{table*}

\begin{table*}
\caption{Binary Mid Eclipse Times for KIC~5653126}
 \label{tab:K5653126ToM}
{\tiny
\begin{tabular}{@{}lrllrllrllrl}
\hline
BJD & Cycle  & std. dev. & BJD & Cycle  & std. dev. & BJD & Cycle  & std. dev. & BJD & Cycle  & std. dev. \\ 
$-2\,400\,000$ & no. &   \multicolumn{1}{c}{$(d)$} & $-2\,400\,000$ & no. &   \multicolumn{1}{c}{$(d)$} & $-2\,400\,000$ & no. &   \multicolumn{1}{c}{$(d)$} & $-2\,400\,000$ & no. &   \multicolumn{1}{c}{$(d)$} \\ 
\hline
54985.880280 &    0.0 & 0.001008 & 55601.819600 &   16.0 & 0.000606 & 55924.515186 &   24.5 & 0.000873 & 56232.501154 &   32.5 & 0.000407 \\
55024.378351 &    1.0 & 0.000919 & 55616.792529 &   16.5 & 0.001050 & 55948.315731 &   25.0 & 0.000611 & 56256.101135 &   33.0 & 0.000512 \\
55062.864010 &    2.0 & 0.000987 & 55655.301329 &   17.5 & 0.001073 & 55962.989074 &   25.5 & 0.000831 & 56270.996996 &   33.5 & 0.000535 \\
55101.342264 &    3.0 & 0.000757 & 55678.740626 &   18.0 & 0.000473 & 55986.817581 &   26.0 & 0.000585 & 56294.580181 &   34.0 & 0.000649 \\
55139.814286 &    4.0 & 0.000768 & 55693.781426 &   18.5 & 0.001091 & 56001.469564 &   26.5 & 0.000736 & 56309.469386 &   34.5 & 0.000546 \\
55178.279810 &    5.0 & 0.001009 & 55717.229270 &   19.0 & 0.000452 & 56025.309280 &   27.0 & 0.000494 & 56333.095304 &   35.0 & 0.000538 \\
55255.190783 &    7.0 & 0.000505 & 55732.241098 &   19.5 & 0.000946 & 56039.959497 &   27.5 & 0.000633 & 56347.910705 &   35.5 & 0.000421 \\
55293.653366 &    8.0 & 0.000773 & 55755.739637 &   20.0 & 0.000627 & 56063.793230 &   28.0 & 0.000680 & 56371.655943 &   36.0 & 0.000631 \\
55332.139298 &    9.0 & 0.000720 & 55794.256460 &   21.0 & 0.000628 & 56102.261335 &   29.0 & 0.001786 & 56386.331661 &   36.5 & 0.000380 \\
55370.669436 &   10.0 & 0.000685 & 55809.139507 &   21.5 & 0.000915 & 56116.966702 &   29.5 & 0.000533 & 56410.251600 &   37.0 & 0.000559 \\
55409.241435 &   11.0 & 0.000620 & 55832.775348 &   22.0 & 0.000647 & 56140.730905 &   30.0 & 0.000483 & 58695.507479 &   96.5 & 0.001907 \\
55447.836369 &   12.0 & 0.000839 & 55847.593106 &   22.5 & 0.000714 & 56155.479722 &   30.5 & 0.000489 & 58720.043332 &   97.0 & 0.000359 \\
55486.412344 &   13.0 & 0.000539 & 55871.292892 &   23.0 & 0.000296 & 56179.187227 &   31.0 & 0.000550 & 58733.938821 &   97.5 & 0.001338 \\
55524.926705 &   14.0 & 0.000669 & 55886.050315 &   23.5 & 0.001052 & 56193.992289 &   31.5 & 0.000461 & 59426.660574 &  115.5 & 0.000410 \\
55578.251428 &   15.5 & 0.001871 & 55909.807881 &   24.0 & 0.000502 & 56217.642274 &   32.0 & 0.000453 &&& \\
\hline
\end{tabular}}

{\it Notes.} Integer and half-integer cycle numbers refer to primary and secondary eclipses, respectively. Most of the eclipses (cycle nos. $0.0$ to $37.0$) were observed by the \textit{Kepler} spacecraft. The last 4 times of minima were determined from the \textit{TESS} observations.
\end{table*}

\begin{table*}
\caption{Binary Mid Eclipse Times for KIC~5731312}
 \label{tab:K5731312ToM}
{\tiny
\begin{tabular}{@{}lrllrllrllrl}
\hline
BJD & Cycle  & std. dev. & BJD & Cycle  & std. dev. & BJD & Cycle  & std. dev. & BJD & Cycle  & std. dev. \\ 
$-2\,400\,000$ & no. &   \multicolumn{1}{c}{$(d)$} & $-2\,400\,000$ & no. &   \multicolumn{1}{c}{$(d)$} & $-2\,400\,000$ & no. &   \multicolumn{1}{c}{$(d)$} & $-2\,400\,000$ & no. &   \multicolumn{1}{c}{$(d)$}\\ 
\hline
54966.175075 &   -0.5 & 0.000764 & 55333.628289 &   46.0 & 0.000015 & 55707.103506 &   93.0 & 0.000406 & 56088.531453 &  141.0 & 0.000024 \\
54968.092761 &    0.0 & 0.000010 & 55339.658250 &   46.5 & 0.000203 & 55713.140117 &   93.5 & 0.000163 & 56094.569256 &  141.5 & 0.000246 \\
54974.122050 &    0.5 & 0.000159 & 55341.574723 &   47.0 & 0.000009 & 55715.049151 &   94.0 & 0.000144 & 56096.477782 &  142.0 & 0.000032 \\
54976.039127 &    1.0 & 0.000011 & 55347.604496 &   47.5 & 0.000183 & 55721.088077 &   94.5 & 0.000125 & 56102.515246 &  142.5 & 0.000216 \\
54982.068903 &    1.5 & 0.000209 & 55349.521137 &   48.0 & 0.000015 & 55722.995113 &   95.0 & 0.000094 & 56104.424325 &  143.0 & 0.000277 \\
54983.985463 &    2.0 & 0.000010 & 55355.550880 &   48.5 & 0.000180 & 55729.035866 &   95.5 & 0.000165 & 56110.461418 &  143.5 & 0.000249 \\
54990.015264 &    2.5 & 0.000161 & 55357.467537 &   49.0 & 0.000008 & 55730.941608 &   96.0 & 0.000013 & 56112.370760 &  144.0 & 0.000270 \\
54991.931847 &    3.0 & 0.000013 & 55363.497299 &   49.5 & 0.000109 & 55736.983159 &   96.5 & 0.000165 & 56118.408837 &  144.5 & 0.000406 \\
54997.961705 &    3.5 & 0.000540 & 55365.413917 &   50.0 & 0.000023 & 55738.889104 &   97.0 & 0.000028 & 56120.317231 &  145.0 & 0.000025 \\
55005.908035 &    4.5 & 0.000126 & 55373.360400 &   51.0 & 0.000011 & 55744.929443 &   97.5 & 0.000188 & 56134.301656 &  146.5 & 0.000408 \\
55007.824646 &    5.0 & 0.000025 & 55379.389844 &   51.5 & 0.000141 & 55746.837036 &   98.0 & 0.000019 & 56136.209982 &  147.0 & 0.000026 \\
55013.854918 &    5.5 & 0.000273 & 55381.306792 &   52.0 & 0.000020 & 55752.873788 &   98.5 & 0.000345 & 56142.247624 &  147.5 & 0.000466 \\
55021.801620 &    6.5 & 0.007533 & 55387.336423 &   52.5 & 0.000239 & 55754.785199 &   99.0 & 0.000017 & 56144.156486 &  148.0 & 0.000273 \\
55023.717443 &    7.0 & 0.000044 & 55389.253239 &   53.0 & 0.000029 & 55760.818335 &   99.5 & 0.000179 & 56150.194516 &  148.5 & 0.000224 \\
55029.747824 &    7.5 & 0.000206 & 55395.282538 &   53.5 & 0.000187 & 55762.732813 &  100.0 & 0.000027 & 56152.103133 &  149.0 & 0.000033 \\
55031.663898 &    8.0 & 0.000009 & 55397.199631 &   54.0 & 0.000012 & 55768.762676 &  100.5 & 0.000294 & 56158.140284 &  149.5 & 0.000224 \\
55037.694218 &    8.5 & 0.000460 & 55403.228913 &   54.5 & 0.000162 & 55776.707519 &  101.5 & 0.000187 & 56160.049328 &  150.0 & 0.000033 \\
55039.610323 &    9.0 & 0.000009 & 55405.146081 &   55.0 & 0.000017 & 55778.625921 &  102.0 & 0.000036 & 56166.086922 &  150.5 & 0.000218 \\
55045.641197 &    9.5 & 0.000200 & 55411.175125 &   55.5 & 0.000177 & 55784.653615 &  102.5 & 0.000248 & 56167.995737 &  151.0 & 0.000016 \\
55047.556735 &   10.0 & 0.000119 & 55413.092493 &   56.0 & 0.000025 & 55786.571836 &  103.0 & 0.000022 & 56174.032976 &  151.5 & 0.000406 \\
55053.587571 &   10.5 & 0.000221 & 55419.121650 &   56.5 & 0.000155 & 55792.598831 &  103.5 & 0.000451 & 56175.942240 &  152.0 & 0.000015 \\
55055.503124 &   11.0 & 0.000457 & 55421.038916 &   57.0 & 0.000016 & 55794.517621 &  104.0 & 0.000195 & 56181.979363 &  152.5 & 0.000332 \\
55061.533981 &   11.5 & 0.000112 & 55427.067802 &   57.5 & 0.000216 & 55800.546393 &  104.5 & 0.000241 & 56183.888641 &  153.0 & 0.000024 \\
55069.480237 &   12.5 & 0.000243 & 55428.985335 &   58.0 & 0.000018 & 55808.493170 &  105.5 & 0.000315 & 56189.925969 &  153.5 & 0.000533 \\
55071.395971 &   13.0 & 0.000022 & 55435.014009 &   58.5 & 0.000425 & 55810.409207 &  106.0 & 0.000021 & 56191.835061 &  154.0 & 0.000049 \\
55077.426517 &   13.5 & 0.007966 & 55436.931662 &   59.0 & 0.000016 & 55816.439608 &  106.5 & 0.000206 & 56197.871736 &  154.5 & 0.000313 \\
55079.342482 &   14.0 & 0.000026 & 55442.960559 &   59.5 & 0.000350 & 55818.355067 &  107.0 & 0.000016 & 56199.781524 &  155.0 & 0.000036 \\
55085.373394 &   14.5 & 0.000164 & 55444.878146 &   60.0 & 0.000010 & 55824.385890 &  107.5 & 0.000171 & 56207.727951 &  156.0 & 0.000266 \\
55087.288850 &   15.0 & 0.000598 & 55450.906860 &   60.5 & 0.000187 & 55826.301094 &  108.0 & 0.000013 & 56213.764372 &  156.5 & 0.000184 \\
55093.319711 &   15.5 & 0.000218 & 55452.824531 &   61.0 & 0.000356 & 55832.333141 &  108.5 & 0.000383 & 56215.674350 &  157.0 & 0.000148 \\
55095.235304 &   16.0 & 0.000139 & 55458.853540 &   61.5 & 0.000541 & 55834.247096 &  109.0 & 0.000033 & 56221.710617 &  157.5 & 0.000579 \\
55101.266198 &   16.5 & 0.000256 & 55460.770949 &   62.0 & 0.000018 & 55840.280563 &  109.5 & 0.000318 & 56223.620809 &  158.0 & 0.000033 \\
55103.181739 &   17.0 & 0.000016 & 55466.799626 &   62.5 & 0.000264 & 55842.193305 &  110.0 & 0.000052 & 56229.657510 &  158.5 & 0.000228 \\
55109.212758 &   17.5 & 0.000274 & 55468.717395 &   63.0 & 0.000011 & 55848.226221 &  110.5 & 0.000337 & 56231.567421 &  159.0 & 0.000091 \\
55111.128177 &   18.0 & 0.000158 & 55474.746069 &   63.5 & 0.000251 & 55850.139471 &  111.0 & 0.000029 & 56239.513593 &  160.0 & 0.000022 \\
55117.159292 &   18.5 & 0.000216 & 55476.663736 &   64.0 & 0.000019 & 55856.174065 &  111.5 & 0.000287 & 56245.550641 &  160.5 & 0.000309 \\
55119.074613 &   19.0 & 0.000016 & 55482.692261 &   64.5 & 0.000151 & 55858.085681 &  112.0 & 0.000254 & 56253.496328 &  161.5 & 0.000612 \\
55125.105809 &   19.5 & 0.000175 & 55484.610170 &   65.0 & 0.000011 & 55864.120557 &  112.5 & 0.000205 & 56255.406420 &  162.0 & 0.000018 \\
55127.021056 &   20.0 & 0.000090 & 55490.638475 &   65.5 & 0.000177 & 55872.067321 &  113.5 & 0.000161 & 56261.443196 &  162.5 & 0.000342 \\
55133.052014 &   20.5 & 0.000150 & 55492.556593 &   66.0 & 0.000023 & 55873.978294 &  114.0 & 0.000024 & 56263.352962 &  163.0 & 0.000023 \\
55134.967491 &   21.0 & 0.000019 & 55498.584800 &   66.5 & 0.000107 & 55880.014242 &  114.5 & 0.000188 & 56269.389212 &  163.5 & 0.000559 \\
55140.998089 &   21.5 & 0.000122 & 55500.503001 &   67.0 & 0.000015 & 55881.924550 &  115.0 & 0.000020 & 56271.299378 &  164.0 & 0.000031 \\
55142.913891 &   22.0 & 0.000010 & 55506.531027 &   67.5 & 0.000187 & 55887.959977 &  115.5 & 0.000384 & 56277.334994 &  164.5 & 0.000493 \\
55148.944450 &   22.5 & 0.000218 & 55508.449363 &   68.0 & 0.000016 & 55889.870913 &  116.0 & 0.000015 & 56279.245843 &  165.0 & 0.000021 \\
55150.860349 &   23.0 & 0.000014 & 55514.477090 &   68.5 & 0.000184 & 55895.906594 &  116.5 & 0.000428 & 56285.282095 &  165.5 & 0.000309 \\
55156.891079 &   23.5 & 0.000476 & 55516.395763 &   69.0 & 0.000059 & 55897.817329 &  117.0 & 0.000032 & 56287.192268 &  166.0 & 0.000043 \\
55158.806794 &   24.0 & 0.000073 & 55522.423206 &   69.5 & 0.000366 & 55905.763674 &  118.0 & 0.000019 & 56295.138601 &  167.0 & 0.000046 \\
55164.837494 &   24.5 & 0.000169 & 55524.342128 &   70.0 & 0.000021 & 55911.800220 &  118.5 & 0.000484 & 56301.174530 &  167.5 & 0.000198 \\
55166.753217 &   25.0 & 0.000650 & 55530.370030 &   70.5 & 0.000226 & 55913.710071 &  119.0 & 0.000023 & 56303.085053 &  168.0 & 0.000357 \\
55172.784071 &   25.5 & 0.000172 & 55532.288492 &   71.0 & 0.000068 & 55919.746814 &  119.5 & 0.000254 & 56309.120862 &  168.5 & 0.000331 \\
55174.699652 &   26.0 & 0.000017 & 55538.316200 &   71.5 & 0.001541 & 55921.656462 &  120.0 & 0.000069 & 56325.014860 &  170.5 & 0.000215 \\
55180.730393 &   26.5 & 0.000828 & 55540.234957 &   72.0 & 0.000013 & 55927.693120 &  120.5 & 0.000269 & 56326.924190 &  171.0 & 0.000031 \\
55188.677403 &   27.5 & 0.001428 & 55546.262841 &   72.5 & 0.000237 & 55929.602879 &  121.0 & 0.000985 & 56332.959752 &  171.5 & 0.000869 \\
55190.592567 &   28.0 & 0.000014 & 55548.181352 &   73.0 & 0.000019 & 55935.639834 &  121.5 & 0.000384 & 56334.870731 &  172.0 & 0.000024 \\
55196.623579 &   28.5 & 0.000365 & 55570.101220 &   75.5 & 0.000264 & 55937.549249 &  122.0 & 0.000019 & 56340.906265 &  172.5 & 0.000344 \\
55198.539015 &   29.0 & 0.000013 & 55572.020390 &   76.0 & 0.000105 & 55943.586154 &  122.5 & 0.000424 & 56342.817103 &  173.0 & 0.000498 \\
55204.569878 &   29.5 & 0.000672 & 55578.047381 &   76.5 & 0.000217 & 55945.495649 &  123.0 & 0.000015 & 56348.852519 &  173.5 & 0.000405 \\
55206.485398 &   30.0 & 0.000012 & 55579.966819 &   77.0 & 0.000012 & 55953.442100 &  124.0 & 0.000063 & 56350.763564 &  174.0 & 0.000030 \\
55212.516022 &   30.5 & 0.000114 & 55585.993701 &   77.5 & 0.000803 & 55961.388436 &  125.0 & 0.000249 & 56356.799018 &  174.5 & 0.000299 \\
55214.431837 &   31.0 & 0.000500 & 55587.913096 &   78.0 & 0.000016 & 55967.426352 &  125.5 & 0.000344 & 56364.745086 &  175.5 & 0.000334 \\
55220.462171 &   31.5 & 0.000183 & 55593.939728 &   78.5 & 0.000169 & 55969.334845 &  126.0 & 0.000017 & 56366.656424 &  176.0 & 0.000300 \\
55222.378315 &   32.0 & 0.000011 & 55601.886558 &   79.5 & 0.005672 & 55975.371437 &  126.5 & 0.000368 & 56372.691072 &  176.5 & 0.000629 \\
55228.408542 &   32.5 & 0.000220 & 55603.805734 &   80.0 & 0.000009 & 55977.281365 &  127.0 & 0.000014 & 56374.602784 &  177.0 & 0.000011 \\
55236.355244 &   33.5 & 0.000116 & 55609.832308 &   80.5 & 0.000192 & 55983.318477 &  127.5 & 0.000854 & 56380.637708 &  177.5 & 0.000240 \\
55238.271157 &   34.0 & 0.000304 & 55611.752134 &   81.0 & 0.000537 & 55985.227795 &  128.0 & 0.000018 & 56382.549227 &  178.0 & 0.000023 \\
55244.301614 &   34.5 & 0.000138 & 55617.778659 &   81.5 & 0.000465 & 55991.265481 &  128.5 & 0.000314 & 56388.584601 &  178.5 & 0.000455 \\
55246.217527 &   35.0 & 0.000124 & 55619.698344 &   82.0 & 0.000035 & 55993.174277 &  129.0 & 0.000026 & 56390.495616 &  179.0 & 0.000040 \\
55252.247660 &   35.5 & 0.000150 & 55625.725315 &   82.5 & 0.009792 & 55999.211256 &  129.5 & 0.000419 & 56398.441961 &  180.0 & 0.000020 \\
55254.163981 &   36.0 & 0.000011 & 55627.644583 &   83.0 & 0.000020 & 56001.120707 &  130.0 & 0.000023 & 56404.476857 &  180.5 & 0.000275 \\
55260.194203 &   36.5 & 0.000210 & 55633.670959 &   83.5 & 0.000139 & 56007.158095 &  130.5 & 0.000242 & 56406.388372 &  181.0 & 0.000014 \\
55262.110499 &   37.0 & 0.000018 & 55641.617754 &   84.5 & 0.000482 & 56009.067095 &  131.0 & 0.000079 & 56412.423649 &  181.5 & 0.000237 \\
55268.140794 &   37.5 & 0.009811 & 55643.536983 &   85.0 & 0.000017 & 56017.013518 &  132.0 & 0.000013 & 56414.334813 &  182.0 & 0.000015 \\
55270.056859 &   38.0 & 0.000016 & 55649.564014 &   85.5 & 0.015367 & 56023.050882 &  132.5 & 0.000179 & 56420.369227 &  182.5 & 0.000304 \\
55278.003269 &   39.0 & 0.000021 & 55651.483076 &   86.0 & 0.000017 & 56024.959968 &  133.0 & 0.000047 & 56422.281258 &  183.0 & 0.000030 \\
55284.033439 &   39.5 & 0.000115 & 55657.509993 &   86.5 & 0.000128 & 56030.997745 &  133.5 & 0.000285 & 58686.989921 &  468.0 & 0.001216 \\
55285.949724 &   40.0 & 0.000012 & 55659.429124 &   87.0 & 0.000014 & 56032.906246 &  134.0 & 0.000048 & 58694.938535 &  469.0 & 0.000897 \\
55291.980354 &   40.5 & 0.000152 & 55665.456850 &   87.5 & 0.000173 & 56040.852813 &  135.0 & 0.000022 & 58702.883844 &  470.0 & 0.002345 \\
55293.896151 &   41.0 & 0.000020 & 55667.375141 &   88.0 & 0.000587 & 56046.890291 &  135.5 & 0.000307 & 58718.776923 &  472.0 & 0.001153 \\
55299.926187 &   41.5 & 0.000130 & 55673.403406 &   88.5 & 0.000450 & 56048.799264 &  136.0 & 0.000012 & 58726.724369 &  473.0 & 0.000759 \\
55301.842577 &   42.0 & 0.000691 & 55675.321082 &   89.0 & 0.000010 & 56054.837060 &  136.5 & 0.000628 & 58734.669429 &  474.0 & 0.002943 \\
55307.872191 &   42.5 & 0.000228 & 55681.349721 &   89.5 & 0.000122 & 56056.745742 &  137.0 & 0.000049 & 58782.349083 &  480.0 & 0.000235 \\
55309.789000 &   43.0 & 0.000017 & 55683.266775 &   90.0 & 0.000063 & 56062.783531 &  137.5 & 0.000344 & 59346.542549 &  551.0 & 0.000103 \\
55315.819174 &   43.5 & 0.000179 & 55689.297065 &   90.5 & 0.000331 & 56064.692037 &  138.0 & 0.000017 & 59426.006304 &  561.0 & 0.000680 \\
55317.735426 &   44.0 & 0.000017 & 55691.212500 &   91.0 & 0.000019 & 56070.729960 &  138.5 & 0.000397 & 59433.954788 &  562.0 & 0.001706 \\
55323.765540 &   44.5 & 0.001238 & 55697.244017 &   91.5 & 0.000159 & 56072.638598 &  139.0 & 0.000040 & 59441.901713 &  563.0 & 0.001086 \\
55325.681840 &   45.0 & 0.000020 & 55699.158067 &   92.0 & 0.000011 & 56080.585082 &  140.0 & 0.000270 &&&\\
55331.711144 &   45.5 & 0.001815 & 55705.191634 &   92.5 & 0.000261 & 56086.622750 &  140.5 & 0.000377 &&&\\
\hline
\end{tabular}}

{\it Notes.} Integer and half-integer cycle numbers refer to primary and secondary eclipses, respectively. Most of the eclipses (cycle nos. $-0.5$ to $183.0$) were observed by the \textit{Kepler} spacecraft. Nine of the last eleven eclipses were observed by \textit{TESS} spacecraft, while the remaining two events at cycle nos. $480.0$ and $551.0$ were measured in the frame of our photometric follow-up campaign ate Baja Astronomical Observatory.
\end{table*}

\begin{table*}
\caption{Binary Mid Eclipse Times for KIC~8023317}
 \label{tab:K8023317ToM}
{\tiny
\begin{tabular}{@{}lrllrllrllrl}
\hline
BJD & Cycle  & std. dev. & BJD & Cycle  & std. dev. & BJD & Cycle  & std. dev. & BJD & Cycle  & std. dev. \\ 
$-2\,400\,000$ & no. &   \multicolumn{1}{c}{$(d)$} & $-2\,400\,000$ & no. &   \multicolumn{1}{c}{$(d)$} & $-2\,400\,000$ & no. &   \multicolumn{1}{c}{$(d)$} & $-2\,400\,000$ & no. &   \multicolumn{1}{c}{$(d)$} \\ 
\hline
54957.481039 &   -1.5 & 0.003882 & 55322.222343 &   20.5 & 0.001913 & 55703.530522 &   43.5 & 0.003163 & 56068.272800 &   65.5 & 0.002545 \\
54963.158283 &   -1.0 & 0.002084 & 55327.891045 &   21.0 & 0.000374 & 55709.208576 &   44.0 & 0.000152 & 56073.945577 &   66.0 & 0.000335 \\
54979.735521 &    0.0 & 0.000667 & 55338.803137 &   21.5 & 0.003081 & 55720.109439 &   44.5 & 0.005567 & 56084.849998 &   66.5 & 0.002950 \\
54990.646340 &    0.5 & 0.002306 & 55344.469324 &   22.0 & 0.000296 & 55725.789237 &   45.0 & 0.000153 & 56090.525492 &   67.0 & 0.000403 \\
54996.312001 &    1.0 & 0.000751 & 55355.382042 &   22.5 & 0.004062 & 55736.691345 &   45.5 & 0.003046 & 56118.010828 &   68.5 & 0.003700 \\
55007.226717 &    1.5 & 0.002444 & 55361.047751 &   23.0 & 0.000435 & 55742.369592 &   46.0 & 0.000195 & 56134.584409 &   69.5 & 0.002488 \\
55012.889006 &    2.0 & 0.000632 & 55377.626530 &   24.0 & 0.000303 & 55753.273202 &   46.5 & 0.003360 & 56140.265744 &   70.0 & 0.000353 \\
55023.805337 &    2.5 & 0.002539 & 55388.537765 &   24.5 & 0.004921 & 55758.949237 &   47.0 & 0.000192 & 56151.162291 &   70.5 & 0.002339 \\
55029.465882 &    3.0 & 0.000580 & 55394.205565 &   25.0 & 0.000313 & 55775.528335 &   48.0 & 0.000181 & 56156.845581 &   71.0 & 0.000440 \\
55040.386336 &    3.5 & 0.002063 & 55405.115258 &   25.5 & 0.004043 & 55786.428708 &   48.5 & 0.005962 & 56167.740271 &   71.5 & 0.020530 \\
55046.044017 &    4.0 & 0.000559 & 55410.784624 &   26.0 & 0.000313 & 55792.107196 &   49.0 & 0.000158 & 56173.424274 &   72.0 & 0.000405 \\
55056.961520 &    4.5 & 0.005357 & 55421.695571 &   26.5 & 0.001658 & 55808.685892 &   50.0 & 0.000161 & 56184.320573 &   72.5 & 0.003389 \\
55062.624482 &    5.0 & 0.000456 & 55427.364175 &   27.0 & 0.000282 & 55819.585934 &   50.5 & 0.004095 & 56190.002597 &   73.0 & 0.000444 \\
55073.535058 &    5.5 & 0.001753 & 55438.275599 &   27.5 & 0.002804 & 55825.264396 &   51.0 & 0.000192 & 56200.921955 &   73.5 & 0.026363 \\
55079.202423 &    6.0 & 0.000258 & 55443.943368 &   28.0 & 0.000278 & 55836.168155 &   51.5 & 0.003206 & 56206.581385 &   74.0 & 0.001063 \\
55090.113963 &    6.5 & 0.003792 & 55454.848437 &   28.5 & 0.002016 & 55841.842568 &   52.0 & 0.000205 & 56217.488623 &   74.5 & 0.004251 \\
55095.786785 &    7.0 & 0.000426 & 55460.523121 &   29.0 & 0.000260 & 55852.743217 &   52.5 & 0.002779 & 56223.155198 &   75.0 & 0.000425 \\
55106.689719 &    7.5 & 0.002709 & 55471.431640 &   29.5 & 0.002633 & 55858.421038 &   53.0 & 0.000161 & 56234.063821 &   75.5 & 0.014746 \\
55112.367385 &    8.0 & 0.000534 & 55477.103342 &   30.0 & 0.000330 & 55869.325089 &   53.5 & 0.003970 & 56239.731504 &   76.0 & 0.000712 \\
55123.266507 &    8.5 & 0.002415 & 55488.008317 &   30.5 & 0.001156 & 55874.999472 &   54.0 & 0.000227 & 56256.309637 &   77.0 & 0.000735 \\
55128.947101 &    9.0 & 0.000418 & 55504.588103 &   31.5 & 0.002551 & 55885.904695 &   54.5 & 0.002017 & 56267.221811 &   77.5 & 0.014309 \\
55139.849036 &    9.5 & 0.002429 & 55510.263108 &   32.0 & 0.000239 & 55891.577273 &   55.0 & 0.000210 & 56272.888721 &   78.0 & 0.000606 \\
55145.527153 &   10.0 & 0.000438 & 55521.163567 &   32.5 & 0.002804 & 55902.488453 &   55.5 & 0.004037 & 56283.789721 &   78.5 & 0.002393 \\
55162.106056 &   11.0 & 0.000355 & 55526.843208 &   33.0 & 0.000242 & 55908.155799 &   56.0 & 0.000245 & 56289.469261 &   79.0 & 0.000913 \\
55173.013238 &   11.5 & 0.002594 & 55537.741007 &   33.5 & 0.002578 & 55919.065074 &   56.5 & 0.002868 & 56306.052051 &   80.0 & 0.001044 \\
55178.684850 &   12.0 & 0.000413 & 55543.423250 &   34.0 & 0.000308 & 55924.734286 &   57.0 & 0.000199 & 56322.631580 &   81.0 & 0.000807 \\
55189.595226 &   12.5 & 0.004625 & 55570.904480 &   35.5 & 0.005396 & 55941.312915 &   58.0 & 0.000181 & 56339.212519 &   82.0 & 0.000724 \\
55195.263418 &   13.0 & 0.000438 & 55576.580449 &   36.0 & 0.000265 & 55952.225777 &   58.5 & 0.006010 & 56355.792119 &   83.0 & 0.000654 \\
55211.842421 &   14.0 & 0.000383 & 55587.485614 &   36.5 & 0.002632 & 55957.891016 &   59.0 & 0.000343 & 56366.682411 &   83.5 & 0.004419 \\
55222.745060 &   14.5 & 0.003732 & 55593.157418 &   37.0 & 0.000287 & 55968.802452 &   59.5 & 0.003377 & 56372.371163 &   84.0 & 0.000999 \\
55228.420428 &   15.0 & 0.000442 & 55604.062571 &   37.5 & 0.001942 & 55974.470044 &   60.0 & 0.000232 & 56388.950337 &   85.0 & 0.000752 \\
55239.330302 &   15.5 & 0.002764 & 55609.734119 &   38.0 & 0.000215 & 55985.380551 &   60.5 & 0.005272 & 56399.846617 &   85.5 & 0.003054 \\
55244.998533 &   16.0 & 0.000331 & 55620.648239 &   38.5 & 0.003179 & 55991.049106 &   61.0 & 0.000317 & 56405.528614 &   86.0 & 0.000884 \\
55255.907944 &   16.5 & 0.004017 & 55626.309496 &   39.0 & 0.000181 & 56001.957595 &   61.5 & 0.003586 & 56422.108527 &   87.0 & 0.000893 \\
55261.577101 &   17.0 & 0.000433 & 55642.887215 &   40.0 & 0.000154 & 56007.627930 &   62.0 & 0.000315 & 58693.430893 &  224.0 & 0.004442 \\
55272.483288 &   17.5 & 0.004255 & 55653.801692 &   40.5 & 0.003299 & 56018.535295 &   62.5 & 0.005302 & 58710.011053 &  225.0 & 0.013546 \\
55278.155506 &   18.0 & 0.000407 & 55659.466322 &   41.0 & 0.000151 & 56024.206950 &   63.0 & 0.000303 & 59406.320603 &  267.0 & 0.011177 \\
55289.067358 &   18.5 & 0.002984 & 55670.377522 &   41.5 & 0.006269 & 56035.117721 &   63.5 & 0.003322 & 59422.914616 &  268.0 & 0.013971 \\
55294.733814 &   19.0 & 0.000283 & 55676.046305 &   42.0 & 0.000144 & 56040.786267 &   64.0 & 0.000426 & 59439.478298 &  269.0 & 0.015880 \\
55305.643953 &   19.5 & 0.003086 & 55686.954996 &   42.5 & 0.002951 & 56051.692989 &   64.5 & 0.008381 &&&\\
55311.312393 &   20.0 & 0.000428 & 55692.627680 &   43.0 & 0.000148 & 56057.366071 &   65.0 & 0.000409 &&&\\
\hline
\end{tabular}}

{\it Notes.} Integer and half-integer cycle numbers refer to primary and secondary eclipses, respectively. Most of the eclipses (cycle nos. $-1.5$ to $87.0$) were observed by the \textit{Kepler} spacecraft. The last 5 times of minima were determined from the \textit{TESS} observations.
\end{table*}

--------------------------------------------------------------------------------------------------------------------------------------------------

\begin{table*}
\caption{Binary Mid Eclipse Times for KIC~6964043}
 \label{tab:K69640431ToM}
{\tiny 
\begin{tabular}{@{}lrllrllrl}
\hline
BJD & Cycle  & std. dev. & BJD & Cycle  & std. dev. & BJD & Cycle  & std. dev. \\ 
$-2\,400\,000$ & no. &   \multicolumn{1}{c}{$(d)$} & $-2\,400\,000$ & no. &   \multicolumn{1}{c}{$(d)$} & $-2\,400\,000$ & no. &   \multicolumn{1}{c}{$(d)$} \\ 
\hline
55190.173588 &    0.0 & 0.000688 & 55603.078800 &   38.5 & 0.000459 & 56005.126585 &   76.0 & 0.000682 \\ 
55195.460530 &    0.5 & 0.000538 & 55608.431983 &   39.0 & 0.000517 & 56010.587115 &   76.5 & 0.000490 \\ 
55200.886326 &    1.0 & 0.000801 & 55613.819266 &   39.5 & 0.000514 & 56015.844256 &   77.0 & 0.000756 \\ 
55206.172452 &    1.5 & 0.000961 & 55619.165008 &   40.0 & 0.000423 & 56021.303286 &   77.5 & 0.000600 \\ 
55211.601150 &    2.0 & 0.000762 & 55624.552763 &   40.5 & 0.000641 & 56026.563994 &   78.0 & 0.000575 \\ 
55222.317438 &    3.0 & 0.001062 & 55629.889603 &   41.0 & 0.000592 & 56032.020670 &   78.5 & 0.001174 \\ 
55227.602391 &    3.5 & 0.000799 & 55635.278777 &   41.5 & 0.000496 & 56037.290799 &   79.0 & 0.000592 \\ 
55238.318601 &    4.5 & 0.000860 & 55646.079263 &   42.5 & 0.001026 & 56042.742439 &   79.5 & 0.000545 \\ 
55243.743685 &    5.0 & 0.000709 & 55651.325473 &   43.0 & 0.000501 & 56053.481243 &   80.5 & 0.000471 \\ 
55249.038230 &    5.5 & 0.000625 & 55656.720776 &   43.5 & 0.000500 & 56058.772697 &   81.0 & 0.000639 \\ 
55254.459530 &    6.0 & 0.001183 & 55662.040660 &   44.0 & 0.000405 & 56064.313580 &   81.5 & 0.000577 \\ 
55259.747089 &    6.5 & 0.000986 & 55667.440537 &   44.5 & 0.000550 & 56069.534880 &   82.0 & 0.000663 \\ 
55265.170662 &    7.0 & 0.000728 & 55672.754895 &   45.0 & 0.000611 & 56075.056447 &   82.5 & 0.000945 \\ 
55270.466723 &    7.5 & 0.000901 & 55683.469003 &   46.0 & 0.000394 & 56080.283777 &   83.0 & 0.000710 \\ 
55281.174261 &    8.5 & 0.001023 & 55688.871939 &   46.5 & 0.000706 & 56085.786810 &   83.5 & 0.000593 \\ 
55286.606988 &    9.0 & 0.001019 & 55694.183544 &   47.0 & 0.000348 & 56091.030676 &   84.0 & 0.001283 \\ 
55291.891233 &    9.5 & 0.001037 & 55699.587626 &   47.5 & 0.000510 & 56096.519692 &   84.5 & 0.000957 \\ 
55297.325645 &   10.0 & 0.000912 & 55704.898973 &   48.0 & 0.000607 & 56107.246770 &   85.5 & 0.000691 \\ 
55302.608118 &   10.5 & 0.000573 & 55710.302091 &   48.5 & 0.000752 & 56112.482348 &   86.0 & 0.000672 \\ 
55313.329315 &   11.5 & 0.000951 & 55715.614288 &   49.0 & 0.000633 & 56117.968120 &   86.5 & 0.001017 \\ 
55318.767221 &   12.0 & 0.000690 & 55721.022439 &   49.5 & 0.000926 & 56133.913187 &   88.0 & 0.000587 \\ 
55324.047906 &   12.5 & 0.000922 & 55726.327539 &   50.0 & 0.000414 & 56139.407798 &   88.5 & 0.000847 \\ 
55329.494681 &   13.0 & 0.000680 & 55731.734106 &   50.5 & 0.000528 & 56144.627374 &   89.0 & 0.000677 \\ 
55334.787302 &   13.5 & 0.000661 & 55737.045969 &   51.0 & 0.000394 & 56150.125700 &   89.5 & 0.000971 \\ 
55345.606170 &   14.5 & 0.000999 & 55742.449747 &   51.5 & 0.000561 & 56155.340121 &   90.0 & 0.000732 \\ 
55351.013111 &   15.0 & 0.000645 & 55747.763620 &   52.0 & 0.000479 & 56160.839199 &   90.5 & 0.000876 \\ 
55356.358888 &   15.5 & 0.000594 & 55753.167304 &   52.5 & 0.000346 & 56166.054369 &   91.0 & 0.000773 \\ 
55361.760234 &   16.0 & 0.000642 & 55758.479819 &   53.0 & 0.000386 & 56171.555981 &   91.5 & 0.000516 \\ 
55367.087933 &   16.5 & 0.000552 & 55763.884715 &   53.5 & 0.000520 & 56176.771913 &   92.0 & 0.000721 \\ 
55372.506219 &   17.0 & 0.000588 & 55769.198047 &   54.0 & 0.000442 & 56182.270350 &   92.5 & 0.000706 \\ 
55377.825983 &   17.5 & 0.000392 & 55774.600402 &   54.5 & 0.000437 & 56187.482710 &   93.0 & 0.000766 \\ 
55383.234593 &   18.0 & 0.000584 & 55779.917753 &   55.0 & 0.000488 & 56192.985929 &   93.5 & 0.000837 \\ 
55388.557970 &   18.5 & 0.000563 & 55785.315934 &   55.5 & 0.000453 & 56198.196368 &   94.0 & 0.000747 \\ 
55393.953470 &   19.0 & 0.000537 & 55790.642026 &   56.0 & 0.000557 & 56203.702472 &   94.5 & 0.001134 \\ 
55399.283659 &   19.5 & 0.000537 & 55796.037501 &   56.5 & 0.000497 & 56208.911051 &   95.0 & 0.000690 \\ 
55404.672129 &   20.0 & 0.000553 & 55801.369553 &   57.0 & 0.000446 & 56214.415334 &   95.5 & 0.001477 \\ 
55415.386026 &   21.0 & 0.000488 & 55806.764662 &   57.5 & 0.000393 & 56219.627442 &   96.0 & 0.000992 \\ 
55420.722963 &   21.5 & 0.000613 & 55812.101279 &   58.0 & 0.000409 & 56225.132630 &   96.5 & 0.000603 \\ 
55426.100080 &   22.0 & 0.000472 & 55822.865967 &   59.0 & 0.001426 & 56230.344551 &   97.0 & 0.000501 \\ 
55436.813775 &   23.0 & 0.000491 & 55828.367561 &   59.5 & 0.000570 & 56235.848362 &   97.5 & 0.000688 \\ 
55442.155904 &   23.5 & 0.000851 & 55839.081236 &   60.5 & 0.000504 & 56241.060220 &   98.0 & 0.000805 \\ 
55447.527523 &   24.0 & 0.001023 & 55844.365770 &   61.0 & 0.000421 & 56251.772460 &   99.0 & 0.001024 \\ 
55452.872094 &   24.5 & 0.000811 & 55849.818469 &   61.5 & 0.000388 & 56257.279768 &   99.5 & 0.000939 \\ 
55458.244107 &   25.0 & 0.000598 & 55855.108304 &   62.0 & 0.000468 & 56262.494709 &  100.0 & 0.000807 \\ 
55463.585359 &   25.5 & 0.000440 & 55860.550658 &   62.5 & 0.000667 & 56267.995214 &  100.5 & 0.000896 \\ 
55468.956843 &   26.0 & 0.000574 & 55876.550659 &   64.0 & 0.000525 & 56273.223127 &  101.0 & 0.000942 \\ 
55474.299841 &   26.5 & 0.000818 & 55881.997269 &   64.5 & 0.000952 & 56278.718490 &  101.5 & 0.000656 \\ 
55479.670071 &   27.0 & 0.000719 & 55887.267179 &   65.0 & 0.000511 & 56283.949424 &  102.0 & 0.000835 \\ 
55485.016731 &   27.5 & 0.000656 & 55892.716930 &   65.5 & 0.000433 & 56289.446880 &  102.5 & 0.000905 \\ 
55490.387630 &   28.0 & 0.000607 & 55897.980067 &   66.0 & 0.000485 & 56294.698965 &  103.0 & 0.002269 \\ 
55495.729755 &   28.5 & 0.000505 & 55908.695388 &   67.0 & 0.000676 & 56300.244529 &  103.5 & 0.000617 \\ 
55501.102486 &   29.0 & 0.000536 & 55914.146946 &   67.5 & 0.000578 & 56305.467755 &  104.0 & 0.001790 \\ 
55506.447625 &   29.5 & 0.000968 & 55919.411396 &   68.0 & 0.000457 & 56321.764224 &  105.5 & 0.001067 \\ 
55511.820059 &   30.0 & 0.000861 & 55924.865240 &   68.5 & 0.000612 & 56326.965378 &  106.0 & 0.001313 \\ 
55517.160827 &   30.5 & 0.000711 & 55930.120478 &   69.0 & 0.000488 & 56337.699957 &  107.0 & 0.001383 \\ 
55522.536941 &   31.0 & 0.000633 & 55935.578562 &   69.5 & 0.000839 & 56348.428551 &  108.0 & 0.001215 \\ 
55527.877431 &   31.5 & 0.000478 & 55940.836280 &   70.0 & 0.000587 & 56353.951312 &  108.5 & 0.001430 \\
55533.256203 &   32.0 & 0.000498 & 55946.298236 &   70.5 & 0.000539 & 56364.669588 &  109.5 & 0.000918 \\ 
55538.593485 &   32.5 & 0.000475 & 55957.010179 &   71.5 & 0.000538 & 56375.390259 &  110.5 & 0.000913 \\ 
55549.311845 &   33.5 & 0.000531 & 55962.264805 &   72.0 & 0.000495 & 56380.578115 &  111.0 & 0.000901 \\ 
55570.764524 &   35.5 & 0.000438 & 55967.726462 &   72.5 & 0.000475 & 56386.110902 &  111.5 & 0.001041 \\ 
55576.159985 &   36.0 & 0.000452 & 55972.979333 &   73.0 & 0.000497 & 56396.827476 &  112.5 & 0.001104 \\ 
55581.552540 &   36.5 & 0.000568 & 55978.439783 &   73.5 & 0.000692 & 56402.001462 &  113.0 & 0.000929 \\ 
55586.935079 &   37.0 & 0.000443 & 55983.693744 &   74.0 & 0.000532 & 56407.547455 &  113.5 & 0.003209 \\ 
55592.366837 &   37.5 & 0.000638 & 55989.154588 &   74.5 & 0.000553 & 56412.719525 &  114.0 & 0.000981 \\
55597.676834 &   38.0 & 0.000492 & 55999.871173 &   75.5 & 0.000480 & 56423.431712 &  115.0 & 0.001415 \\
\hline
\end{tabular}}

{\it Notes.} Integer and half-integer cycle numbers refer to primary and secondary eclipses, respectively. 
\end{table*}

\end{document}